\documentclass[preprint,3p,11pt,authoryear]{elsarticle}

\newif\ifblind
\blindfalse

\usepackage{amsmath}
\usepackage{amssymb}
\usepackage{booktabs}
\usepackage{graphicx}
\usepackage{multirow}
\usepackage{array}
\usepackage{longtable}
\usepackage{algorithm}
\usepackage{algpseudocode}
\usepackage{xcolor}
\usepackage{url}
\usepackage[T1]{fontenc}
\usepackage{textcomp}
\usepackage[colorlinks=true,linkcolor=blue,citecolor=blue,urlcolor=blue]{hyperref}

\newcommand{\aum}{\textsc{aum}}
\newcommand{\aumr}{\textsc{aum-r}}
\newcommand{\shellone}{Shell~1}
\newcommand{\shelltwo}{Shell~2}
\newcommand{\shellthree}{Shell~3}
\newcommand{\shellfour}{Shell~4}
\newcommand{\ci}[2]{$[#1,\,#2]$}

\begin{document}

\begin{frontmatter}

\makeatletter
\@ifundefined{ifblind}{\newif\ifblind\blindfalse}{}
\ifblind
  \@ifundefined{doubleblindtrue}{}{\doubleblindtrue}
\else
  \@ifundefined{doubleblindfalse}{}{\doubleblindfalse}
\fi
\makeatother

\title{Creating an Atomic User Model for Personality-Aware
       Large Language Model Interaction}

\ifblind
  \author[]{Anonymous author(s)}
\else
  \author[mech]{B. Sankar\corref{cor1,cor2}}
  \ead{sankarb@iisc.ac.in}
  \ead[orcid]{0000-0001-5844-6273}

  \author[cpdm]{Deepthika S}
  \ead{deepthikas123@gmail.com}
  \ead[orcid]{0009-0009-4386-6510}

  \author[cpdm]{Pawni Yadav}
  \ead{pawniyadav3435@gmail.com}

  \author[cpdm]{Amogh A S}
  \ead{amogh.setty07@gmail.com}

  \cortext[cor1]{Corresponding author: B. Sankar}
  \cortext[cor2]{Corresponding author email: sankarb@iisc.ac.in}

  \affiliation[mech]{organization={Department of Mechanical Engineering,
      Indian Institute of Science (IISc)},
    city={Bengaluru},
    postcode={560012},
    state={Karnataka},
    country={India}}

  \affiliation[cpdm]{organization={Department of Design and Manufacturing,
      Indian Institute of Science (IISc)},
    city={Bengaluru},
    postcode={560012},
    state={Karnataka},
    country={India}}

 \affiliation[corres]{organization={Corresponding Author: B. Sankar \\ Corresponding Author Email: sankarb@iisc.ac.in}}
\fi

\begin{abstract}
Assistants built on large language models are increasingly expected to
write as their user would write, and the dominant approach to that
expectation operates on a single channel: past preferences are
summarised out of conversation history and reinserted into the context
window. We argue that this inverts the natural order of inference.
Preferences are the task-dependent surface of an underlying personality
structure that is comparatively stable, so a system that stores only
preferences must relearn the person whenever the task changes. This
paper makes four contributions. First, we characterise a phenomenon we
call \emph{personality seepage}, in which the linguistic surface of a
prompt carries a personality fingerprint that the assistant mirrors
without any access to the personality that produced it. Second, we
propose the Atomic User Model (AUM), a structured, human-readable
representation that organises a person as a stable identity nucleus
surrounded by four interpretable shells covering psychological,
cognitive and experiential, behavioural, and social content, together
with cross-shell entries recording internal conflict and authenticity.
Third, we treat AUM as a retrieval index over a person rather than a
prompt prefix, and specify a personality-aware pipeline in which a task
classifier, a component-selection function and a budgeted retriever
return a small payload of fields at generation time. Fourth, we evaluate
the pipeline in a simulation study with sixteen language-model-simulated
participants, six style-sensitive tasks and three seeds, and in a
synthetic scaling study of the retriever itself. Retrieving eight fields
matched the style fidelity of injecting the entire user model while using
23\% of the context (211 tokens against 915), improved on flat
preference notes by 0.24 points on a five-point scale
($p < 0.001$, $d_z = 0.50$), and raised forced-choice identification of
the participant's own voice from 14.9\% to 42.7\% against 25\% chance.
Four pre-registered controls returned null, including the swarm
retriever against forward greedy, which locates the effect in the
structured representation rather than in the search over it. A
supplementary analysis shows the benefit is largest for participants
whose voice the un-personalised assistant reproduces worst
($\rho = -0.61$, $p = 0.013$), which is a design result rather than a
statistical one: personalisation is worth most to the people the default
serves least. We discuss what a budgeted payload buys for
privacy-by-architecture, and what a simulation can and cannot establish
about whether real people recognise themselves.
\end{abstract}




\begin{keyword}
user modelling \sep personalisation \sep personality \sep
large language models \sep retrieval augmented generation \sep
context budget \sep scrutability \sep privacy by architecture \sep
simulation study \sep artificial fish swarm algorithm
\end{keyword}

\end{frontmatter}

\section{Introduction}
\label{sec:intro}

Adaptive systems have modelled their users for as long as they have
adapted to them. The generic user modelling tradition in
human-computer interaction established that a user model is a reusable
component with an explicit schema, maintained separately from the
application that consumes it \citep{kobsa2001generic,brusilovsky2007user},
and personalised search built profiles from observed behaviour: interests
and activities harvested from a desktop index
\citep{teevan2005personalizing}, click and query histories evaluated at
scale \citep{dou2007largescale}, and short-term session signals combined
with long-term interest \citep{bennett2012modeling}. Assistants built on
large language models (LLMs) have inherited this design in a compressed
form. Deployed memory features distil past conversations into short
preference notes and reinsert them into later contexts, and the research
literature has followed with memory architectures
\citep{packer2023memgpt,zhou2024cognitive,liao2026profilememory},
per-user parameter-efficient adaptation \citep{tan2024democratizing},
and surveys that map the resulting space
\citep{tan2023usermodeling,zhang2025personalization}.

What these approaches share is not an algorithm but a unit of storage.
They store what a user has done, or has said they want. They do not
store the person from whom those wants emerge. The difference is easiest
to see when a stored preference goes stale.

\begin{quote}
A user plans a six-day conference trip to Paris and discusses cost of
living and itineraries with an assistant. Two weeks later the user
privately decides not to travel, and does not say so. A month after
that, while drafting a project timeline, the assistant blocks out the
original travel week as an absence.
\end{quote}

\noindent The system is not wrong about its source: it remembers the
preference faithfully, and the retrieved item was exactly the right
item. It is wrong about the person. It holds no representation of the
user as a decision maker whose plans get revised, so it has no basis on
which to flag a stale assumption. Retrieval quality is not the failure
here, and better retrieval over the same unit would not have prevented
it.

This paper argues that the missing layer is a different unit of storage.
The argument rests on a claim borrowed from personality psychology
rather than from information retrieval. Personality traits show
substantial rank-order stability across the lifespan, with
test-retest correlations rising from roughly 0.3 in childhood to 0.7 in
later adulthood \citep{roberts2000rankorder}, and mean-level change,
while real, is gradual and patterned \citep{roberts2006patterns,specht2011stability}.
Preferences, by contrast, are situational: they are what personality
produces when it meets a particular decision context. If the stable
layer is the one that generalises across tasks, then a system that
stores only the unstable layer is storing the wrong thing.

We make four contributions.

\begin{enumerate}
\item \textbf{Personality seepage.} We characterise a phenomenon in
  which a user's linguistic and behavioural patterns enter the
  assistant through the surface of their prompt, and the assistant
  mirrors that surface while having no access to the personality that
  produced it (Section~\ref{sec:seepage}). Seepage is why generic
  assistants can appear briefly personalised and then fail on exactly
  the parts of a task the user did not spell out.

\item \textbf{The Atomic User Model.} We propose a structured,
  human-readable representation that organises a person as a stable
  identity Nucleus surrounded by four shells covering psychological,
  cognitive and experiential, behavioural, and social content, plus
  cross-shell entries recording internal conflict, growth trajectory and
  an authenticity index (Section~\ref{sec:aum}). The metaphor is not
  decorative: it encodes two design commitments, differential stability
  and differential observability, that a flat schema cannot express.

\item \textbf{Personality-aware retrieval.} We treat \aum{} as an index
  over a person rather than a prompt prefix, and specify a pipeline in
  which a task classifier, a component-selection function $\sigma(t)$
  and a budgeted retriever return a payload of $k$ fields at generation
  time (Section~\ref{sec:pipeline}). The retrieval objective carries an
  explicit redundancy penalty, which is what makes payload selection a
  combinatorial problem rather than a sort.

\item \textbf{Two studies.} We evaluate the pipeline with sixteen
  LLM-simulated participants across six style-sensitive tasks and three
  seeds (Sections~\ref{sec:method} and \ref{sec:results}), and we study
  the retrieval problem on its own in a synthetic sweep over instrument
  size, budget and redundancy weight (Section~\ref{sec:scaling}).
\end{enumerate}

The headline empirical result is a context result. Eight retrieved
fields matched the style fidelity of injecting the entire 32-field user
model while using 23\% of the injected context, and improved on flat
preference notes by 0.24 points on a five-point scale. Forced-choice
identification of the participant's own voice rose from 14.9\% under
preference notes, which is not distinguishable from chance, to 42.7\%
against a 25\% baseline.

Four controls returned null, and we report them in full because
together they bound the claim. Selecting the right $k$ fields was not
distinguishable from selecting $k$ at random from the same admissible
pool; component selection contributed nothing beyond retrieval; the
swarm retriever neither beat forward greedy on generated style nor
reached a higher value of its own objective; and the redundancy term did
not earn its place against a relevance-only top-$k$. The scaling study
of Section~\ref{sec:scaling} explains why, and the explanation is
specific: at the operating point the simulation used, forward greedy is
already within 0.1\% of the exhaustive optimum, so there was nothing for
a population method to recover. What the positive results measure is the
structured representation, not the search over it.

A supplementary analysis produced the finding we think matters most for
a human-computer interaction audience. The benefit of personalisation
was largest exactly for those participants whose voice the
un-personalised assistant reproduced worst
(Spearman $\rho = -0.61$, $p = 0.013$). Personalisation is worth most to
the people the default serves least, which is a statement about who a
system is for rather than about how well it scores.

\begin{figure}[t]
\centering
\includegraphics[width=0.96\textwidth]{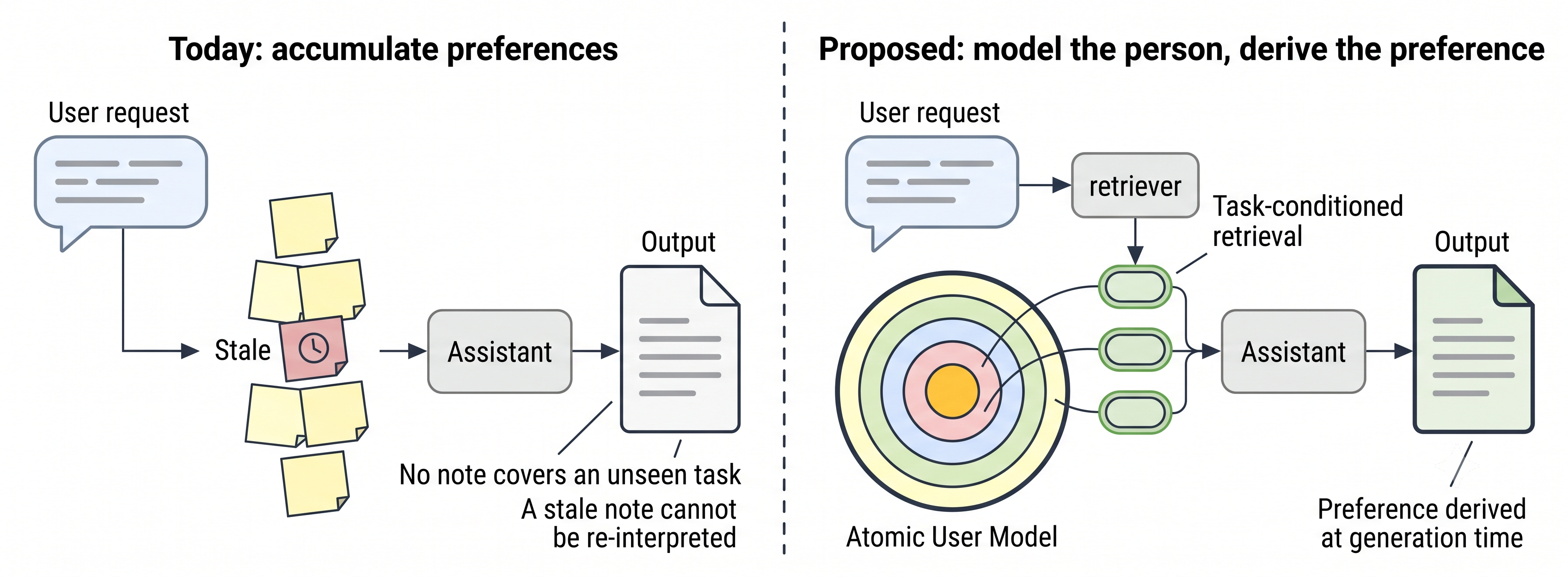}
\caption{The inversion this paper proposes. Left: current
  personalisation accumulates preferences observed at the surface and
  reinserts them, so an unseen task has no covering note and a stale
  note cannot be re-interpreted. Right: \aum{} stores the comparatively
  stable personality structure and derives the situational preference at
  generation time, retrieving only the fields the task needs.}
\label{fig:inversion}
\end{figure}

Figure~\ref{fig:inversion} states the inversion in one picture.

The remainder of the paper is organised as follows.
Section~\ref{sec:related} situates the work.
Section~\ref{sec:seepage} defines personality, preferences and persona
and introduces seepage. Section~\ref{sec:aum} specifies \aum{}.
Section~\ref{sec:pipeline} formalises personality-aware retrieval.
Sections~\ref{sec:method} and \ref{sec:results} present the simulation
study and its results. Section~\ref{sec:scaling} reports the scaling
study. Section~\ref{sec:discussion} discusses implications and
limitations, and Section~\ref{sec:conclusion} concludes. Appendices give
the full specification, every prompt used, the persona set and the
complete statistical tables.

\section{Related Work}
\label{sec:related}

\subsection{User models as components in adaptive systems}

The idea that a user model should be a separable, inspectable component
predates the systems this paper is about. \citet{kobsa2001generic}
surveyed generic user modelling systems and argued for a shared server
holding assumptions about the user, queried by applications rather than
duplicated inside them. \citet{brusilovsky2007user} developed the same
position for adaptive hypermedia and educational systems, where an
overlay model over a domain structure gives the adaptation something
explicit to reason about. Information retrieval developed a parallel but
distinct tradition, in which the user model is a statistical summary of
behaviour rather than a schema: term vectors and topical distributions
built from a desktop index \citep{teevan2005personalizing}, click and
query histories whose value for ranking turns out to vary sharply across
queries \citep{dou2007largescale}, and a decomposition of behavioural
evidence into short-term session signals and long-term interest
\citep{bennett2012modeling}. Diversity-aware ranking developed
alongside, first as maximal marginal relevance
\citep{carbonell1998mmr}, then as an explicit evaluation concern
\citep{clarke2008novelty} and an optimisation target
\citep{agrawal2009diversifying}.

\aum{} belongs to the first tradition and borrows from the second. It is
a schema, in Kobsa's sense, whose slots are declared in advance rather
than induced from data. What it takes from information retrieval is the
insight that a profile is only useful to the extent that a system can
select from it, and that selecting well means trading relevance against
redundancy.

\subsection{Memory and personalisation in LLM assistants}

Recent work recasts personalisation of LLM assistants as memory
management. MemGPT treats the model as an operating system with paged
memory \citep{packer2023memgpt}; generative agents combine memory,
reflection and planning to produce believable behaviour
\citep{park2023generative}; medical and lifelong assistants coordinate
short-term and long-term stores
\citep{zhang2024llmmedical,wang2024aipersona}. Closer to retrieval,
cognitive personalised search couples an LLM with an efficient memory
mechanism over query logs \citep{zhou2024cognitive}, profile memory
organises per-user summaries for reuse \citep{liao2026profilememory},
and agent-driven retrieval personalises search results directly
\citep{chhetri2026spark}. A separate branch adapts model parameters per
user rather than the context, through per-user parameter-efficient
fine-tuning \citep{tan2024democratizing}; a recent survey maps the
whole space \citep{zhang2025personalization}.

Evaluation has developed in step. LaMP \citep{salemi2024lamp} and its
long-form extension \citep{salemi2024longlamp} define personalisation as
task-level retrieval over user history. PersonaBench shows that
retrieval-augmented systems answer questions about private user data
poorly even when given direct document access
\citep{tan2025personabench}, and rating-prediction studies find that
LLMs capture user preferences but in opaque ways
\citep{kang2023llms,huang2025simulator}.

These systems answer how to store and fetch user-relevant content.
\aum{} answers a prior question: what the stored items should be, and
how they should be organised so that fetching a few of them is enough.
The two are complementary, and the pipeline of
Section~\ref{sec:pipeline} is deliberately built from standard
components so that the representation, rather than the machinery, is
what is under test.

\subsection{Context budgets and compression}

A structured user model of sixty fields cannot be injected on every
turn. The problem of fitting useful conditioning into a bounded context
has been attacked from the compression side: LLMLingua and its long-context
successor learn to drop tokens from a prompt while preserving downstream
performance \citep{jiang2023llmlingua,jiang2024longllmlingua}, and
selective-context methods prune low-information spans
\citep{li2023compressing}. Our approach is complementary and operates a
level up. Rather than compressing a fixed payload, we select which
fields enter the payload at all, conditioned on the task. The two could
compose: compression could shorten the eight fields that selection
returns.

\subsection{Personality: structure, stability, and its trace in language}

The Five-Factor Model provides the trait vocabulary most widely used in
computational work \citep{mccrae1992intro}, with modern instruments
including the BFI-2 \citep{soto2017bfi2}, public-domain item pools
\citep{goldberg2006ipip} and very brief measures for settings where
length is prohibitive \citep{gosling2003tipi}. Cross-cultural
generalisation of the factor structure has been examined but remains a
live concern \citep{mccrae2002crosscultural}. We include Myers-Briggs
type as a field inside \shellone{} because users often know their type,
while noting that its psychometric properties are weak: the type
dichotomies do not correspond to natural categories, and reliability
across retests is poor
\citep{pittenger2005cautionary,stein2019mbti,mccrae1989mbti}. This is
precisely the argument for a schema in which an inventory is one field
among many rather than the representation itself.

That personality leaves a measurable trace in language is well
established. \citet{pennebaker1999linguistic} showed that function-word
use is a stable individual difference; \citet{yarkoni2010personality}
related Big Five scores to word use across 100,000 words of blog text;
\citet{schwartz2013openvocab} moved from closed-vocabulary counting to
open-vocabulary differential language analysis over social media, and
the LIWC line of work supplies the standard closed-vocabulary
instrument \citep{tausczik2010liwc,boyd2017language}. Prediction of
traits from digital footprints reaches useful accuracy
\citep{kosinski2013private,park2015automatic}, sometimes exceeding human
judges given enough behavioural evidence \citep{youyou2015computerbased},
though meta-analysis places the typical correlation more modestly
\citep{azucar2018predicting}, and earlier attempts to model personality
from social media profile data illustrate how far the field has moved
\citep{goyal2022research}. Within LLM research, trait recognition
from text is feasible \citep{ji2023chatgpt,rao2023chatgpt} and implicit
persona can be modelled as a latent variable in dialogue
\citep{cho2022personalized}, but the resulting representations are trait
scores or latent vectors: neither modular nor inspectable.

The reliance on self-report in Section~\ref{sec:method} inherits a known
limitation. Self and other observers know different things about a
person, with the self better on internal states and observers better on
externally visible traits \citep{vazire2010soka,connelly2010other}, and
self-report has long been criticised as a substitute for observed
behaviour \citep{baumeister2007selfreports}. We return to this in
Section~\ref{sec:limitations}.

\subsection{Structured user representations and human digital twins}

The closest contemporary architecture is the General User Model, which
accumulates confidence-weighted natural-language propositions about a
user from screen observations \citep{shaikh2025gum}. Generative agents
grounded in self-report interviews simulate individual humans better
than demographic conditioning alone \citep{park2024generative}, which is
both the closest precedent for our representation and the precedent for
our evaluation method. Both populate a representation bottom up, from
observation or interview. \aum{} is top down: it specifies the slots
into which such observations should be organised, which is what makes a
component-selection function over shells definable at all.

The human digital twin literature pursues a superficially similar goal
from an engineering direction, and a recent survey maps its scope
\citep{lin2024digitaltwin}. The personal informatics tradition in HCI
supplies the complementary user-centred account of what it is like to
live with a model of oneself: staged models of collection and reflection
\citep{li2010stagebased}, the lived-informatics revision that treats
tracking as episodic rather than linear \citep{epstein2015lived}, and
evidence on how people without prior self-tracking experience engage
with their own data \citep{rapp2016personal}. \aum{} differs from a
digital twin in purpose. It is built for style-sensitive task
completion, not identity substitution, and Section~\ref{sec:discussion}
argues that the distinction has consequences the twin framing obscures.

\subsection{Scrutability, control, and the privacy of a user model}

If a system holds a model of a person, the person should be able to read
it. \citet{kay2012scrutinize} set out the drivers and principles for
personalised systems that users can scrutinise and control, building on
earlier work on scrutable adaptive hypertext
\citep{czarkowski2002scrutable} and on interfaces for inspecting
semantic user models \citep{bakalov2010introspective}.
\citet{cramer2008transparency} found that transparency affected
acceptance of a content-based recommender, and \citet{harper2015control}
showed that giving users direct control over their recommendations
changed both behaviour and satisfaction. \citet{jeromela2022scrutability}
extends the question to intelligent personal assistants specifically.
Mental-model work supplies the reason this matters: users arrive with
expectations of conversational agents that the systems do not meet
\citep{luger2016badpa,mahmood2025breakdowns}, and their models of what
an agent knows shape how they use it \citep{gero2020mentalmodels,ngo2020mentalmodels}.

The privacy dimension is sharper for \aum{} than for a preference list,
because the content is more intimate. Contextual integrity supplies the
governing frame: what matters is not secrecy but whether an information
flow matches the norms of the context it came from
\citep{nissenbaum2011contextual}. Empirically, users of LLM-based
conversational agents disclose a great deal and reason about the risks
only partially \citep{zhang2024fairgame}, and language models can leak
personal information present in training data
\citep{huang2022leaking}. On-device and federated approaches offer one
architectural answer \citep{chen2019federated}. Our position, developed
in Sections~\ref{sec:pipeline} and \ref{sec:ethics}, is that a budgeted
payload converts local residence from a policy promise into a
measurable property: if eight fields suffice, the other twenty-four
never leave the device.

\subsection{Writing assistance and the user's voice}

The application setting for this work is writing assistance, where the
question of whose voice appears in the output is not incidental.
CoAuthor documented human-LLM collaborative writing at scale
\citep{lee2022coauthor}; Wordcraft explored story writing with a model
in the loop \citep{yuan2022wordcraft}; a recent design space synthesises
the field \citep{lee2024designspace} and a companion study aligns
research directions with what writers actually ask for
\citep{reza2025cowriting}. The finding that most directly motivates
\aum{} is \citet{jakesch2023opinionated}: co-writing with an opinionated
language model shifts the user's own expressed views. If a default
assistant voice can move what a person says, then a system that instead
conditions on a model of that person is not merely a convenience
feature.

\subsection{Selecting a subset under a budget}

Choosing $k$ fields to maximise a payload score with a redundancy
penalty is a constrained subset-selection problem. When such an
objective is monotone submodular, the forward greedy algorithm carries
the classical $(1 - 1/e)$ guarantee \citep{nemhauser1978analysis}, which
is why greedy is the right baseline rather than a straw man. Our
objective, defined in Section~\ref{sec:pipeline}, uses a mean rather
than a sum and subtracts a mean pairwise similarity, so it is not
submodular in general and the guarantee does not transfer; the empirical
question of how close greedy comes is therefore live, and
Section~\ref{sec:scaling} answers it directly.

Population-based metaheuristics are widely applied to subset selection,
particularly feature selection
\citep{xue2016evolutionary,nguyen2020swarmfs}. The artificial fish swarm
algorithm \citep{li2002afsa} models a population of agents executing
prey, swarm and follow behaviours, and has accumulated a substantial
literature of variants and applications
\citep{neshat2014afsa,pourpanah2023afsa}. We use it, and we report that
at the scale our simulation ran it did not earn its cost. We think
reporting that is more useful than dropping the comparison, and
Section~\ref{sec:scaling} identifies the regime in which the answer
changes.

\subsection{Simulated participants and model-based judgement}

Using language models in place of human participants is now a small
field with a sharp internal debate. \citet{argyle2023outofone} showed
that conditioning on demographic backstories reproduces aggregate
patterns in survey data, and \citet{park2024generative} showed that
grounding agents in structured self-report improves individual-level
simulation. The critical literature is equally developed:
\citet{bisbee2024synthetic} document instability and misestimation in
synthetic survey responses, and \citet{wang2025flatten} show that
replacing human participants can misportray and flatten identity groups.
We adopt simulation as a feasibility method with explicit limits, stated
in Section~\ref{sec:whysimulate}, and we do not present any result here
as evidence about human users.

Model-based judgement carries its own hazards, all of which we control
for by construction. Judges exhibit position bias
\citep{wang2024notfair}, self-preference for their own generations
\citep{panickssery2024selfpreference}, and imperfect but usable
agreement with human raters
\citep{zheng2023judging,chiang2023alternative,liu2023geval}.
Section~\ref{sec:measures} describes the specific controls: a judge from
a different model family than the generator, per-trial option shuffling,
and a judge that never sees the user model it is implicitly evaluating.

For the automatic style measure we use authorship representations rather
than semantic embeddings. LUAR learns universal authorship
representations by contrastive training over authors
\citep{riverasoto2021luar}, extending earlier invariant representations
of social-media users \citep{andrews2019invariant}. The distinction
matters here: a semantic embedding scores any two deadline-extension
emails as similar because they are about the same thing, which is
exactly the wrong invariance. Work on what authorship representations
actually encode \citep{wang2023authorship,wegmann2022sameauthor}, on
making style embeddings interpretable \citep{patel2023interpretable},
and on robustness \citep{man2024counterfactual} informs how we read the
resulting numbers. Style-transfer evaluation supplies the broader
methodological caution that automatic style metrics are easy to
misreport \citep{mir2019evaluating,ostheimer2023standardization}, and
fine-grained linguistic control offers an alternative route to the same
goal \citep{alhafni2024finegrained}.

\subsection{Cognitive architectures}

Finally, the design philosophy behind a fixed, modular schema comes from
the cognitive architecture tradition. ACT-R
\citep{anderson2004integrated} and Soar \citep{laird2012soar} are
long-standing arguments that explicit structure buys something that
implicit competence does not, and \citet{sun2024cogarch} makes the case
that such architectures and LLMs are complementary rather than
competing. \aum{} applies the philosophy not to cognition in general but
to the narrower problem of representing the user inside an LLM-mediated
system.

\section{Personality, Preferences, Persona, and Seepage}
\label{sec:seepage}

\subsection{Three terms used in a specific way}

Work on personalisation uses \emph{personality}, \emph{preference} and
\emph{persona} loosely and often interchangeably. Because the argument
of this paper is precisely about the relation between them, we fix the
terms.

\begin{description}
\item[Personality] is the structured set of stable, task-independent
  traits spanning the cognitive, affective, behavioural and
  social-contextual dimensions of a person. Stability is meant in the
  rank-order sense established empirically by
  \citet{roberts2000rankorder}: a person's position relative to others
  changes slowly, even where absolute levels drift with age and life
  events \citep{roberts2006patterns,specht2011stability}.

\item[Preferences] are the task-dependent outputs of personality applied
  to a particular decision context. ``Prefers bullet points in status
  updates'' is a preference. It is downstream of conscientiousness, of a
  professional register learned in a particular workplace, and of a
  belief about what respects a reader's time; it is not itself any of
  those things.

\item[Persona] is the externally perceived projection of personality
  through one channel: a professional profile, a social media account, a
  conference biography. Goffman's account of self-presentation
  \citep{goffman1959presentation} is the reference point, and the key
  property is that a persona is a selective and audience-dependent
  rendering rather than a compressed copy.
\end{description}

\noindent \aum{} models personality. Preferences are derived from
personality and context at generation time. Personas are situational
projections and are represented, when they are represented at all, as
fields inside \shellfour{}.

The ordering matters for a practical reason. If preferences are the
stored unit, then every new task requires either a stored preference
that happens to cover it or a guess. If personality is the stored unit,
a preference for an unseen task is derivable, and a stale stored
preference can be re-interpreted against a structure that did not
change. The Paris example of Section~\ref{sec:intro} is exactly a case
where the second operation was needed and unavailable.

\subsection{Personality seepage}

\begin{table}[t]
\centering
\caption{Personality seepage on a single task: an email requesting a
  deadline extension from a colleague. Three users phrase the same
  underlying request differently and the assistant mirrors each surface.
  The third column is the point: what the generic response misses is in
  each case a property of the user, not of the prompt.}
\label{tab:seepage}
\small
\renewcommand{\arraystretch}{1.25}
\begin{tabular}{@{}p{0.16\textwidth} p{0.36\textwidth} p{0.40\textwidth}@{}}
\toprule
\textbf{User} & \textbf{Prompt surface (abridged)} &
\textbf{What the generic response misses} \\
\midrule
U1: organised, meticulous &
``Current deadline Friday 5pm; requested Monday 5pm. Reason: data
validation overran scope. Offer: draft by Saturday.'' &
The structure is right and the dates are carried through, but the reply
drops the acknowledgement of the reviewer's effort that U1 includes in
every message they send. Nothing in the prompt asks for it. \\
U2: empathetic, verbose &
``Could you help me write a friendly note to a colleague? I feel a bit
bad asking, but maybe we could push the Q3 deadline.'' &
The reply amplifies the apology into self-deprecation. U2's phrasing
invites it; U2's stated values reject it. Mirroring the surface
overshoots the person. \\
U3: concise, informal &
``email to ask for more time on a report, colleague, Q3 stuff. make it
not too formal.'' &
Brevity is preserved and so is the omission: no date and no reason,
the two things U3 habitually leaves out of a first draft and reliably
needs by the second. \\
\bottomrule
\end{tabular}
\end{table}

Table~\ref{tab:seepage} illustrates a phenomenon we call
\emph{personality seepage}. When a person writes a prompt, their
linguistic and behavioural patterns go into the prompt with them, and
the model amplifies those patterns in its output. That the trace is
there is not in doubt: function-word use is a stable individual
difference \citep{pennebaker1999linguistic}, and personality is
recoverable from text at useful accuracy
\citep{yarkoni2010personality,schwartz2013openvocab}.

The trouble is what the assistant does with it. It matches the surface
without access to what produced the surface. Three consequences follow,
and each appears in Table~\ref{tab:seepage}.

\begin{enumerate}
\item \textbf{Gaps are not filled.} Whatever the user habitually omits
  from a prompt is also absent from the output, because the model has no
  independent source for it. U3's missing deadline is the example.

\item \textbf{Surface features are over-extrapolated.} A hedging
  register in the prompt becomes a hedging register in the output,
  amplified, because the model treats the surface as the target rather
  than as evidence. U2's apology is the example.

\item \textbf{Stable habits are not reproduced.} Regularities the user
  applies across every instance of a genre, which never appear in any
  single prompt because the user assumes them, are invisible. U1's
  crediting of collaborators is the example.
\end{enumerate}

\noindent Seepage also explains a common subjective experience: a
generic assistant feels briefly personalised, because it does echo
something real about the user, and then fails on exactly the parts of a
task the user did not spell out. The echo is real; the model beneath it
is absent.

\subsection{Why accumulating preferences treats the symptom}

The remedy that deployed systems offer for seepage is to accumulate
preference notes from past conversations and reinsert them. This is a
reasonable engineering response to the observation that the assistant
forgets, and it does address forgetting. It does not address seepage,
because the notes are observed at the level of preferences while the
cause of variation across users sits one layer below.

Two properties of accumulated notes make this concrete. First, they are
\emph{topically bound}: a note distilled from a conversation about
choosing a laptop is about laptops, and generalises to an unrelated
writing task only by accident. Section~\ref{sec:results} gives an
empirical form of this observation, in that the preference-notes
condition was not distinguishable from chance on forced-choice
identification of the user's own voice. Second, they are
\emph{uninterpretable against each other}: a flat list has no structure
that would let a system notice that two notes conflict, or that one has
gone stale, or that a third is a surface expression of the same
underlying disposition as the first two. Structure is what makes those
operations definable, which is the argument for a schema and the subject
of the next section.

\section{The Atomic User Model}
\label{sec:aum}

\subsection{The metaphor, and what it commits us to}

\begin{figure}[t]
\centering
\includegraphics[width=0.62\textwidth]{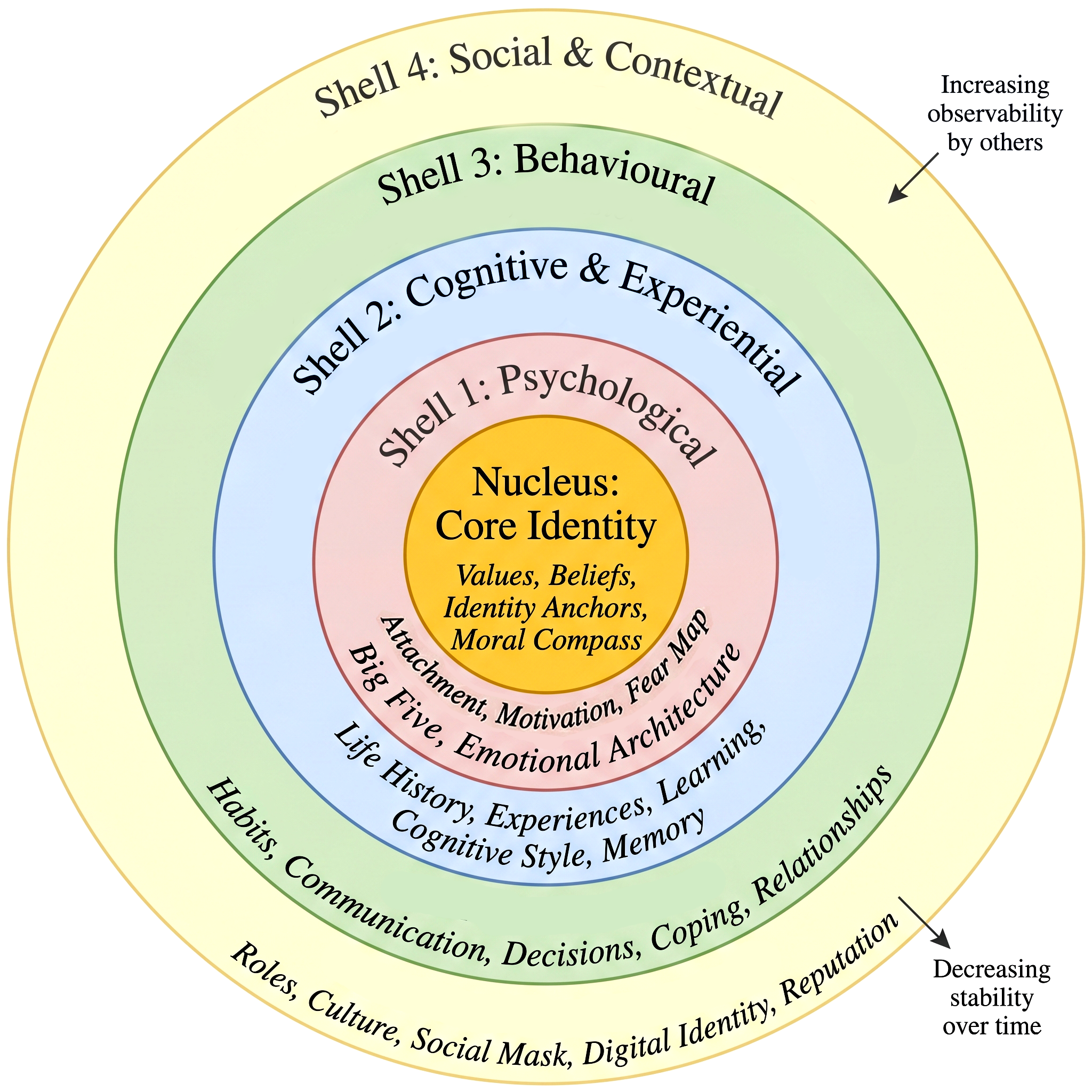}
\caption{The Atomic User Model. A Nucleus of core identity is surrounded
  by four shells. Example fields per shell are shown in italics.
  Stability decreases and observability by others increases from the
  centre outwards.}
\label{fig:aum-arch}
\end{figure}

\aum{} represents a person as a Nucleus surrounded by four shells, on
the metaphor of an atom with a stable centre and reactive outer layers
(Figure~\ref{fig:aum-arch}). The metaphor is doing work rather than
decorating, and it commits the representation to two properties that a
flat schema cannot express.

\begin{description}
\item[Differential stability.] Identity is stable while behaviour
  adapts. The Nucleus changes on the timescale of years, the inner
  shells on months, the outer shells on weeks. A representation that
  treats every field as equally durable will either refresh core values
  as often as it refreshes a communication habit, which is wasteful and
  invasive, or refresh a communication habit as rarely as it refreshes
  core values, which makes it wrong.

\item[Differential observability.] A stranger sees the outermost shell,
  a colleague sees more, a close friend or family member sees much more,
  and the Nucleus is visible only to the self. This layering echoes the
  onion model of social penetration theory
  \citep{altman1973social,carpenter2016social}, and it has a direct
  computational consequence: the fields most useful for deep
  personalisation are exactly the fields most costly to expose. A
  representation that does not encode the gradient cannot reason about
  the trade-off.
\end{description}

\noindent Both properties bear on Section~\ref{sec:pipeline}. The first
motivates differential update rates as a field attribute. The second is
why a budgeted payload is a privacy mechanism and not only an efficiency
one.

\begin{figure}[t]
\centering
\includegraphics[width=0.72\textwidth]{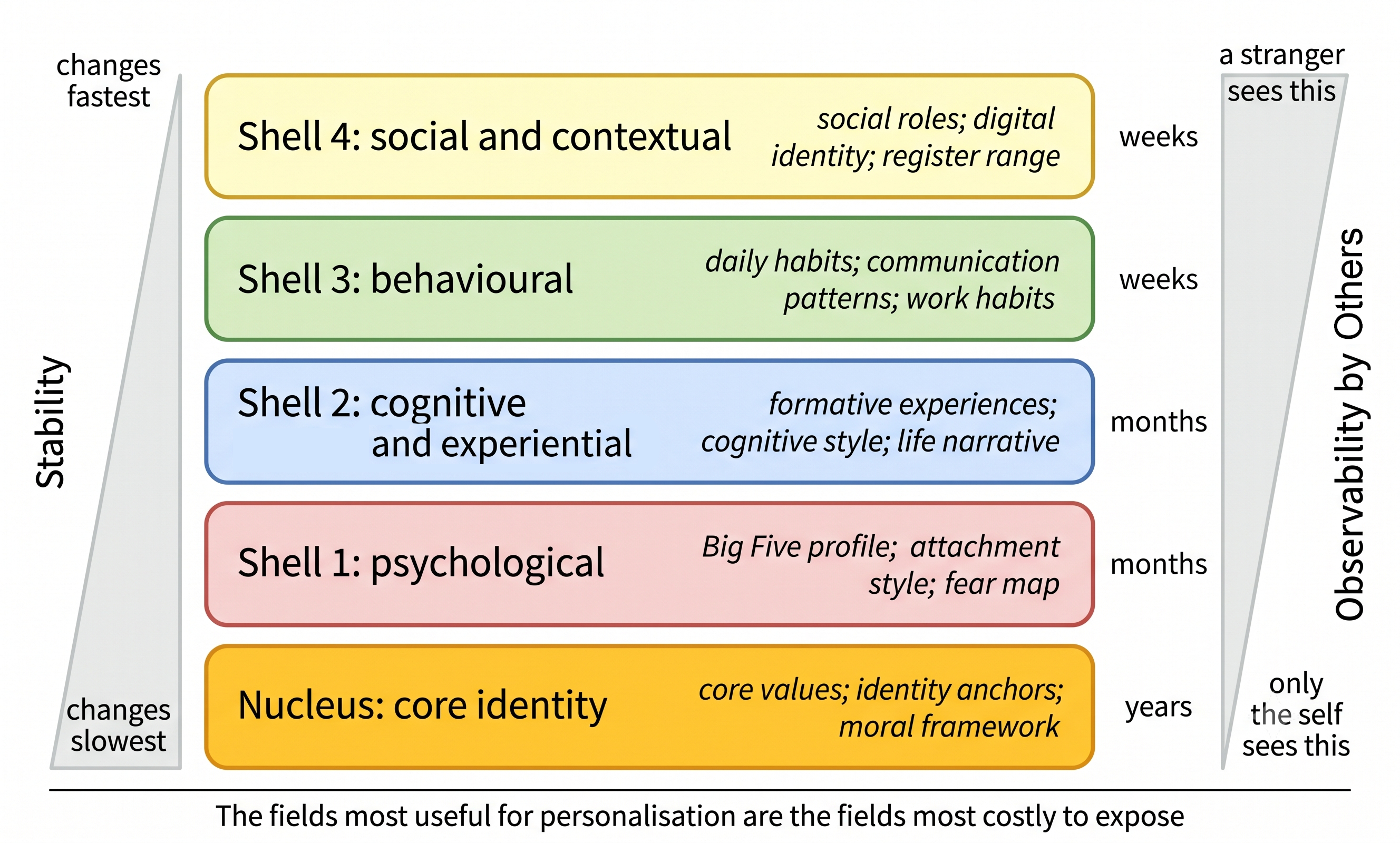}
\caption{Differential stability across tiers. The nominal update period
  $\tau$ falls from years at the Nucleus to weeks at \shellfour{}, while
  observability by others rises in the opposite direction. A schema that
  does not encode this gradient cannot express a refresh policy, and
  cannot reason about the cost of exposing a field.}
\label{fig:updaterates}
\end{figure}

\subsection{Formal definition}

The \aum{} of a user $u$ is a set of typed fields
\begin{equation}
A_u = \{f_1, f_2, \ldots, f_n\},
\label{eq:aum}
\end{equation}
where each field $f_i$ is a tuple
\begin{equation}
f_i = \langle \, \mathrm{shell}(f_i),\; \mathrm{name}(f_i),\;
                \mathrm{value}(f_i),\; \tau(f_i) \, \rangle .
\label{eq:field}
\end{equation}
Here $\mathrm{shell}(f_i) \in \mathcal{S}$ with
\begin{equation}
\mathcal{S} = \{\text{Nucleus},\, \shellone,\, \shelltwo,\,
                \shellthree,\, \shellfour,\, \text{CrossShell}\},
\end{equation}
$\mathrm{name}(f_i)$ is a fixed slot name drawn from the specification,
$\mathrm{value}(f_i)$ is a short natural-language string, and
$\tau(f_i)$ is a nominal update period. Three properties of this
definition are load-bearing.

\begin{enumerate}
\item \textbf{Values are natural language, not vectors.} A field reads
  ``I credit whoever reviewed a draft before I ask them for anything
  else'', not a coordinate. This is what makes the model scrutable in
  the sense of \citet{kay2012scrutinize}: the user can read, edit or
  delete any field without an interpretation layer.

\item \textbf{Slots are declared in advance.} The set of
  $(\mathrm{shell}, \mathrm{name})$ pairs is fixed by the
  specification, so two users' models are structurally comparable and a
  selection function over shells is definable. This is the property that
  distinguishes \aum{} from bottom-up accumulations of free-form
  propositions \citep{shaikh2025gum}.

\item \textbf{Update rate is a field attribute.} $\tau$ ranges over
  years for Nucleus fields, months for \shellone{} and \shelltwo{}, and
  weeks for \shellthree{} and \shellfour{}. We do not specify an update
  mechanism in this paper and return to the omission in
  Section~\ref{sec:limitations}.
\end{enumerate}

\subsection{The five tiers}

\paragraph{Nucleus: core identity.}
The Nucleus holds what changes most slowly and what, if changed, makes
the person a different agent: core values, fundamental beliefs about
self, others and life, identity anchors of the form ``I am the kind of
person who \ldots'', a moral framework, existential orientation, and
life purpose. The framing follows Erikson's account of identity
\citep{erikson1968identity}. Operationally, a Nucleus field is one whose
violation the user would describe as being untrue to themselves rather
than as an inconsistency.

\paragraph{\shellone{}: psychological core.}
\shellone{} explains how the Nucleus is expressed. It holds personality
structure, incorporating but not reducing to inventories such as the
Five-Factor Model \citep{mccrae1992intro,soto2017bfi2}; emotional
architecture; the attachment system \citep{bowlby1969attachment};
motivational drivers; a fear map; and stress response. Trait
inventories, including Myers-Briggs type where a user knows it, are
fields inside \shellone{}, not substitutes for the model. Given the
psychometric objections to type-based instruments
\citep{pittenger2005cautionary,stein2019mbti}, a schema that demotes an
inventory to one field among thirty-two is a feature rather than a
compromise.

\paragraph{\shelltwo{}: cognitive and experiential.}
\shelltwo{} is life history as it bears on cognition: formative
experiences, a record of difficult episodes, learning style, cognitive
style, the personal narrative the user tells about their own life, and
knowledge base. It is the tier that explains why a given \shellone{}
profile produces the particular \shellthree{} behaviours it does in this
particular person.

\paragraph{\shellthree{}: behavioural patterns.}
\shellthree{} holds daily habits, communication patterns, decision
behaviour, coping mechanisms, relationship behaviour and work habits.
This is the tier most often mistaken for personality in deployed
systems, because it is the tier that observation reaches. It is the
output of the inner shells, not their cause, which is why a system that
models only \shellthree{} can predict what a user did and not what they
would do in a situation it has not seen.

\paragraph{\shellfour{}: social and contextual.}
\shellfour{} is the interface to the world: social roles, cultural
conditioning, digital identity, reputation, the social mask a person
presents \citep{goffman1959presentation}, and register range. Most LLM
personalisation today operates on the outer half of \shellfour{},
because that is the part chat history exposes.

\paragraph{Cross-shell integration.}
Three entries sit across tiers rather than inside one.
\emph{Internal conflicts} record cases where a \shellthree{} behaviour
contradicts a Nucleus value, which is the machinery that lets a system
notice that a stored preference is out of character rather than merely
old. \emph{Growth trajectory} records the direction of deliberate
change. The \emph{authenticity index} measures alignment between what
the inner shells hold and what the outer shells present, and is the
field a system would consult before, for example, matching a user's
professional register in a private message.

\subsection{The full specification and the reduced instrument}

The full specification holds approximately sixty fields across the five
tiers plus cross-shell entries; Appendix~\ref{app:spec} lists it. The
studies in this paper use a reduced 32-field instrument, six fields per
tier plus two cross-shell entries, shown in Table~\ref{tab:instrument}.
The reduction was made for a practical reason: a simulated participant
completing sixty fields in character produces noticeably thinner
answers per field than one completing thirty-two, and we preferred
fewer, denser fields. Section~\ref{sec:scaling} examines what happens to
retrieval as the instrument grows towards the full size.

\begin{table}[t]
\centering
\caption{The reduced 32-field instrument used in both studies. Six
  fields per tier plus two cross-shell entries. Update period $\tau$ is
  the nominal refresh interval for the tier.}
\label{tab:instrument}
\small
\renewcommand{\arraystretch}{1.2}
\begin{tabular}{@{}p{0.20\textwidth} p{0.09\textwidth} p{0.63\textwidth}@{}}
\toprule
\textbf{Tier} & \textbf{$\tau$} & \textbf{Fields} \\
\midrule
Nucleus &
years &
Core Values; Fundamental Beliefs; Identity Anchors; Moral Framework;
Existential Orientation; Life Purpose \\
\shellone{} psychological &
months &
Big Five Profile; Emotional Architecture; Attachment Style;
Motivational Drivers; Fear Map; Stress Response \\
\shelltwo{} cognitive &
months &
Formative Experiences; Difficult Episodes; Learning Style;
Cognitive Style; Life Narrative; Knowledge Base \\
\shellthree{} behavioural &
weeks &
Daily Habits; Communication Patterns; Decision Behaviour;
Coping Mechanisms; Relationship Behaviour; Work Habits \\
\shellfour{} social &
weeks &
Social Roles; Cultural Conditioning; Digital Identity; Reputation;
Social Mask; Register Range \\
CrossShell &
months &
Internal Conflicts; Authenticity Index \\
\bottomrule
\end{tabular}
\end{table}

\subsection{Two design commitments}

\paragraph{Local residence.}
\aum{} contains inner-shell content of a kind that has no analogue in a
preference list. It should reside on the user's device or in a
user-controlled vault, never in a third-party log. We state this as an
architectural commitment rather than a policy preference because the
retrieval design of Section~\ref{sec:pipeline} makes it enforceable:
only the selected payload need ever leave the device, and
Section~\ref{sec:results} measures how small that payload can be
without loss. Contextual integrity \citep{nissenbaum2011contextual}
gives the frame for what such a boundary is for.

\paragraph{User authorship.}
Every field is a human-inspectable string that the user can read, edit
or delete. This places \aum{} in the scrutable-personalisation tradition
\citep{kay2012scrutinize,czarkowski2002scrutable,bakalov2010introspective}
and answers the interpretability objection to trait-vector user models
\citep{ji2023chatgpt,kang2023llms}. A budgeted payload strengthens the
commitment: eight short strings are few enough to show the user before
they are sent, where thirty-two are not.

\section{Personality-Aware Retrieval}
\label{sec:pipeline}

\aum{} is a representation of a person; it requires a pipeline to be
useful. This section specifies one, in which \aum{} is treated as an
index over a person and queried under a context budget rather than
injected wholesale.

\begin{figure}[t]
\centering
\includegraphics[width=0.98\textwidth]{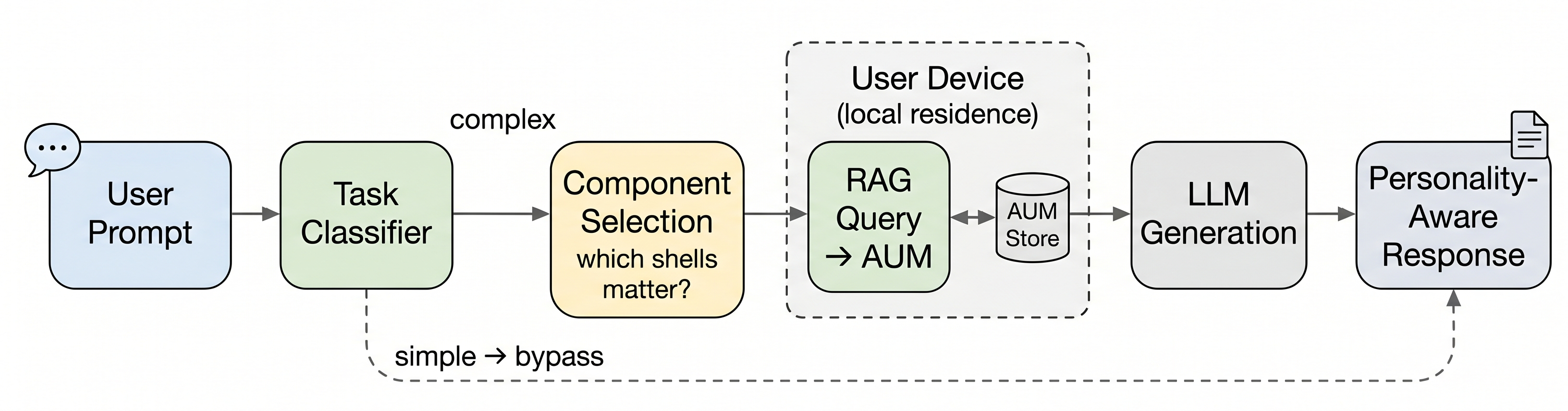}
\caption{The personality-aware retrieval pipeline. A task classifier
  routes simple requests around the user model entirely. Complex
  requests pass through component selection, which restricts the search
  to the shells the task type touches, and then through a budgeted
  retrieval that returns $k$ fields. The \aum{} store remains on the
  user's device; only the payload leaves it.}
\label{fig:pipeline}
\end{figure}

\subsection{The four stages}

Given a prompt $q$ from user $u$, the pipeline of
Figure~\ref{fig:pipeline} proceeds as follows.

\paragraph{(i) Task identification.}
A classifier assigns $q$ a task type $t$ and a complexity label. A
request is \emph{simple} if a correct answer is the same regardless of
who asked, such as a unit conversion or a factual lookup, and
\emph{complex} if the form of an acceptable answer depends on who is
asking. Simple requests bypass the user model, which matters both for
latency and for disclosure: a system that consults an intimate user
model to convert kilometres into miles is leaking for no benefit.
Section~\ref{sec:robustness} reports how accurately this stage runs.

\paragraph{(ii) Component selection.}
A selection function $\sigma: \mathcal{T} \rightarrow 2^{\mathcal{S}}$
restricts the search space to the shells that a task type touches. The
mapping used throughout is given in Table~\ref{tab:sigma}; CrossShell is
always admissible. The admissible pool is
\begin{equation}
\mathrm{pool}(t) = \{\, i : \mathrm{shell}(f_i) \in
                     \sigma(t) \cup \{\text{CrossShell}\} \,\}.
\label{eq:pool}
\end{equation}
If $|\mathrm{pool}(t)| < k + 4$ the pool is topped up with the most
query-relevant fields from outside $\sigma(t)$. Without this
top-up a large $k$ would silently collapse to ``every field $\sigma(t)$
admits'', which would make the $k = 16$ condition a different
experiment from the $k = 4$ and $k = 8$ conditions rather than a third
point on one curve.

\begin{table}[t]
\centering
\caption{The component-selection function $\sigma(t)$. CrossShell is
  admissible for every complex task type. Factual requests are routed
  around the user model entirely, which is what the empty set denotes.}
\label{tab:sigma}
\small
\renewcommand{\arraystretch}{1.15}
\begin{tabular}{@{}l l@{}}
\toprule
\textbf{Task type $t$} & \textbf{$\sigma(t)$} \\
\midrule
communication          & \shellthree{}, \shellfour{} \\
self-presentation      & \shellfour{}, Nucleus, \shellone{} \\
relational-evaluative  & \shellone{}, \shellthree{}, Nucleus \\
planning-decision      & \shellthree{}, \shellone{}, \shelltwo{} \\
explanatory            & \shelltwo{}, \shellthree{} \\
creative               & \shellone{}, \shelltwo{} \\
factual (simple)       & $\emptyset$ (bypass) \\
\bottomrule
\end{tabular}
\end{table}

\paragraph{(iii) Retrieval.}
A retriever selects a payload $P(q, u, k) \subseteq A_u$ with
$|P| = k \ll n$ by maximising the objective of
Section~\ref{sec:objective} over $\mathrm{pool}(t)$.

\paragraph{(iv) Generation.}
The model is conditioned on $(q, P)$. The payload is inserted as a
delimited block ahead of the user's request; the exact template is in
Appendix~\ref{app:prompts}.

\subsection{The payload objective}
\label{sec:objective}

Let $\mathbf{f}_i$ denote an embedding of field $i$ and $\mathbf{q}$ an
embedding of the query, both $\ell_2$-normalised. The obvious objective
is mean relevance,
\begin{equation}
F_{\mathrm{rel}}(S) = \frac{1}{|S|}\sum_{i \in S}
                      \cos(\mathbf{f}_i, \mathbf{q}),
\label{eq:rel}
\end{equation}
but this is separable: it decomposes into per-item scores, so its exact
maximiser over a pool is obtained by sorting, and no search is required
or useful. Under a context budget, mean relevance is also the wrong
objective. Eight near-duplicate fields waste seven slots, and a
budgeted payload should therefore trade relevance against diversity in
the manner of maximal marginal relevance \citep{carbonell1998mmr} and
of result diversification more broadly
\citep{clarke2008novelty,agrawal2009diversifying}. We use
\begin{equation}
F(S) \;=\; \underbrace{\frac{1}{|S|}\sum_{i \in S}
      \cos(\mathbf{f}_i, \mathbf{q})}_{\text{relevance}}
      \;-\; \lambda \,
      \underbrace{\frac{1}{|S|\,(|S|-1)}\sum_{\substack{i, j \in S \\ i \neq j}}
      \cos(\mathbf{f}_i, \mathbf{f}_j)}_{\text{redundancy}},
\label{eq:objective}
\end{equation}
with $\lambda \geq 0$ the redundancy weight, and the payload is
\begin{equation}
P(q, u, k) \;=\; \arg\max_{\substack{S \subseteq \mathrm{pool}(t) \\ |S| = k}}
                 F(S).
\label{eq:argmax}
\end{equation}

At $\lambda = 0$, Equation~\ref{eq:objective} reduces to
Equation~\ref{eq:rel} and the problem is a sort. For $\lambda > 0$ the
redundancy term couples the items and the problem becomes
combinatorial, with $\binom{|\mathrm{pool}(t)|}{k}$ candidate payloads
and no closed form. We use $\lambda = 0.15$ throughout, and
Section~\ref{sec:scaling} reports what changes as $\lambda$ varies.

Two remarks on the structure of the objective. First, because it uses a
mean rather than a sum and subtracts a mean pairwise similarity, it is
not monotone submodular in general, so the classical greedy guarantee of
\citet{nemhauser1978analysis} does not transfer; how close greedy comes
is an empirical question, answered in Section~\ref{sec:scaling}. Second,
the redundancy term is defined over the payload only, not over the
user's history, which is what distinguishes it from novelty in a search
result list.

\subsection{Retrievers}

Four retrievers are implemented, plus exhaustive search wherever the
search space is small enough to enumerate.

\paragraph{Dense top-$k$.}
Take the $k$ highest-relevance fields, ignoring redundancy. This is the
retrieval-augmented generation default and is the exact maximiser of
Equation~\ref{eq:rel}. Cost: one objective evaluation.

\paragraph{Forward greedy.}
Start from $S = \emptyset$ and repeatedly add the field that most
increases $F$. Cost: $O(k \cdot |\mathrm{pool}|)$ objective
evaluations. Strong, cheap and deterministic; it is the baseline that
matters.

\paragraph{Random $k$.}
Draw $k$ fields uniformly from $\mathrm{pool}(t)$. This is a control
rather than a method. Without it, an improvement for a retrieved payload
is equally consistent with the hypothesis that any $k$ short strings
about the user help, which is not the claim under test.

\paragraph{Artificial fish swarm.}
A population of $p$ fish, each a $k$-subset of the pool, evolves for $T$
iterations under three behaviours drawn from
\citet{li2002afsa}. Algorithm~\ref{alg:afsa} gives the procedure. In
\emph{prey}, a fish attempts up to $\mathit{try}$ random local swaps and
accepts the first improvement, falling back to a random walk. In
\emph{swarm}, a fish moves toward the centre of its visual
neighbourhood if that centre is better and the neighbourhood is not
crowded. In \emph{follow}, it moves toward the best neighbour under the
same conditions. Distance between fish is
$d(a,b) = 1 - |a \cap b| / k$, and a move toward a target replaces the
lowest-relevance member of the current subset with a member of the
target. The global best is retained, so the returned payload never
degrades across iterations. In the simulation study we use
$p = 12$, $T = 8$, visual $= 6$, $\mathit{try} = 4$ and crowd factor
$0.75$; Section~\ref{sec:scaling} shows that this fixed budget is the
source of the algorithm's weakness rather than the behaviours are.

\begin{algorithm}[!ht]
\caption{Artificial fish swarm over $k$-subsets}
\label{alg:afsa}
\begin{algorithmic}[1]
\Require pool $\mathcal{P}$, budget $k$, objective $F$, population $p$,
         iterations $T$, visual $v$, tries $\mathit{try}$, crowd $\delta$
\Ensure payload $S^\star$ with $|S^\star| = k$
\State $\mathrm{Pop} \gets \{\,S_1, \ldots, S_p\,\}$, each a uniform
       random $k$-subset of $\mathcal{P}$
\State $S^\star \gets \arg\max_{S \in \mathrm{Pop}} F(S)$
\For{$\iota = 1$ \textbf{to} $T$}
  \ForAll{$S \in \mathrm{Pop}$}
    \State $C \gets \{\, \Call{Prey}{S} \,\}$
    \State $N \gets \{\, S' \in \mathrm{Pop} : S' \neq S,\;
           1 - |S \cap S'|/k \le v/k \,\}$
    \If{$N \neq \emptyset$ \textbf{and} $|N|/p \le \delta$}
      \State $S_c \gets$ the $k$ most frequent members across $N$
             \Comment{swarm centre}
      \If{$F(S_c) > F(S)$}
        $C \gets C \cup \{\, \Call{MoveToward}{S, S_c} \,\}$
      \EndIf
      \State $S_b \gets \arg\max_{S' \in N} F(S')$
      \If{$F(S_b) > F(S)$}
        $C \gets C \cup \{\, \Call{MoveToward}{S, S_b} \,\}$
      \EndIf
    \EndIf
    \State $S \gets \arg\max_{S' \in C \cup \{S\}} F(S')$
  \EndFor
  \State $S^\star \gets \arg\max\{\, F(S^\star),\;
         \max_{S \in \mathrm{Pop}} F(S) \,\}$
         \Comment{elitism}
\EndFor
\State \Return $S^\star$
\Statex
\Function{Prey}{$S$}
  \For{$r = 1$ \textbf{to} $\mathit{try}$}
    \State $S' \gets$ $S$ with one or two members replaced at random
           from $\mathcal{P} \setminus S$
    \If{$F(S') > F(S)$} \Return $S'$ \EndIf
  \EndFor
  \State \Return $S$ with one member replaced at random
         \Comment{random walk}
\EndFunction
\Statex
\Function{MoveToward}{$S$, $S_t$}
  \State replace $\arg\min_{i \in S} \cos(\mathbf{f}_i, \mathbf{q})$
         with a uniformly chosen member of $S_t \setminus S$
  \State \Return the result
\EndFunction
\end{algorithmic}
\end{algorithm}

\subsection{Why local residence becomes measurable}

The design objective is to maximise style fidelity subject to
$|P| \le k$. That constraint is what separates \aum{} from a prompt
prefix, and it is where the commitments of Section~\ref{sec:aum} stop
being rhetorical. If $k$ fields suffice, the remaining $n - k$ never
leave the device, and the disclosure per query is bounded by
construction rather than by policy. The size of the achievable $k$ is
therefore not only an efficiency number but a privacy number, and
Section~\ref{sec:results} measures it.

\section{Simulation Study: Design}
\label{sec:method}

\subsection{Why simulate, and what a simulation can establish}
\label{sec:whysimulate}

Participants in this study are language-model-simulated, not human. We
state this first, and repeat it in the abstract, because it governs how
every number in Section~\ref{sec:results} should be read.

The method has a precedent and a critique, and both are relevant.
\citet{argyle2023outofone} showed that conditioning a language model on
demographic backstories reproduces aggregate patterns found in human
survey data, and \citet{park2024generative} showed that grounding agents
in structured self-report improves individual-level simulation over
demographic conditioning alone, which is the closest precedent for what
we do here. Against this, \citet{bisbee2024synthetic} document
instability and misestimation in synthetic survey responses, and
\citet{wang2025flatten} show that replacing human participants can
misportray and flatten identity groups.

We therefore fix what the study is for. A simulation of this kind
\emph{can} establish that a pipeline runs end to end; that a budgeted
retrieval over a structured user model recovers fields that a
downstream generator uses; that the resulting text differs measurably
from text produced without the model; and that this holds under
controls which rule out the most obvious alternative explanations. It
\emph{cannot} establish that a real person recognises their own voice,
because the entity doing the recognising here is a judge model reading a
persona description, which is a considerably easier task. It also cannot
speak to how a person would feel about a system holding such a model,
which is a question for the human study described in
Section~\ref{sec:future}.

There is a second and less obvious reason to simulate first. The human
study this substitutes for would require participants to disclose
inner-shell content about themselves: fears, difficult episodes, moral
commitments. Running that study before knowing whether the architecture
works at all would spend real disclosure on a question that a simulation
can answer. We regard establishing feasibility in simulation as the
ethically prior step, not merely the cheaper one.

\subsection{Simulated participants}

Sixteen personas are declared in code rather than sampled by a model, so
that a run is reproducible from a seed and the population is auditable.
Each persona carries Big Five $z$-scores in $[-2, 2]$, a one-line life
context, a formative-experience seed on which \shelltwo{} can draw, and
concrete idiolect markers, which are what make style measurable rather
than merely asserted. The $z$-scores were spread deliberately: no two
personas share a sign pattern across all five traits, and every trait
has both tails represented, so that style differences are not confounded
with a single dominant axis. Table~\ref{tab:personas} lists the set.
Appendix~\ref{app:personas} gives the full declarations.

\begin{table}[t]
\centering
\caption{The sixteen simulated participants. Big Five $z$-scores are
  shown as O/C/E/A/N. Idiolect markers are abbreviated; the full
  declarations are in Appendix~\ref{app:personas}.}
\label{tab:personas}
\footnotesize
\renewcommand{\arraystretch}{1.15}
\begin{tabular}{@{}r p{0.19\textwidth} p{0.22\textwidth} p{0.42\textwidth}@{}}
\toprule
\textbf{\#} & \textbf{Label} & \textbf{O/C/E/A/N} & \textbf{Idiolect markers (abridged)} \\
\midrule
0  & meticulous planner     & $+0.2/+1.8/-0.6/+0.4/+0.9$ & numbered points, explicit dates, ask in first sentence \\
1  & warm over-explainer    & $+0.6/+0.1/+1.1/+1.7/+0.5$ & long pleasantries, hedges, pre-emptive apology, ``we'' for ``I'' \\
2  & blunt minimalist       & $+0.3/+0.8/-0.9/-1.4/-0.7$ & lowercase, fragments, no greeting or sign-off, two lines \\
3  & associative creative   & $+1.9/-0.8/+0.7/+0.6/+0.3$ & starts mid-thought, parentheses, concrete images, ends early \\
4  & analytical flat        & $+0.9/+1.3/-1.2/-0.5/-0.9$ & conclusion first, quantifies, ``approximately'', no exclamations \\
5  & bright energetic       & $+1.0/+0.2/+1.9/+1.2/-0.4$ & exclamations, short punchy sentences, direct address \\
6  & dry sceptic            & $+1.1/+0.6/-0.5/-1.1/+0.8$ & understatement, undercutting asides, rhetorical questions \\
7  & formal correct         & $-0.3/+1.5/-0.2/+0.9/+0.2$ & full salutations, passives, ``I should be grateful if'', no contractions \\
8  & scattered fast         & $+1.4/-1.6/+1.3/+0.3/+1.1$ & mid-paragraph topic changes, follow-up corrections, abbreviations \\
9  & nurturing relational   & $+0.4/+0.3/+0.5/+1.9/+0.7$ & names the other's feeling first, softens every imperative \\
10 & competitive driver     & $+0.5/+1.4/+1.2/-1.3/-0.5$ & power verbs, deadlines as commitments, reframes asks as opportunities \\
11 & reflective introvert   & $+1.2/+0.9/-1.8/+0.8/+0.6$ & short paragraphs with pauses, precise word choice, no small talk \\
12 & storytelling extrovert & $+1.3/-0.2/+1.7/+1.0/-0.2$ & opens with a scene, second person, builds to the ask, quotes dialogue \\
13 & pragmatic efficient    & $-0.4/+1.6/+0.1/+0.2/-1.0$ & imperatives, concrete nouns, one line per item \\
14 & quirky metaphorical    & $+1.8/+0.4/+0.3/+0.7/+0.4$ & cross-domain comparisons, semicolons, deliberate archaisms \\
15 & stoic calm             & $+0.1/+1.0/-0.7/+0.5/-1.8$ & even sentence length, no intensifiers, facts then action \\
\bottomrule
\end{tabular}
\end{table}

\subsection{Instrument population}

Each persona completes the 32-field instrument of
Table~\ref{tab:instrument} in character, by guided self report: the
agent is instructed to fill every field with one short first-person
sentence of roughly ten to twenty-five words, specific and concrete
rather than generic. Responses are parsed against the declared schema
and rejected if any shell is missing or any field is empty, with up to
six retries and an explicit corrective instruction on each. The exact
prompt is in Appendix~\ref{app:prompts}.

Populating a user model by self report inherits a known limitation: self
and other observers know different things about a person, and the self
is not the better source for every trait
\citep{vazire2010soka,connelly2010other}. For a simulated participant
whose ground truth is a declared persona this is less severe than it
would be for a human, since the agent has access to its own
declaration, but the resulting instrument is still a self-description
rather than an observation. Section~\ref{sec:limitations} returns to
what this means for a human deployment.

\subsection{Task battery, prompt surfaces, and references}

Six style-sensitive tasks form the evaluation battery, chosen to span
the task types of Table~\ref{tab:sigma}: requesting a deadline extension
(communication), declining an invitation from a colleague
(communication), writing a short public post about one's work
(self-presentation), giving written feedback on a junior collaborator's
draft (relational-evaluative), planning a two-day trip
(planning-decision), and explaining a technical result to a friend
outside the field (explanatory). Four additional simple tasks, all
factual lookups, exist only to exercise and score the classifier's
simple-versus-complex branch and are excluded from the main results.
Table~\ref{tab:tasks} lists the battery.

\begin{table}[t]
\centering
\caption{The task battery. The six complex tasks form the evaluation
  set. The four simple tasks are used only to score the classifier's
  routing decision and never enter the main results.}
\label{tab:tasks}
\small
\renewcommand{\arraystretch}{1.15}
\begin{tabular}{@{}l p{0.52\textwidth} l@{}}
\toprule
\textbf{ID} & \textbf{Task} & \textbf{Type} \\
\midrule
T1 & Requesting a deadline extension from a colleague & communication \\
T2 & Declining an invitation from a colleague & communication \\
T3 & Writing a short public post about your work & self-presentation \\
T4 & Giving written feedback on a junior collaborator's draft & relational-evaluative \\
T5 & Planning a two-day trip & planning-decision \\
T6 & Explaining a technical result to a friend outside the field & explanatory \\
\midrule
S1 to S4 & Four factual lookups (boiling point, unit conversion, date, arithmetic) & factual \\
\bottomrule
\end{tabular}
\end{table}

For each complex task an agent produces two texts, in an order that is
the single most important correctness property of the design.

\begin{enumerate}
\item The \textbf{prompt surface}: the message this person would type to
  an assistant, in their own voice, at their own level of detail and
  formality, capped at sixty words. This is the personality-seepage
  surface of Section~\ref{sec:seepage}, and it is the \emph{only}
  user-derived text the retriever ever sees.

\item The \textbf{reference}: what this person would write themselves,
  with no assistant involved, matching their usual length, structure and
  habits. This is the ground truth against which every condition is
  scored, and it is generated without sight of the prompt surface.
\end{enumerate}

\noindent Passing the reference to the retriever would give every
\aumr{} condition an oracle that the Full \aum{} and preference-notes
baselines never receive, and would make the comparison meaningless. The
separation is enforced in code.

\subsection{The preference-notes baseline}

The baseline that matters is not the generic assistant but the flat
preference notes that deployed memory features actually store, so it has
to be built the way such features build it. Each agent generates four
short transcripts of past assistant conversations, four to six turns
each, on topics drawn from a held-out list that shares no item with the
evaluation battery: choosing between two laptops, a recurring appliance
problem, a week of simple meals, what to do with a free Saturday,
picking a book, and replying to an awkward group message. A separate
summariser, with no access to the persona and no access to the \aum{},
then reads those transcripts and writes at most twelve flat preference
bullets.

Two decisions here are deliberate. The topics are held out so that the
baseline cannot contain a rehearsal of the evaluation tasks. The
summariser is blind to the persona and the user model so that the
baseline is a lossy distillation of observed conversation, which is what
a deployed system has, rather than a partial copy of the treatment,
which is what concatenating \shellthree{} and \shellfour{} fields would
produce.

\subsection{Conditions}

Ten conditions, all sharing a base model and a prompt, split into a main
block and a control block (Table~\ref{tab:conditions}).

\begin{table}[t]
\centering
\caption{Conditions. The main block appears in Table~\ref{tab:main};
  the control block isolates one component at a time and appears in
  Table~\ref{tab:contrasts}. Only the four conditions marked
  \textsc{id} are shown to the identification judge.}
\label{tab:conditions}
\small
\renewcommand{\arraystretch}{1.15}
\begin{tabular}{@{}p{0.20\textwidth} l p{0.50\textwidth}@{}}
\toprule
\textbf{Condition} & \textbf{Block} & \textbf{Injected beyond the prompt} \\
\midrule
Generic \textsc{id}          & main    & Nothing \\
Preference notes \textsc{id} & main    & Flat bullets distilled from four held-out prior conversations \\
Full \aum{} \textsc{id}      & main    & All 32 fields as JSON \\
\aumr{}, $k=4$               & main    & $\sigma(t)$ + swarm retrieval, four fields \\
\aumr{}, $k=8$ \textsc{id}   & main    & $\sigma(t)$ + swarm retrieval, eight fields \\
\aumr{}, $k=16$              & main    & $\sigma(t)$ + swarm retrieval, sixteen fields \\
\midrule
Random $k=8$                 & control & Eight fields drawn uniformly from the same admissible pool \\
No $\sigma(t)$, $k=8$        & control & Swarm retrieval over all shells, no component selection \\
Dense top-$k$, $k=8$         & control & Relevance-only top eight, redundancy term removed \\
Greedy, $k=8$                & control & Forward greedy on the same objective \\
\bottomrule
\end{tabular}
\end{table}

Each control removes exactly one component, so that a null result
localises. Random $k$ separates ``the right fields'' from ``any $k$
fields about the user''. No $\sigma(t)$ asks whether component selection
contributes beyond retrieval. Dense top-$k$ asks whether the redundancy
term earns its place. Greedy asks whether the swarm earns its cost.

\subsection{Measures}
\label{sec:measures}

\paragraph{Fidelity.}
A judge model rates, on a five-point scale, the statement that a
candidate text reads like something the person would have written. The
judge sees the persona and one text the person genuinely wrote, and
never sees the \aum{}. Withholding the user model from the judge is not
incidental: showing it would present the judge with the same fields the
\aumr{} payload contained, allowing it to reward surface agreement with
the payload rather than similarity to the person.

\paragraph{Identification.}
A forced choice among the four \textsc{id} conditions, presented in an
order shuffled per trial, with chance at 25\%. The judge again sees the
persona and the reference. Position bias in LLM judges is documented
\citep{wang2024notfair}, so option order is randomised per trial and
tested post hoc in Section~\ref{sec:robustness}.

\paragraph{Style similarity.}
The cosine between LUAR authorship embeddings
\citep{riverasoto2021luar} of the candidate and the participant's own
reference, computed locally. Authorship representations are used rather
than semantic embeddings for a specific reason: a semantic embedding
scores any two deadline-extension emails as highly similar because they
are about the same thing, and the property we need is sensitivity to
\emph{who} wrote a text and insensitivity to \emph{what} it is about
\citep{wegmann2022sameauthor,wang2023authorship}. The pipeline raises
rather than falling back if the authorship model cannot be loaded,
because a silent substitution would change what the column means while
leaving its heading intact. A semantic cosine is computed in addition
and reported separately, never as style similarity.

\paragraph{Context.}
The number of tokens in exactly the string prepended to the prompt,
counted with the generator's own tokeniser. This is a measurement, not a
word count, so that the ratio reported in
Section~\ref{sec:results} is exact.

\paragraph{Judge independence.}
The judge model differs from the generator by construction, and the
pipeline refuses to run otherwise. LLM evaluators favour their own
generations \citep{panickssery2024selfpreference}, so a run in which
judge and generator coincide produces a fidelity column that mostly
measures self-preference.

\subsection{Models, seeds and cost}

Table~\ref{tab:repro} records the configuration. Every API call is
cached on disk under a hash of its arguments, so a re-run after a crash
or a code change costs nothing for work already completed, and the run
is reproducible from the manifest plus a seed.

\begin{table}[t]
\centering
\caption{Reproducibility record for the simulation study. Token and cost
  figures are read from the run ledger, not estimated.}
\label{tab:repro}
\small
\renewcommand{\arraystretch}{1.15}
\begin{tabular}{@{}p{0.46\textwidth} p{0.46\textwidth}@{}}
\toprule
\textbf{Item} & \textbf{Value} \\
\midrule
Generator model (also plays the participant) & \texttt{gpt-5.6-luna} \\
Judge model (fidelity and identification)    & \texttt{gpt-5.6-terra} \\
Field and query embeddings                   & \texttt{text-embedding-3-small} \\
Authorship-style embeddings                  & \texttt{rrivera1849/LUAR-MUD}, run locally on CPU \\
Generation temperature                       & 0.7 \\
Persona temperature                          & 0.8 \\
Judge temperature                            & 0.0 \\
Participants $\times$ tasks $\times$ seeds   & $16 \times 6 \times 3$ \\
Seeds                                        & 0, 1, 2 \\
Conditions                                   & 10 \\
Instrument size $n$                          & 32 fields \\
Redundancy weight $\lambda$                  & 0.15 \\
Swarm population, iterations                 & 12, 8 \\
Total API calls                              & 13{,}281 \\
Input tokens                                 & 3{,}869{,}382 \\
Output tokens                                & 760{,}598 \\
Total cost                                   & USD 4.81 \\
\bottomrule
\end{tabular}
\end{table}

\subsection{Analysis plan}
\label{sec:analysis}

The unit of analysis is the (participant, task) cell averaged over
seeds, giving $16 \times 6 = 96$ paired observations per condition.
Seeds are repeated measures of the same cell rather than additional
participants, so averaging before testing is what keeps the degrees of
freedom honest; treating three seeds $\times$ 96 cells as 288
independent observations would inflate every test.

For each planned contrast we report the mean paired difference, a 95\%
bootstrap confidence interval over 10{,}000 resamples, a paired $t$
statistic, a Wilcoxon signed-rank statistic, Cohen's $d_z$ and Cliff's
delta. The Wilcoxon test is primary for fidelity, because a five-point
rating scale is ordinal; the paired $t$ is primary for the continuous
style measure. The Holm procedure corrects within metric.
Identification uses an exact binomial test against 25\% chance with a
Wilson interval.

Contrasts were fixed before the run and are those listed in
Section~\ref{sec:results}. Supplementary analyses reported in
Section~\ref{sec:robustness} were specified after inspecting the main
results and are labelled as such; we do not present them as
confirmatory.

\section{Simulation Study: Results}
\label{sec:results}

\subsection{Main comparison}

Table~\ref{tab:main} gives the main block, and Table~\ref{tab:contrasts}
the planned contrasts. Three findings carry the section.

\begin{table}[t]
\centering
\caption{Main results. Fidelity and identification are judge ratings on
  behalf of simulated participants; style is LUAR authorship cosine
  against the participant's own reference; semantic is a general-purpose
  embedding cosine, reported for comparison only and never as style;
  context is tokens injected beyond the prompt. Means over
  $16 \times 6 = 96$ cells and three seeds, with 95\% bootstrap
  intervals on fidelity.}
\label{tab:main}
\small
\renewcommand{\arraystretch}{1.2}
\begin{tabular}{@{}l c c c c r@{}}
\toprule
\textbf{Condition} & \textbf{Fidelity} & \textbf{95\% CI} &
\textbf{Ident.} & \textbf{Style} & \textbf{Context} \\
 & (1 to 5) & & (\%) & (LUAR) & (tokens) \\
\midrule
Generic (prompt only)        & 2.65 & [2.45, 2.84] & \phantom{0}6.9 & 0.674 & \phantom{00}0 \\
Preference notes             & 2.68 & [2.48, 2.88] & 14.9 & 0.684 & \phantom{0}164 \\
Full \aum{} (all 32 fields)  & 2.90 & [2.71, 3.08] & 35.4 & 0.705 & 915 \\
\addlinespace
\aumr{}, $k=4$               & 2.86 & [2.68, 3.06] & n/a  & 0.701 & \phantom{0}111 \\
\aumr{}, $k=8$ \textbf{(ours)} & \textbf{2.91} & [2.74, 3.09] & \textbf{42.7} & 0.698 & \phantom{0}211 \\
\aumr{}, $k=16$              & 2.98 & [2.78, 3.17] & n/a  & 0.701 & \phantom{0}406 \\
\midrule
\multicolumn{6}{@{}l}{\emph{Controls, all at $k=8$}} \\
Random $k$ fields            & 2.89 & [2.69, 3.07] & n/a  & 0.697 & \phantom{0}211 \\
No $\sigma(t)$               & 2.95 & [2.75, 3.14] & n/a  & 0.703 & \phantom{0}209 \\
Dense top-$k$                & 2.92 & [2.74, 3.10] & n/a  & 0.699 & \phantom{0}210 \\
Greedy                       & 2.95 & [2.75, 3.14] & n/a  & 0.700 & \phantom{0}210 \\
\bottomrule
\end{tabular}
\end{table}

\begin{figure}[t]
\centering
\includegraphics[width=0.92\textwidth]{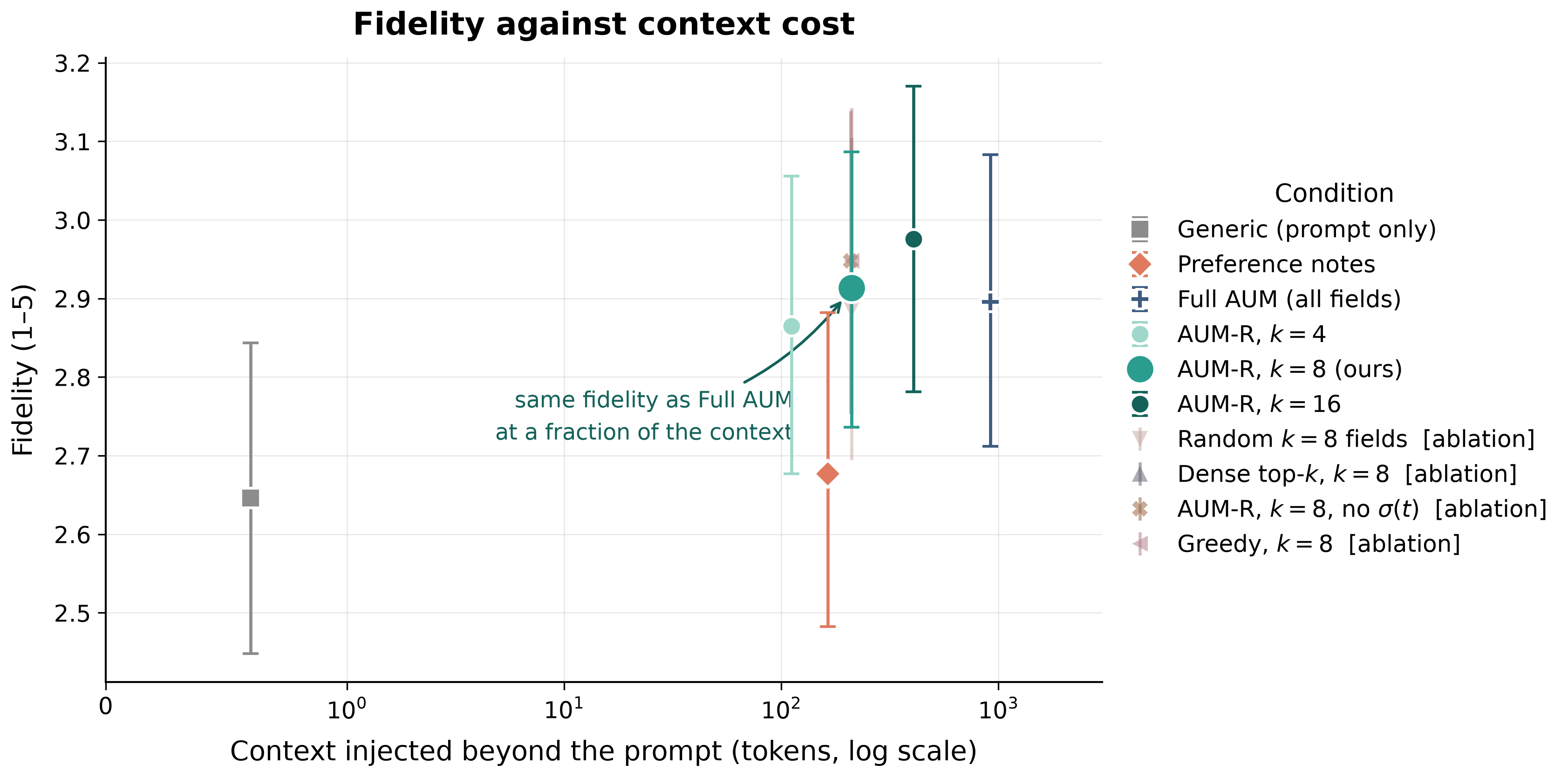}
\caption{Fidelity against injected context on a logarithmic axis. Eight
  retrieved fields reach the fidelity of the entire 32-field user model
  at 23\% of its context. Error bars are 95\% bootstrap intervals over
  the 96 (participant, task) cells.}
\label{fig:pareto}
\end{figure}

\paragraph{Finding 1: a budgeted payload matches the whole model at
under a quarter of the context.}
\aumr{} at $k = 8$ was not distinguishable from Full \aum{} on fidelity
($\Delta = +0.02$, 95\% CI \ci{-0.07}{+0.11}, $p = 1$ after Holm
correction) while injecting 211 tokens against 915, that is 23.0\%.
Figure~\ref{fig:pareto} plots the whole condition set against injected
context on a logarithmic axis; the point of the figure is the vertical
gap between preference notes and \aumr{} at comparable cost, and the
horizontal gap between \aumr{} and Full \aum{} at comparable fidelity.

\paragraph{Finding 2: structure beats flat notes, by a moderate effect.}
\aumr{} at $k = 8$ exceeded preference notes on fidelity by $+0.24$
(95\% CI \ci{+0.14}{+0.33}, Wilcoxon $W = 376.5$,
$p = 7.2 \times 10^{-5}$ after Holm correction, $d_z = +0.50$,
Cliff's $\delta = 0.35$, $n = 96$ cells), and exceeded the generic
baseline by $+0.27$ ($d_z = +0.57$). On LUAR style similarity the same
ordering holds with a smaller effect: $+0.013$ over preference notes
($p = 0.006$) and $+0.024$ over generic
($p < 10^{-5}$, $d_z = +0.53$). Full \aum{} also beat preference notes
($+0.22$, $p = 4 \times 10^{-4}$), which is the expected ordering: the
gain comes from having a structured model at all, and the retrieval
stage recovers it cheaply rather than adding to it.

\begin{figure}[t]
\centering
\includegraphics[width=0.70\textwidth]{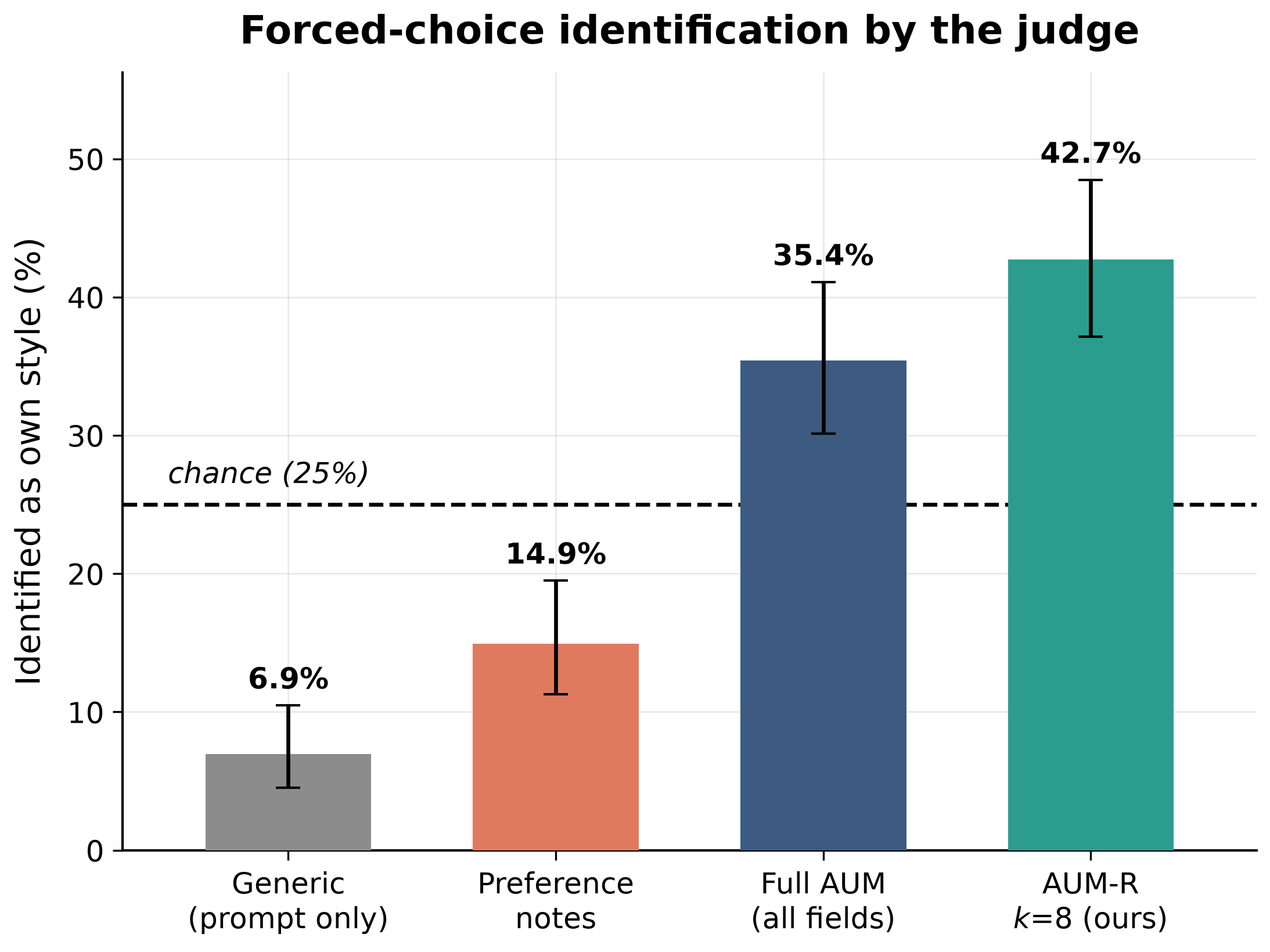}
\caption{Forced-choice identification by the judge, with Wilson
  intervals and the 25\% chance line. Both user-model conditions sit
  above chance; both baselines sit below it.}
\label{fig:ident}
\end{figure}

\paragraph{Finding 3: identification separates the conditions sharply.}
Forced-choice identification of the participant's own voice reached
42.7\% for \aumr{} at $k = 8$ (95\% CI \ci{37.1}{48.5},
$p = 4.2 \times 10^{-11}$ against 25\% chance) and 35.4\% for Full
\aum{} (\ci{30.1}{41.1}, $p = 5.3 \times 10^{-5}$). Preference notes
reached 14.9\% and generic 6.9\%, neither above chance
(Table~\ref{tab:ident}, Figure~\ref{fig:ident}). That both
user-model conditions sit above chance while both baselines sit below it
is the cleanest separation in the study, and it is the result most
directly relevant to the inversion argument of
Section~\ref{sec:seepage}: notes distilled from unrelated past
conversations carry topic, not person.

\begin{figure}[t]
\centering
\includegraphics[width=0.72\textwidth]{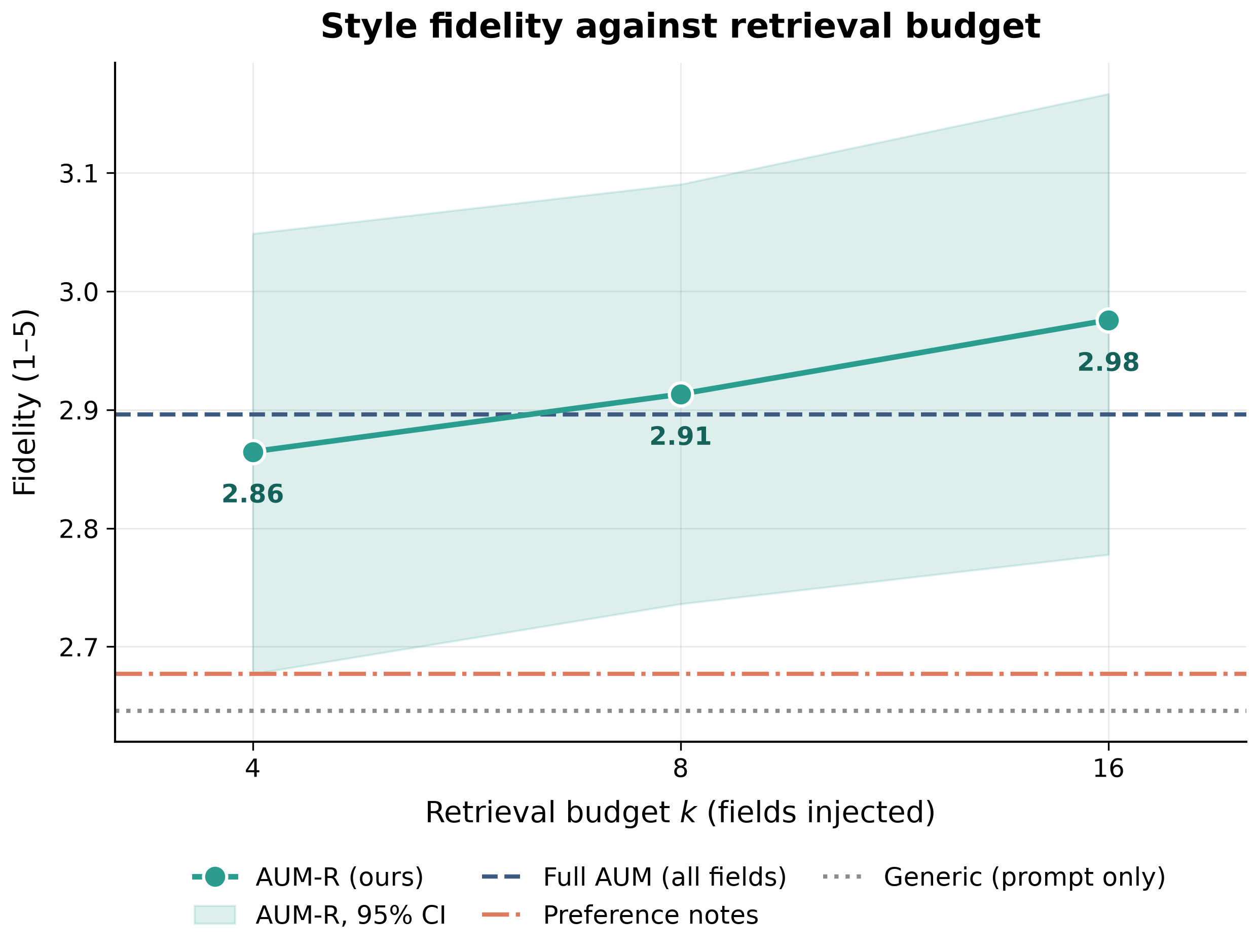}
\caption{Fidelity against retrieval budget $k$, with the Full \aum{} and
  preference-notes levels marked. Fidelity is still rising at the
  largest budget tested, so $k = 8$ is a trade rather than a saturation
  point.}
\label{fig:kcurve}
\end{figure}

\begin{table}[t]
\centering
\caption{Forced-choice identification. Chance is 25\%; $n = 288$
  presentations per condition (96 cells $\times$ 3 seeds). Intervals are
  Wilson; $p$ is an exact binomial test against chance in the upper
  tail.}
\label{tab:ident}
\small
\renewcommand{\arraystretch}{1.2}
\begin{tabular}{@{}l r r c r@{}}
\toprule
\textbf{Condition} & \textbf{Hits} & \textbf{\%} &
\textbf{95\% CI} & \textbf{$p$ vs 25\%} \\
\midrule
\aumr{}, $k=8$ (ours)   & 123 & 42.7 & [37.1, 48.5] & $4.2 \times 10^{-11}$ \\
Full \aum{} (all fields) & 102 & 35.4 & [30.1, 41.1] & $5.3 \times 10^{-5}$ \\
Preference notes         & \phantom{0}43 & 14.9 & [11.3, 19.5] & 1.00 \\
Generic (prompt only)    & \phantom{0}20 & \phantom{0}6.9 & [\phantom{0}4.5, 10.5] & 1.00 \\
\bottomrule
\end{tabular}
\end{table}

\subsection{The budget curve}

Fidelity rose monotonically across the budgets tested: 2.86 at $k = 4$,
2.91 at $k = 8$, 2.98 at $k = 16$ (Figure~\ref{fig:kcurve}). Neither
adjacent step was individually significant after correction
($k=16$ against $k=8$: $\Delta = +0.06$, $p = 0.74$; $k=8$ against
$k=4$: $\Delta = +0.05$, $p = 1$), so the curve is best read as a
gradual improvement rather than a threshold. Two things follow. First,
$k = 8$ is not a saturation point, and a system with a looser budget
should use more fields. Second, and more usefully for the privacy
argument, there is no evidence of degradation as further fields enter
the payload: the concern that irrelevant fields would dilute a payload
and hurt output quality is not borne out at these budgets, so the
argument for a small $k$ rests on cost and disclosure rather than on
quality.

\subsection{Style similarity}

\begin{figure}[t]
\centering
\includegraphics[width=0.92\textwidth]{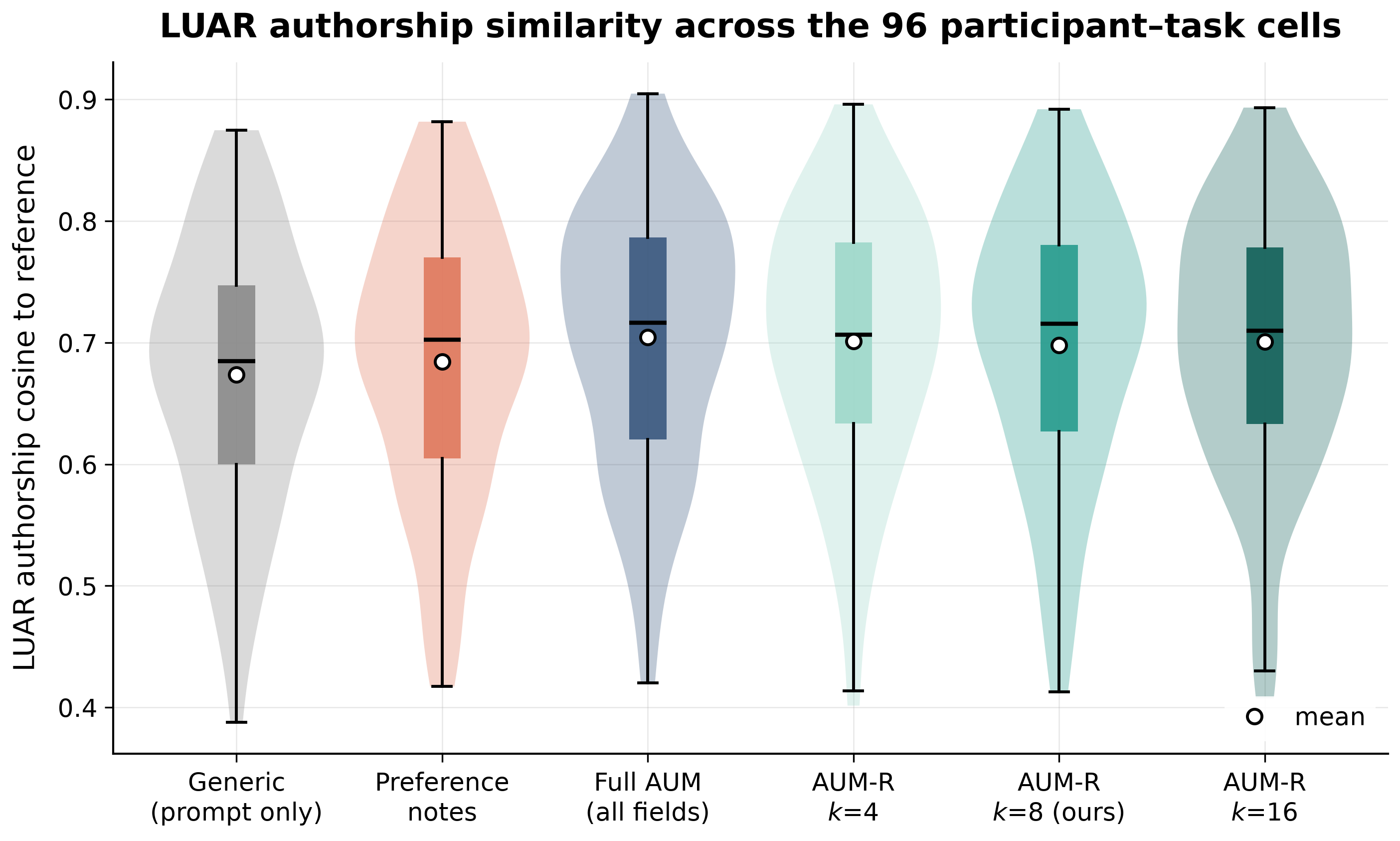}
\caption{Distribution of LUAR authorship-embedding cosine between each
  generated text and the participant's own reference, by condition.}
\label{fig:styledist}
\end{figure}

\begin{figure}[t]
\centering
\includegraphics[width=0.92\textwidth]{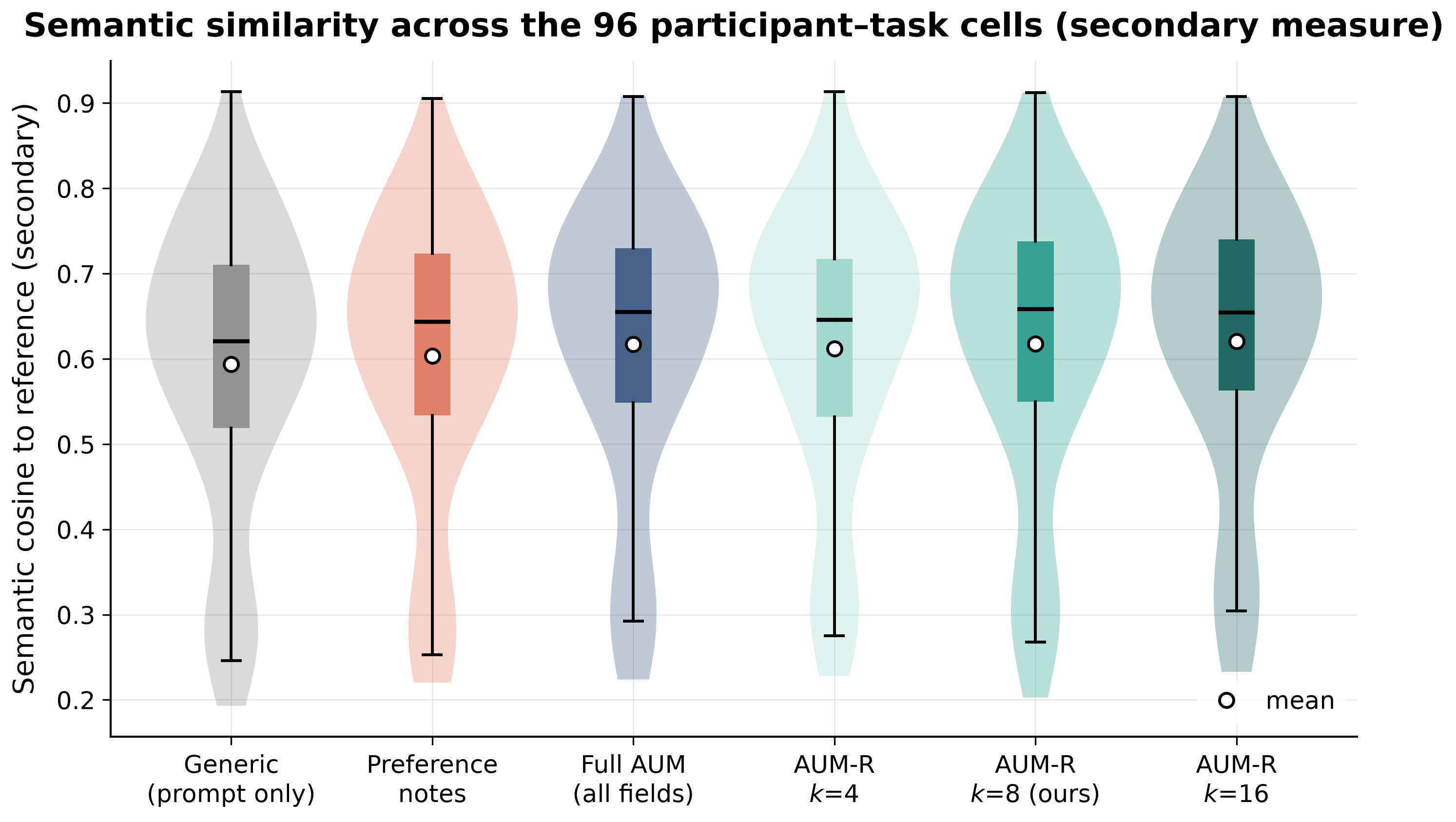}
\caption{The same trials under a general-purpose semantic embedding.
  The compression relative to Figure~\ref{fig:styledist} is the point:
  a semantic metric mostly measures that every candidate addresses the
  same task, which is the wrong invariance for a style claim.}
\label{fig:styledistsem}
\end{figure}

Figure~\ref{fig:styledist} shows the distribution of LUAR authorship
cosine by condition, and Figure~\ref{fig:styledistsem} the
general-purpose semantic cosine for the same trials. The two behave
differently in an instructive way. The authorship metric separates the
conditions in the same order as fidelity, with a compressed range
(0.674 to 0.705 across all ten conditions). The semantic metric
separates them far less (0.593 to 0.620), which is what one expects
given that every candidate for a given trial is an attempt at the same
task: semantic similarity is dominated by topic, and topic is held
constant by design. We report the semantic column only to make that
contrast visible, and we would regard a paper that reported it as
``style similarity'' as having measured the wrong thing.

\subsection{Four controls return null}
\label{sec:nulls}

\begin{table}[t]
\centering
\caption{Planned contrasts, paired over the 96 (participant, task)
  cells. $\Delta$ is the mean paired difference; the interval is a 95\%
  bootstrap interval over 10{,}000 resamples; $p$ is Holm-corrected
  within metric, with Wilcoxon primary for fidelity and paired $t$
  primary for style.}
\label{tab:contrasts}
\footnotesize
\renewcommand{\arraystretch}{1.15}
\begin{tabular}{@{}l l r c r r@{}}
\toprule
\textbf{Metric} & \textbf{Contrast} & \textbf{$\Delta$} &
\textbf{95\% CI} & \textbf{$d_z$} & \textbf{$p$} \\
\midrule
fidelity & \aumr{} $k{=}8$ vs Preference notes      & $+0.236$ & [$+0.142$, $+0.330$] & $+0.50$ & $0.0001$ \\
fidelity & \aumr{} $k{=}8$ vs Generic               & $+0.267$ & [$+0.177$, $+0.365$] & $+0.57$ & $0.0001$ \\
fidelity & Full \aum{} vs Preference notes          & $+0.219$ & [$+0.115$, $+0.323$] & $+0.42$ & $0.0004$ \\
fidelity & \aumr{} $k{=}8$ vs Full \aum{}           & $+0.017$ & [$-0.069$, $+0.108$] & $+0.04$ & $1.0000$ \\
fidelity & \aumr{} $k{=}16$ vs \aumr{} $k{=}8$      & $+0.063$ & [$-0.031$, $+0.156$] & $+0.13$ & $0.7436$ \\
fidelity & \aumr{} $k{=}8$ vs \aumr{} $k{=}4$       & $+0.049$ & [$-0.031$, $+0.125$] & $+0.12$ & $1.0000$ \\
\addlinespace
fidelity & \aumr{} $k{=}8$ vs Random $k{=}8$        & $+0.028$ & [$-0.052$, $+0.108$] & $+0.07$ & $1.0000$ \\
fidelity & \aumr{} $k{=}8$ vs no $\sigma(t)$        & $-0.035$ & [$-0.111$, $+0.045$] & $-0.09$ & $0.9813$ \\
fidelity & \aumr{} $k{=}8$ vs Dense top-$k$         & $-0.007$ & [$-0.083$, $+0.073$] & $-0.02$ & $1.0000$ \\
fidelity & \aumr{} $k{=}8$ vs Greedy                & $-0.035$ & [$-0.111$, $+0.042$] & $-0.09$ & $1.0000$ \\
\midrule
style & \aumr{} $k{=}8$ vs Preference notes         & $+0.013$ & [$+0.006$, $+0.021$] & $+0.36$ & $0.0057$ \\
style & \aumr{} $k{=}8$ vs Generic                  & $+0.024$ & [$+0.015$, $+0.033$] & $+0.53$ & $0.0000$ \\
style & Full \aum{} vs Preference notes             & $+0.020$ & [$+0.012$, $+0.029$] & $+0.47$ & $0.0001$ \\
style & \aumr{} $k{=}8$ vs Full \aum{}              & $-0.007$ & [$-0.014$, $+0.001$] & $-0.18$ & $0.5889$ \\
style & \aumr{} $k{=}8$ vs Random $k{=}8$           & $+0.001$ & [$-0.005$, $+0.007$] & $+0.04$ & $1.0000$ \\
style & \aumr{} $k{=}8$ vs no $\sigma(t)$           & $-0.006$ & [$-0.012$, $+0.001$] & $-0.17$ & $0.6055$ \\
style & \aumr{} $k{=}8$ vs Dense top-$k$            & $-0.001$ & [$-0.007$, $+0.005$] & $-0.03$ & $1.0000$ \\
style & \aumr{} $k{=}8$ vs Greedy                   & $-0.002$ & [$-0.009$, $+0.005$] & $-0.06$ & $1.0000$ \\
\bottomrule
\end{tabular}
\end{table}

All four controls returned null, on both metrics, and we take each in
turn because each bounds the claim differently.

\paragraph{The right $k$ fields, or merely $k$ fields?}
\aumr{} at $k = 8$ was not distinguishable from eight fields drawn
uniformly at random from the same admissible pool
($\Delta = +0.028$, $p = 1$). This is the most consequential null in the
study. It says that once component selection has narrowed the pool to
roughly seventeen fields, which eight of them are injected did not
measurably matter. The gain in Findings 1 to 3 is therefore attributable
to injecting structured, task-plausible fields about the person, not to
ranking within that set.

\paragraph{Does $\sigma(t)$ contribute beyond retrieval?}
Removing component selection and retrieving over all thirty-two fields
did not hurt; it was very slightly better and not significantly so
($\Delta = -0.035$, $p = 0.98$). Given the preceding null this is
coherent rather than surprising: if ranking within the pool does not
matter, restricting the pool cannot matter either, so long as the
restriction does not exclude something essential.
Section~\ref{sec:shellusage} shows that $\sigma(t)$ nonetheless selects
sensibly, which is a weaker but still useful property.

\paragraph{Does the redundancy term earn its place?}
A relevance-only top-$k$ was indistinguishable from the full objective
($\Delta = -0.007$, $p = 1$). Section~\ref{sec:scaling} shows this is a
property of the operating point rather than of the objective: at
$\lambda = 0.15$ and a pool of seventeen, the redundancy penalty rarely
changes which eight fields win.

\begin{figure}[t]
\centering
\includegraphics[width=0.92\textwidth]{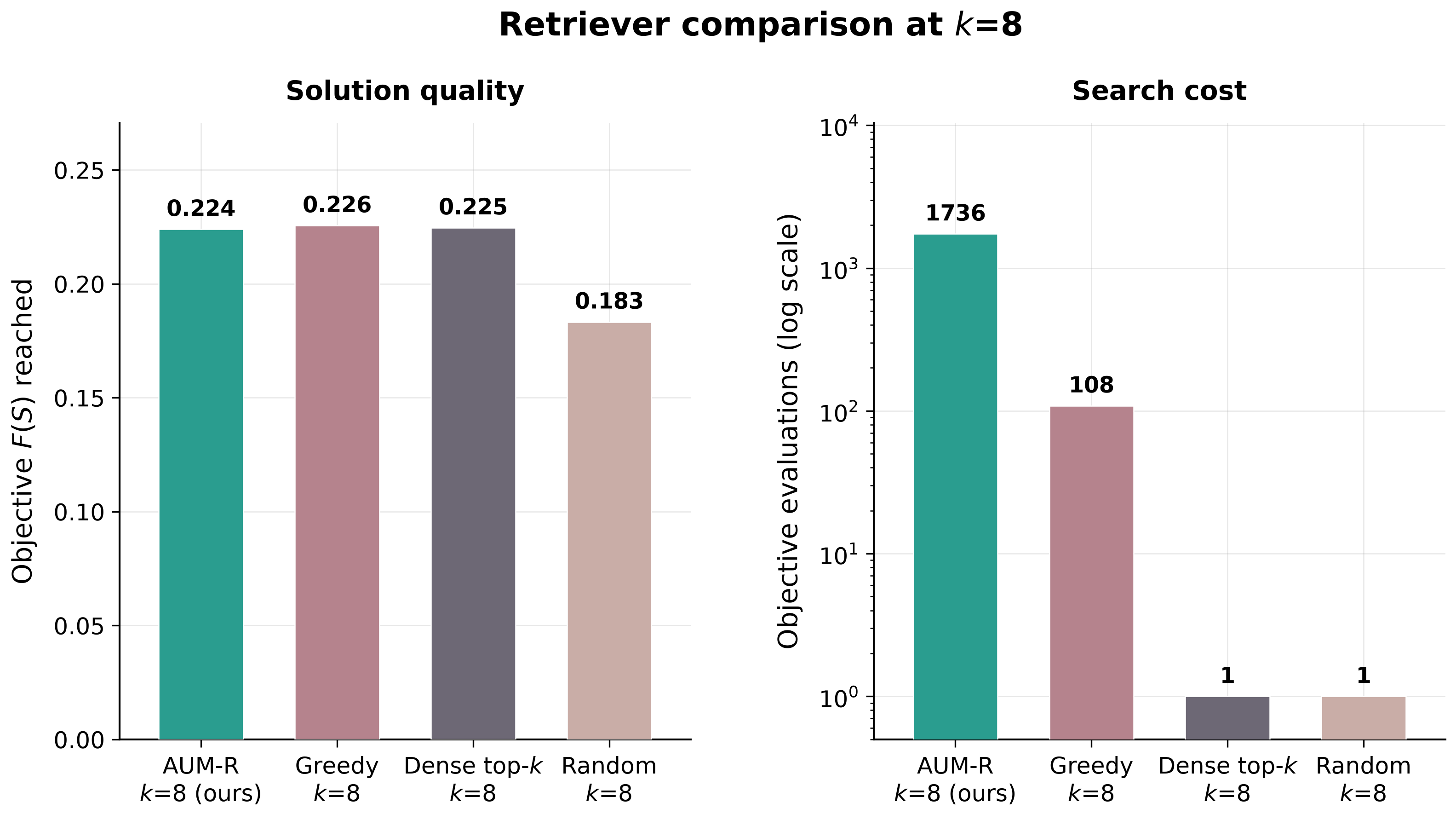}
\caption{Retriever comparison at $k = 8$: objective value reached and
  objective evaluations spent. The swarm is both slightly worse and
  sixteen times more expensive than forward greedy at this scale.}
\label{fig:retrievalcost}
\end{figure}

\paragraph{Does the swarm earn its cost?}
It did not. \aumr{} did not beat forward greedy on generated style
($\Delta = -0.035$, $p = 1$), and it did not reach a higher value of its
own objective. Table~\ref{tab:retrieval} gives the retrieval
diagnostics: the swarm reached $F = 0.224$ using 1{,}736 objective
evaluations, where greedy reached $F = 0.226$ using 108 and dense
top-$k$ reached $F = 0.225$ using one. Random reached $F = 0.183$,
which confirms that the objective is meaningful even though optimising
it harder did not help downstream. Figure~\ref{fig:retrievalcost} shows
the quality and cost side by side.

\begin{table}[t]
\centering
\caption{Retrieval diagnostics at $k = 8$, averaged over all trials.
  $F(S)$ is the value of Equation~\ref{eq:objective} reached; evaluations
  counts calls to $F$; the pool is the size of $\mathrm{pool}(t)$ after
  component selection. The swarm spent sixteen times greedy's budget to
  reach a marginally lower objective value.}
\label{tab:retrieval}
\small
\renewcommand{\arraystretch}{1.2}
\begin{tabular}{@{}l r r r r@{}}
\toprule
\textbf{Retriever} & \textbf{$F(S)$} & \textbf{Evaluations} &
\textbf{Seconds} & \textbf{Pool} \\
\midrule
Greedy, $k=8$          & 0.2256 & \phantom{0,}108.5 & 0.0014 & 16.9 \\
Dense top-$k$, $k=8$   & 0.2246 & \phantom{0,00}1.0 & 0.0001 & 16.9 \\
\aumr{} (swarm), $k=8$ & 0.2240 & 1{,}735.7 & 0.0196 & 16.9 \\
Random, $k=8$          & 0.1833 & \phantom{0,00}1.0 & 0.0001 & 16.9 \\
\addlinespace
\aumr{}, $k=4$         & 0.2512 & 1{,}748.5 & 0.0166 & 16.9 \\
\aumr{}, $k=16$        & 0.2030 & 1{,}746.8 & 0.0213 & 20.0 \\
\aumr{}, $k=8$, no $\sigma(t)$ & 0.2376 & 1{,}791.6 & 0.0193 & 32.0 \\
\bottomrule
\end{tabular}
\end{table}

\paragraph{What the four nulls jointly establish.}
They localise the effect. Something about injecting a structured,
task-plausible slice of a user model changes generated style
substantially; nothing about which slice, or about how hard one searches
for it, changes it further at this scale. That is a narrower claim than
the one we set out to test, and it is the claim the data supports.
Section~\ref{sec:scaling} takes the optimisation question out of the
language-model setting entirely and asks where, if anywhere, the search
does start to matter.

\subsection{Which shells each task type draws on}
\label{sec:shellusage}

\begin{figure}[t]
\centering
\includegraphics[width=0.88\textwidth]{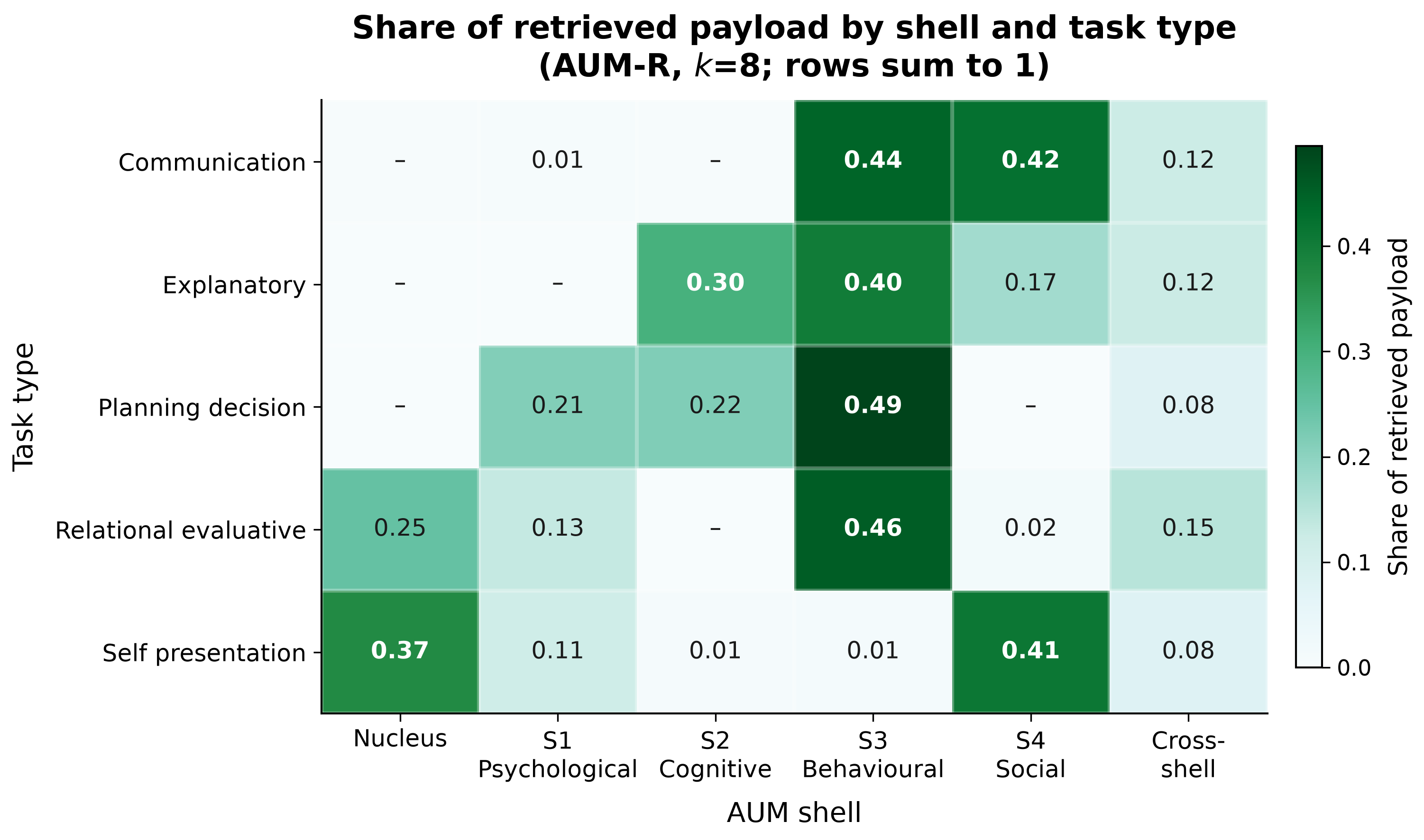}
\caption{Share of the retrieved payload drawn from each shell, by task
  type. Rows sum to 100\%. The composition tracks
  Table~\ref{tab:sigma} without having been constrained to.}
\label{fig:shellusage}
\end{figure}

Although component selection did not change downstream quality, the
payloads it produced are interpretable, and inspecting them is a check
that the pipeline is doing what the specification says.
Figure~\ref{fig:shellusage} shows the share of retrieved payload drawn
from each shell, broken down by task type, and
Table~\ref{tab:shellusage} gives the same figures numerically.

The pattern matches the mapping in Table~\ref{tab:sigma} without having
been forced to. Communication tasks drew 44.4\% of their payload from
\shellthree{} and 42.3\% from \shellfour{}. Self-presentation drew most
heavily on the Nucleus (37.2\%) and \shellfour{} (41.2\%), which is what
one would expect of a task in which a person decides what to project.
Planning and decision tasks drew 49.5\% from \shellthree{} and split the
rest between \shellone{} and \shelltwo{}. Relational-evaluative tasks,
which involve judging another person's work, drew 24.7\% from the
Nucleus, the second-highest Nucleus share in the study, alongside 45.6\%
from \shellthree{}. Explanatory tasks drew 30.0\% from \shelltwo{},
which holds knowledge base and cognitive style, the highest \shelltwo{}
share of any task type. Cross-shell fields appeared in every task type
at between 7.6\% and 14.8\%.

This is a weak result and we present it as one. It shows that the
representation is legible and that retrieval over it behaves sensibly,
not that legibility improves output. It matters mainly for the
scrutability argument: a user inspecting why the assistant produced a
particular draft can be shown a payload whose composition is defensible
in ordinary language.

\subsection{Which tasks the effect holds for}

\begin{table}[t]
\centering
\caption{Effect of \aumr{} at $k=8$ over preference notes, by task.
  $\Delta$ is the mean paired difference in fidelity over the sixteen
  participants; ``favouring'' is the share of participants for whom the
  difference is positive. $p$ is an uncorrected Wilcoxon signed-rank
  test within task.}
\label{tab:bytask}
\small
\renewcommand{\arraystretch}{1.2}
\begin{tabular}{@{}l l r c r r@{}}
\toprule
\textbf{ID} & \textbf{Task} & \textbf{$\Delta$} & \textbf{95\% CI} &
\textbf{Favouring} & \textbf{$p$} \\
\midrule
T6 & Explain to a friend  & $+0.375$ & [$+0.042$, $+0.708$] & 56.3\% & 0.045 \\
T2 & Decline invitation   & $+0.312$ & [$+0.062$, $+0.604$] & 50.0\% & 0.061 \\
T5 & Trip planning        & $+0.250$ & [$+0.083$, $+0.396$] & 62.5\% & 0.024 \\
T3 & Public post          & $+0.229$ & [$+0.063$, $+0.396$] & 68.8\% & 0.018 \\
T4 & Feedback on draft    & $+0.229$ & [$\phantom{+}0.000$, $+0.479$] & 50.0\% & 0.116 \\
T1 & Deadline extension   & $+0.021$ & [$-0.063$, $+0.104$] & 18.8\% & 0.783 \\
\bottomrule
\end{tabular}
\end{table}

\begin{figure}[t]
\centering
\includegraphics[width=0.90\textwidth]{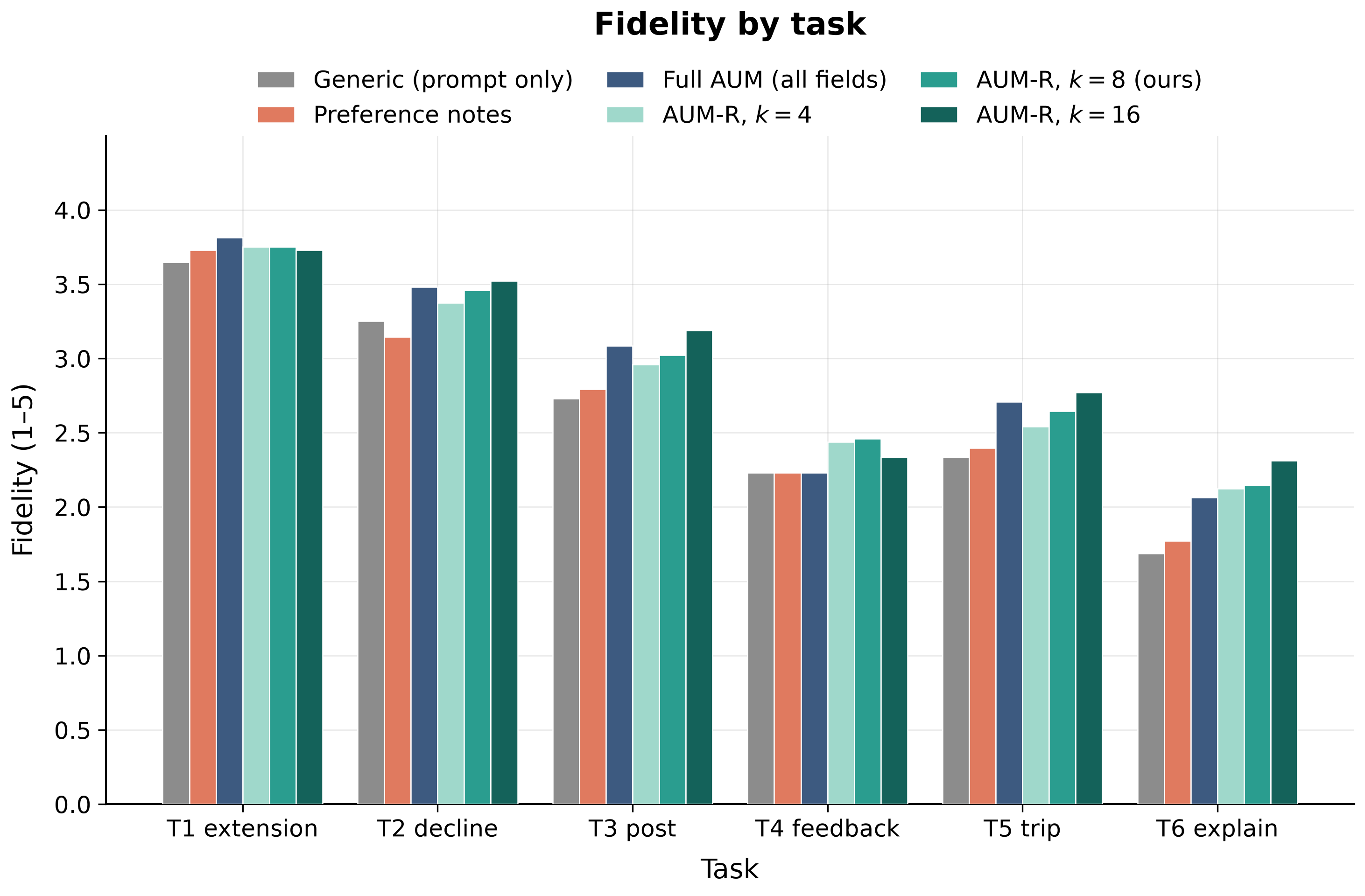}
\caption{Fidelity by task and condition. The deadline-extension task,
  the most conventionalised in the battery, shows the smallest spread
  across conditions.}
\label{fig:pertask}
\end{figure}

All six tasks favoured \aumr{} in the mean (Table~\ref{tab:bytask},
Figure~\ref{fig:pertask}), but the sizes differ in a way worth naming.
The largest effects are on tasks with the loosest genre conventions:
explaining something to a friend outside the field ($+0.375$) and
declining an invitation ($+0.312$). The smallest, and effectively zero,
is the deadline-extension email ($+0.021$, favouring in only 18.8\% of
participants). That is the most conventionalised task in the battery: a
deadline-extension email has a widely shared expected form, and a
generic assistant already produces it. Where the genre supplies the
structure, the user model has little left to contribute; where it does
not, the model fills the gap. This is the task-level counterpart of the
participant-level pattern in Section~\ref{sec:priordistance}.

\subsection{Robustness}
\label{sec:robustness}

Four supplementary analyses address the objections a reader is most
likely to raise. All were specified after the main results and are
reported as exploratory.

\begin{figure}[t]
\centering
\includegraphics[width=0.80\textwidth]{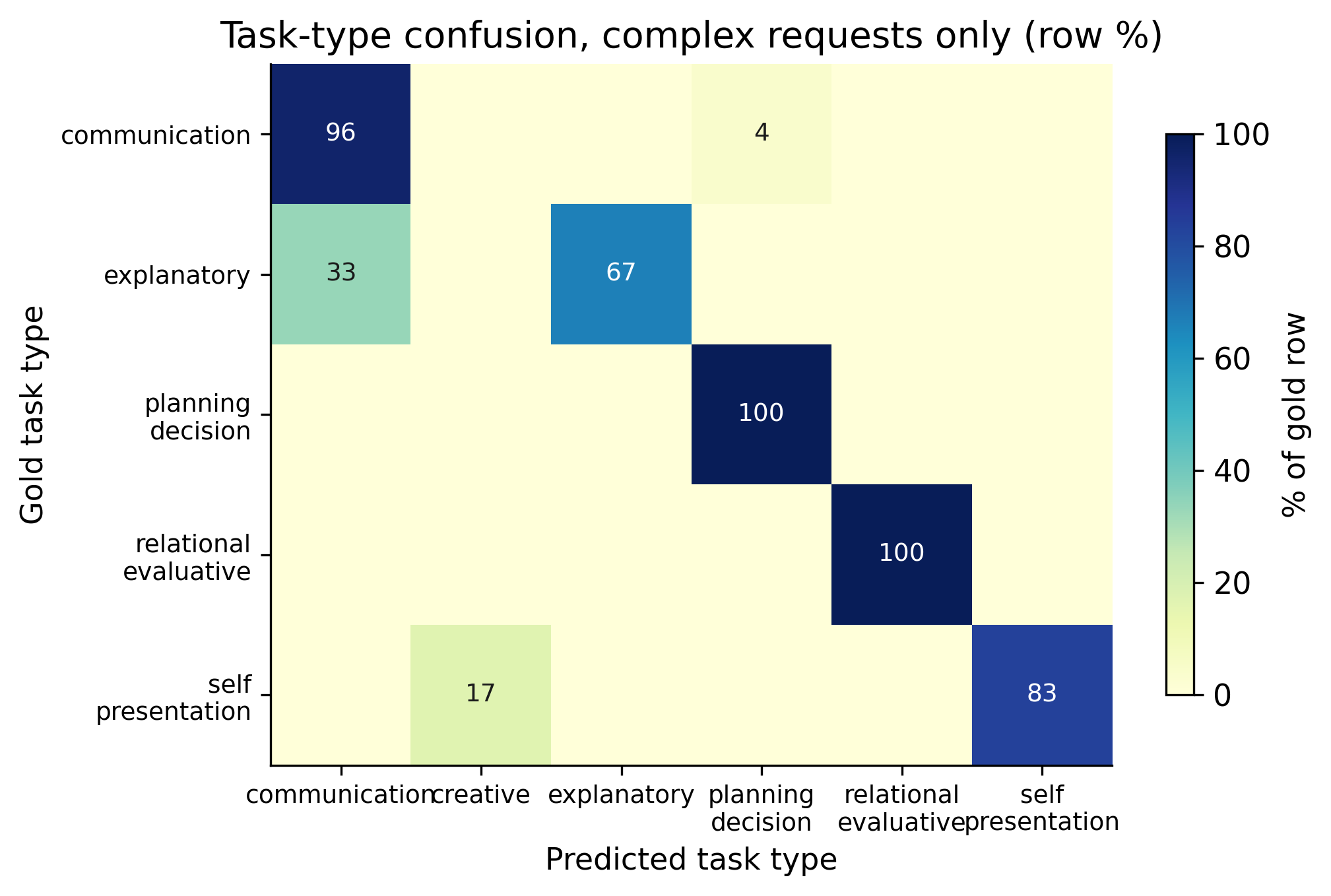}
\caption{Task-type confusion for complex requests, as row percentages.
  The one systematic error, explanatory read as communication, reflects
  a genuine overlap between the two categories.}
\label{fig:confusion}
\end{figure}

\paragraph{The classifier stage works.}
Section~\ref{sec:pipeline} posits a routing stage, and an unevaluated
first stage is a weak point in any pipeline. Over the full battery the
classifier assigned the correct complexity label in 98.8\% of trials
(83 of 84, Wilson interval \ci{93.6}{99.8}), with all twelve simple
trials routed to bypass and 71 of 72 complex trials routed to the user
model. Six-way task-type accuracy on complex trials was 90.3\%
(65 of 72, \ci{81.3}{95.2}) against 16.7\% chance.
Table~\ref{tab:classifier} gives per-type recall and
Figure~\ref{fig:confusion} the confusion matrix. The single systematic
error is that explanatory requests are read as communication in a third
of cases, which is intelligible: explaining a result to a friend
\emph{is} a communication act, and the two $\sigma(t)$ sets share
\shellthree{}.

\begin{table}[t]
\centering
\caption{Task classifier performance, pooled over three seeds. Recall
  is within gold task type on complex trials only.}
\label{tab:classifier}
\small
\renewcommand{\arraystretch}{1.2}
\begin{tabular}{@{}l r r r@{}}
\toprule
\textbf{Gold type} & \textbf{$n$} & \textbf{Recall} & \textbf{Dominant confusion} \\
\midrule
planning-decision      & 12 & 100.0\% & n/a \\
relational-evaluative  & 12 & 100.0\% & n/a \\
communication          & 24 & \phantom{0}95.8\% & planning-decision (1) \\
self-presentation      & 12 & \phantom{0}83.3\% & creative (2) \\
explanatory            & 12 & \phantom{0}66.7\% & communication (4) \\
\midrule
\multicolumn{2}{@{}l}{Overall, six-way} & \phantom{0}90.3\% & chance 16.7\% \\
\multicolumn{2}{@{}l}{Simple vs complex} & \phantom{0}98.8\% & chance 50\% \\
\bottomrule
\end{tabular}
\end{table}

\paragraph{The effect is not a length artefact.}
A judge asked whether a text reads like a person's own writing might be
rewarded largely by length agreement. It is: pooled across all trials,
the absolute log ratio of candidate length to reference length
correlates strongly and negatively with fidelity
(Spearman $\rho = -0.511$, $p = 3 \times 10^{-191}$, $n = 2880$).
Matching the reference's length is a large part of reading like the
person, which is unsurprising and not itself a confound.

The question is whether the treatment effect survives adjustment for it.
It does. Regressing the paired fidelity difference (\aumr{} $k=8$ minus
preference notes) on the paired difference in absolute log length ratio
leaves an intercept of $+0.227$ (95\% CI \ci{+0.127}{+0.327},
$p = 1.9 \times 10^{-5}$) against a raw difference of $+0.236$, and the
covariate slope is not significant ($-0.155$, $p = 0.519$). In other
words the two conditions did not differ appreciably in length agreement,
and the effect is not carried by it. Table~\ref{tab:length} gives the
per-condition figures. The generic condition is the outlier: it produced
the shortest outputs (96 words against a 142-word reference mean), the
worst length agreement (0.007 on the absolute log ratio scale relative
to the next condition) and the highest rate of degenerate outputs
(7.6\% of generic trials produced fewer than fifteen words, typically a
clarifying question rather than an attempt at the task).

\begin{table}[t]
\centering
\caption{Output length by condition, against a reference mean of 141.6
  words. ``Agreement'' is the mean absolute log ratio of candidate to
  reference length, where lower is closer. ``Stub'' is the share of
  trials producing fewer than fifteen words.}
\label{tab:length}
\small
\renewcommand{\arraystretch}{1.2}
\begin{tabular}{@{}l r c r r@{}}
\toprule
\textbf{Condition} & \textbf{Words} & \textbf{95\% CI} &
\textbf{Agreement} & \textbf{Stub} \\
\midrule
Generic (prompt only)   & \phantom{0}96.3 & [71.4, 124.9] & 1.007 & 7.6\% \\
Preference notes        & 112.7 & [85.6, 143.6] & 0.920 & 5.9\% \\
Full \aum{}             & 114.5 & [88.5, 143.6] & 0.844 & 3.5\% \\
\aumr{}, $k=4$          & 112.2 & [86.3, 140.9] & 0.881 & 5.2\% \\
\aumr{}, $k=8$ (ours)   & 114.0 & [86.9, 145.0] & 0.862 & 4.5\% \\
\aumr{}, $k=16$         & 114.1 & [86.6, 145.3] & 0.840 & 2.4\% \\
Greedy, $k=8$           & 109.7 & [83.6, 139.3] & 0.859 & 4.9\% \\
Dense top-$k$, $k=8$    & 109.3 & [83.8, 138.2] & 0.860 & 4.2\% \\
\bottomrule
\end{tabular}
\end{table}

\begin{figure}[t]
\centering
\includegraphics[width=0.90\textwidth]{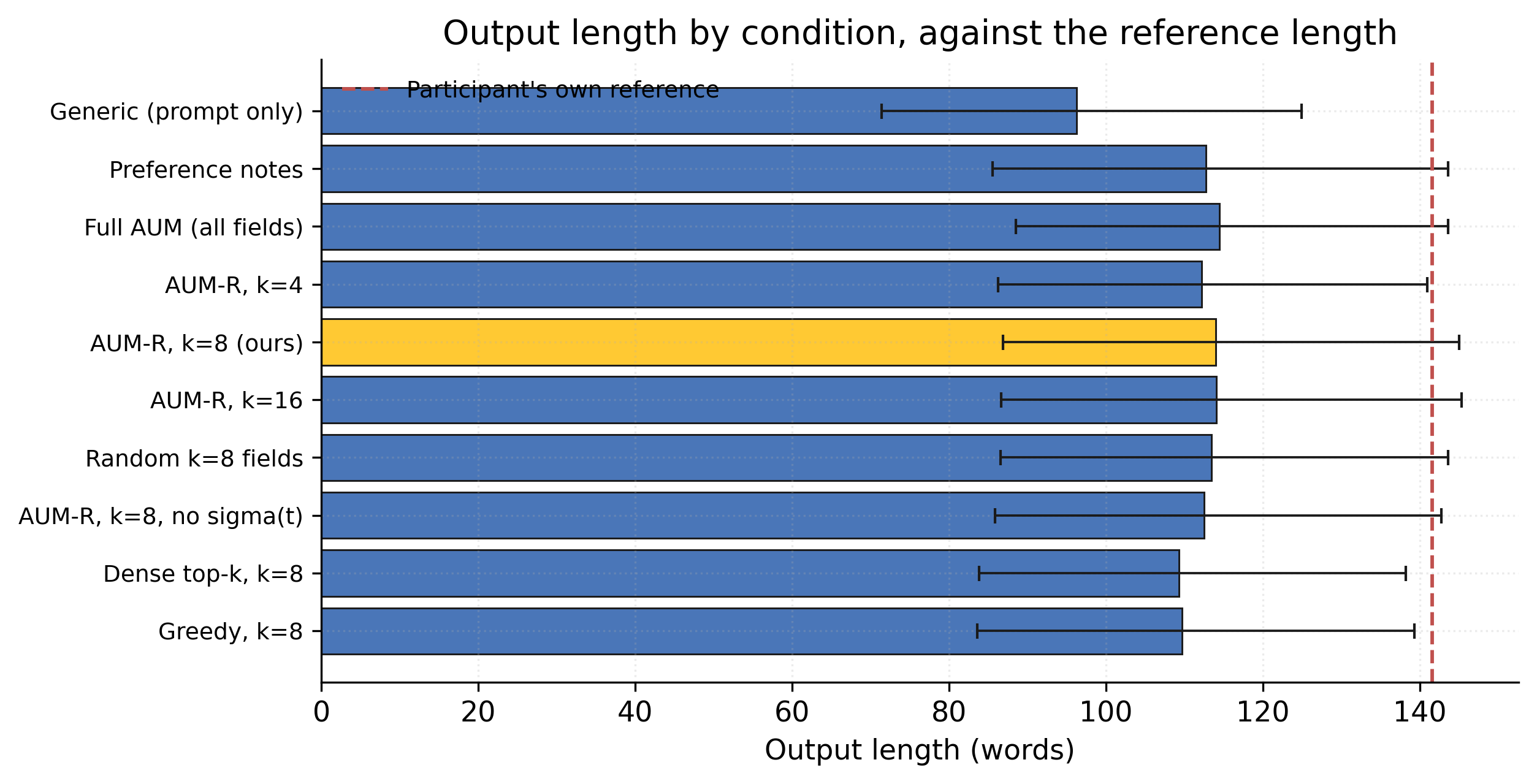}
\caption{Output length by condition against the participants' own
  reference length. Every user-model condition moves output length
  towards the reference; the generic condition is furthest from it.}
\label{fig:length}
\end{figure}

\begin{figure}[t]
\centering
\includegraphics[width=0.88\textwidth]{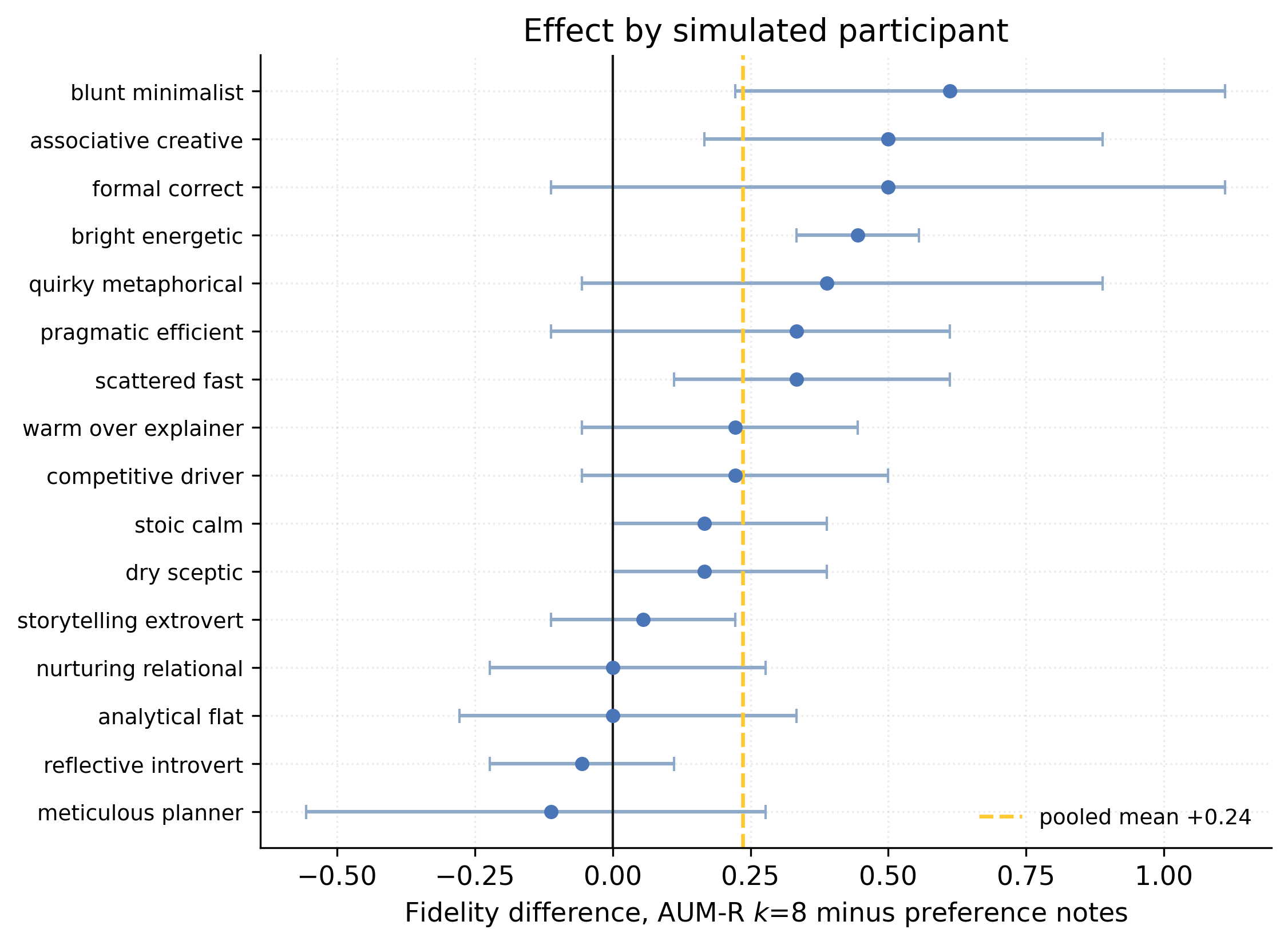}
\caption{Effect of \aumr{} at $k=8$ over preference notes, by simulated
  participant, with 95\% bootstrap intervals over that participant's six
  tasks. The dashed line is the pooled mean.}
\label{fig:byperson}
\end{figure}

\paragraph{The effect is broad, not driven by outliers.}
Twelve of sixteen participants and six of six tasks favoured \aumr{} in
the mean. Of the 96 cells, 51.0\% favoured \aumr{}, 33.3\% were exact
ties on the ordinal scale, and 15.6\% favoured preference notes.
Figure~\ref{fig:byperson} gives a per-participant forest plot, and
Table~\ref{tab:byperson} the numbers. Two participants show small
negative effects, the meticulous planner ($-0.111$) and the reflective
introvert ($-0.056$), and neither interval excludes zero.

\begin{figure}[t]
\centering
\includegraphics[width=0.92\textwidth]{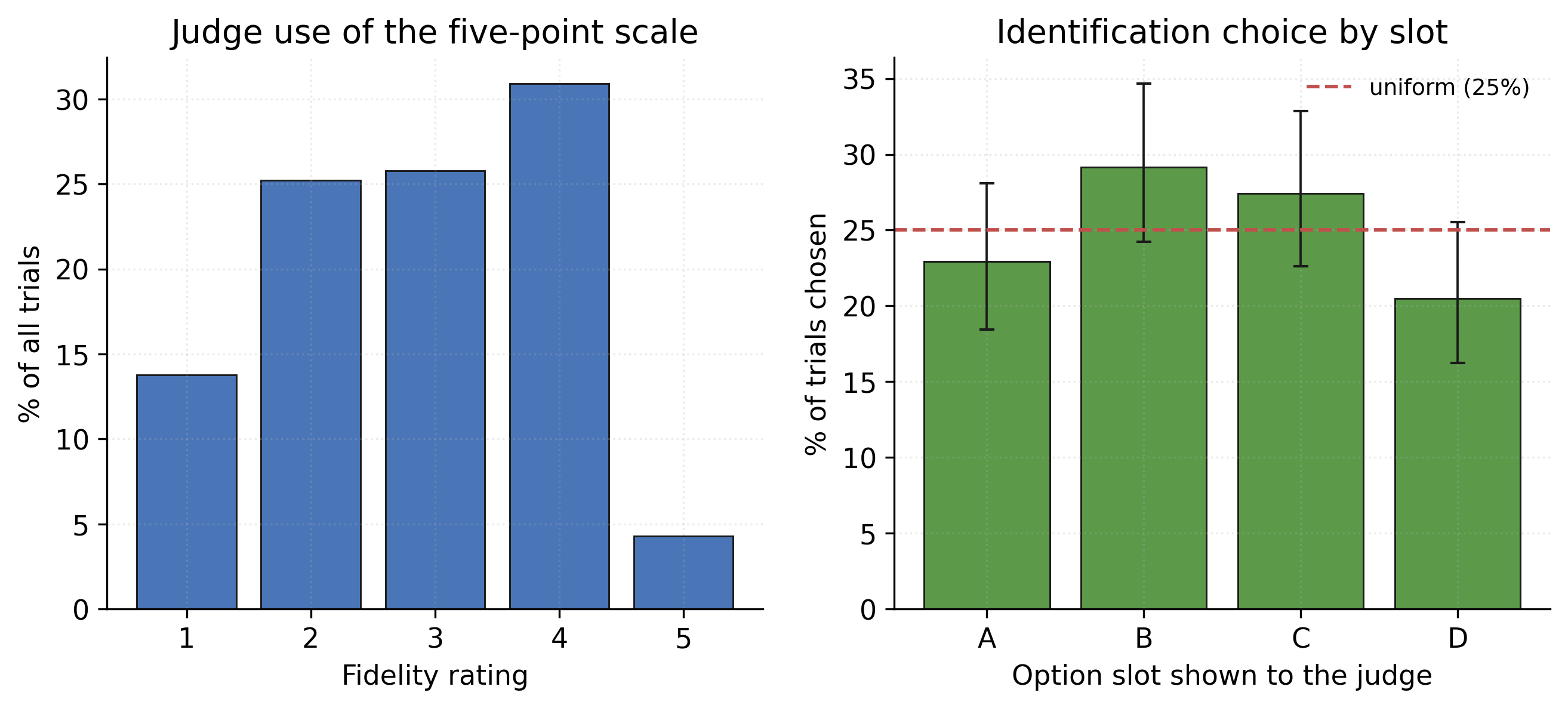}
\caption{Judge diagnostics. Left: use of the five-point scale across all
  trials. Right: identification choices by option slot, against the
  uniform 25\% expectation, with Wilson intervals.}
\label{fig:judge}
\end{figure}

\paragraph{The judge is usable but not sharp, and is not position-biased.}
Across the three seeds the fidelity judge achieved
ICC(2,1) $= 0.57$ over 960 targets, which by the conventional bands is
moderate reliability \citep{shrout1979icc,koo2016icc}. We report the
intraclass correlation rather than a chance-corrected agreement
coefficient because the ratings are ordinal and the three seeds are
exchangeable rather than fixed coders, which is the condition under
which the coefficient families diverge
\citep{krippendorff2004reliability,artstein2008agreement}. Exact agreement
between seed pairs ranged from 42.0\% to 43.2\% and agreement within one
scale point from 85.6\% to 87.4\%, with Spearman correlations of 0.55 to
0.60. The judge used the whole scale rather than collapsing to the
middle: 13.8\% of ratings were 1, 25.2\% were 2, 25.8\% were 3, 30.9\%
were 4 and 4.3\% were 5. On identification, choices did not cluster
significantly by option slot despite the per-trial shuffling
($\chi^2 = 7.37$, 3 d.f., $p = 0.061$), with slot shares of 22.9\%,
29.2\%, 27.4\% and 20.5\%. That $p$ is close enough to the conventional
threshold that we would not describe position bias as absent, only as
not detected at this sample size. Figure~\ref{fig:judge} shows both
diagnostics.

\subsection{Personalisation buys most where the default voice is furthest away}
\label{sec:priordistance}

The per-participant breakdown suggested a pattern that we then tested
directly. Fidelity in the generic condition is a direct measure of how
well the un-personalised assistant already writes like a given
participant: it is the score the default voice earns without any user
model at all. If personalisation is doing what it claims, its benefit
should be largest where that baseline is lowest.

\begin{figure}[t]
\centering
\includegraphics[width=0.82\textwidth]{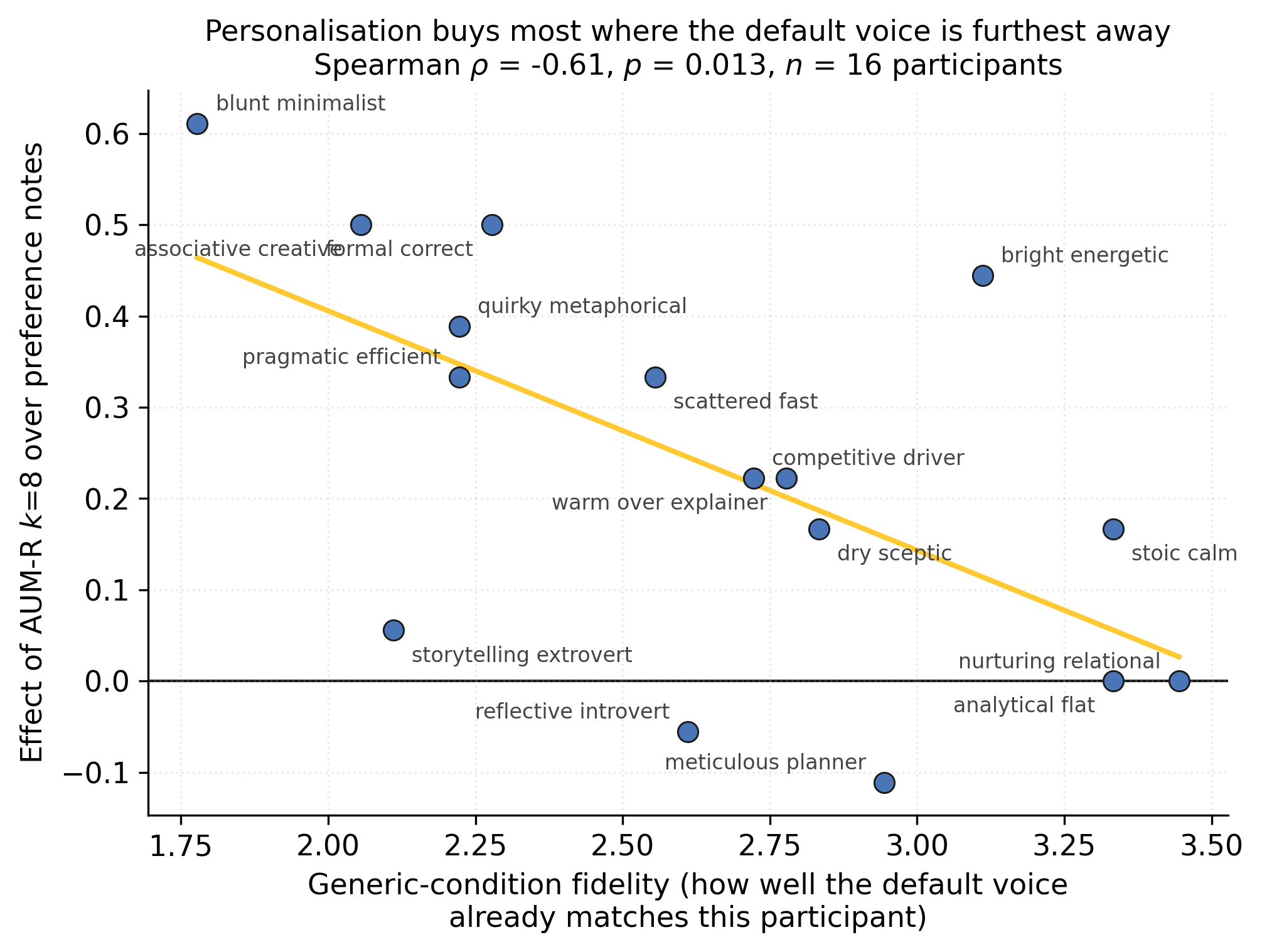}
\caption{The treatment effect against how well the un-personalised
  assistant already matches each participant. Points are the sixteen
  simulated participants; the line is an ordinary least squares fit.}
\label{fig:priordistance}
\end{figure}

It is. Across the sixteen participants, generic-condition fidelity
correlates negatively with the treatment effect
(Spearman $\rho = -0.61$, $p = 0.013$; Pearson $r = -0.60$,
$p = 0.013$), and the relationship holds at cell level with more power
and a smaller coefficient ($\rho = -0.27$, $p = 0.007$, $n = 96$).
Splitting participants at the median generic fidelity, those the default
serves worst gain $+0.333$ (95\% CI \ci{+0.181}{+0.486}) and those it
serves best gain $+0.139$ (\ci{+0.035}{+0.243}); the difference between
the two groups is in the expected direction but does not itself reach
significance (Mann-Whitney $p = 0.096$).
Figure~\ref{fig:priordistance} plots the relationship.

The extremes are interpretable. The blunt minimalist, whose lowercase
fragments and absent sign-offs are about as far from a default assistant
register as the battery contains, scores 1.78 under generic and gains
$+0.611$, the largest effect in the study. The meticulous planner, whose
numbered points and explicit dates are close to what an assistant
produces unprompted, scores 2.94 under generic and gains nothing
($-0.111$). The default voice already was that person.

One caveat is important and we state it rather than burying it. The same
correlation computed on the LUAR style metric rather than on judge
fidelity is in the same direction but does not reach significance
($\rho = -0.12$, $p = 0.242$). The pattern is therefore established in
the judge's ratings and not, on this sample, in the automatic measure.
We report it because its design implication is substantial and because
suppressing a result that did not replicate across both metrics would be
worse than reporting it with its caveat attached.

\begin{table}[t]
\centering
\caption{Effect of \aumr{} at $k=8$ over preference notes, by simulated
  participant, ordered by effect size. ``Generic'' is that
  participant's mean fidelity in the prompt-only condition, which
  measures how well the default voice already matches them.}
\label{tab:byperson}
\footnotesize
\renewcommand{\arraystretch}{1.15}
\begin{tabular}{@{}l r r c r@{}}
\toprule
\textbf{Participant} & \textbf{Generic} & \textbf{$\Delta$} &
\textbf{95\% CI} & \textbf{Favouring} \\
\midrule
blunt minimalist       & 1.78 & $+0.611$ & [$+0.222$, $+1.111$] & 83.3\% \\
associative creative   & 2.06 & $+0.500$ & [$+0.167$, $+0.889$] & 66.7\% \\
formal correct         & 2.28 & $+0.500$ & [$-0.111$, $+1.111$] & 66.7\% \\
bright energetic       & 3.11 & $+0.444$ & [$+0.333$, $+0.556$] & 100.0\% \\
quirky metaphorical    & 2.22 & $+0.389$ & [$-0.056$, $+0.889$] & 50.0\% \\
pragmatic efficient    & 2.22 & $+0.333$ & [$-0.111$, $+0.611$] & 83.3\% \\
scattered fast         & 2.56 & $+0.333$ & [$+0.111$, $+0.611$] & 66.7\% \\
warm over-explainer    & 2.78 & $+0.222$ & [$-0.056$, $+0.444$] & 66.7\% \\
competitive driver     & 2.72 & $+0.222$ & [$-0.056$, $+0.500$] & 50.0\% \\
stoic calm             & 3.33 & $+0.167$ & [$\phantom{+}0.000$, $+0.389$] & 33.3\% \\
dry sceptic            & 2.83 & $+0.167$ & [$\phantom{+}0.000$, $+0.389$] & 33.3\% \\
storytelling extrovert & 2.11 & $+0.056$ & [$-0.111$, $+0.222$] & 33.3\% \\
nurturing relational   & 3.44 & $\phantom{+}0.000$ & [$-0.222$, $+0.278$] & 16.7\% \\
analytical flat        & 3.33 & $\phantom{+}0.000$ & [$-0.278$, $+0.333$] & 33.3\% \\
reflective introvert   & 2.61 & $-0.056$ & [$-0.222$, $+0.111$] & 16.7\% \\
meticulous planner     & 2.94 & $-0.111$ & [$-0.556$, $+0.278$] & 16.7\% \\
\bottomrule
\end{tabular}
\end{table}

\section{Scaling Study of the Retriever}
\label{sec:scaling}

\subsection{Separating two questions}

Section~\ref{sec:nulls} reported that the swarm retriever neither
improved generated style over forward greedy nor reached a higher value
of its own objective, and that it spent roughly sixteen times as many
objective evaluations doing so. That is a result about one operating
point: a reduced 32-field instrument, a pool of about seventeen
admissible fields after component selection, and $k = 8$. It is not a
result about payload selection in general, and treating it as one would
be as much of an error as the reverse.

This section separates the two questions. It removes the language model
entirely and studies the optimisation problem on its own, sweeping the
instrument size $n$, the context budget $k$ and the redundancy weight
$\lambda$. The question it answers is narrow and answerable: for which
$(n, k, \lambda)$ does a population method reach a better value of
Equation~\ref{eq:objective} than forward greedy, and at what cost. No
API call is involved, so the whole sweep is free, deterministic from a
seed, and can be re-run by a reader.

\subsection{Instance model}

Field embeddings from real text are not isotropic Gaussian vectors.
Fields belonging to the same shell talk about related things and
therefore cluster, and that clustering is precisely what makes the
redundancy term non-trivial. Each synthetic instance is generated with
an explicit cluster structure. For $n$ fields distributed over $m$
shells in $d$ dimensions,
\begin{align}
\mathbf{c}_s &\sim \mathrm{Uniform}(\mathcal{S}^{d-1}),
  \qquad s = 1, \ldots, m, \\
\mathbf{f}_i &= \mathrm{normalise}\!\left(
   \rho\, \mathbf{c}_{s(i)} +
   \sqrt{1 - \rho^{2}}\; \boldsymbol{\varepsilon}_i \right),
   \qquad \boldsymbol{\varepsilon}_i \sim \mathcal{N}(\mathbf{0}, \sigma^2 I), \\
\mathbf{q} &= \mathrm{normalise}\!\left(
   \sum_{s \in \mathcal{A}} w_s \mathbf{c}_s +
   \sigma_q \boldsymbol{\varepsilon}_q \right),
\end{align}
where $s(i)$ is the shell of field $i$, $\mathcal{A}$ is a random subset
of shells of size two, and $w_s \sim \mathrm{Uniform}(0.5, 1.5)$. The
query construction mimics a task that draws on two shells, which is
what Table~\ref{tab:sigma} specifies. The parameter $\rho$ controls how
tightly a shell clusters; $\rho = 0$ recovers the isotropic case and
makes the redundancy term nearly inert. We use $\rho = 0.65$, $d = 64$,
$m = \max(2, \lfloor n/6 \rfloor)$ and 25 independent instances per
cell.

\subsection{Methods and budgets}

Six retrievers are compared, plus exhaustive search wherever
$\binom{n}{k} \le 60{,}000$.

\begin{description}
\item[Dense top-$k$] the relevance-only maximiser of
  Equation~\ref{eq:rel}, one objective evaluation.
\item[Greedy] forward greedy on the full objective.
\item[Random] $k$ fields drawn uniformly, the floor.
\item[AFSA (study budget)] the swarm with exactly the population and
  iteration counts used in Section~\ref{sec:method}, $p = 12$ and
  $T = 8$, held fixed as $n$ grows. This is what the simulation
  actually ran.
\item[AFSA (scaled budget)] population and iterations grown with the
  search space:
  $p = \mathrm{clip}(\lceil 2 \log \binom{n}{k} \rceil, 12, 60)$ and
  $T = \mathrm{clip}(\lceil 1.2 \log \binom{n}{k} \rceil, 8, 40)$. The
  constants are set so that $(n, k) = (32, 8)$ reproduces roughly the
  study's own budget and everything larger receives proportionally
  more.
\item[AFSA (greedy warm start)] the scaled-budget swarm with one member
  of the initial population replaced by the greedy solution. This is
  the obvious hybrid and the fairest test of whether a population
  method can add anything on top of greedy.
\end{description}

\noindent The sweep covers $n \in \{16, 32, 64, 128, 256\}$,
$k \in \{4, 8, 16, 32\}$ with $k \le n/2$, and
$\lambda \in \{0, 0.15, 0.30, 0.60\}$, giving 68 cells and 1{,}700
instances. Quality is reported as the relative gap to the best solution
found for that instance by any method, which equals the exhaustive
optimum wherever that is available.

\subsection{Results}

\begin{table}[t]
\centering
\caption{Relative gap to the best solution found, in percent, averaged
  over $k$ and 25 instances per cell, at $\lambda = 0.15$. Lower is
  better. Forward greedy is within 0.1\% of the best solution at every
  instrument size; the swarm at the study's fixed budget degrades
  steeply.}
\label{tab:scalinggap}
\small
\renewcommand{\arraystretch}{1.2}
\begin{tabular}{@{}r r r r r r@{}}
\toprule
\textbf{$n$} & \textbf{Greedy} & \textbf{Dense top-$k$} &
\textbf{AFSA study} & \textbf{AFSA scaled} & \textbf{AFSA warm} \\
\midrule
16  & 0.10 & 0.65 & \phantom{0}0.44 & \phantom{0}0.04 & 0.00 \\
32  & 0.08 & 0.46 & \phantom{0}5.66 & \phantom{0}0.34 & 0.00 \\
64  & 0.01 & 0.35 & 20.94 & \phantom{0}0.57 & 0.01 \\
128 & 0.07 & 0.47 & 33.51 & \phantom{0}5.46 & 0.05 \\
256 & 0.03 & 0.53 & 43.23 & 12.59 & 0.01 \\
\bottomrule
\end{tabular}
\end{table}

\begin{figure}[t]
\centering
\includegraphics[width=0.98\textwidth]{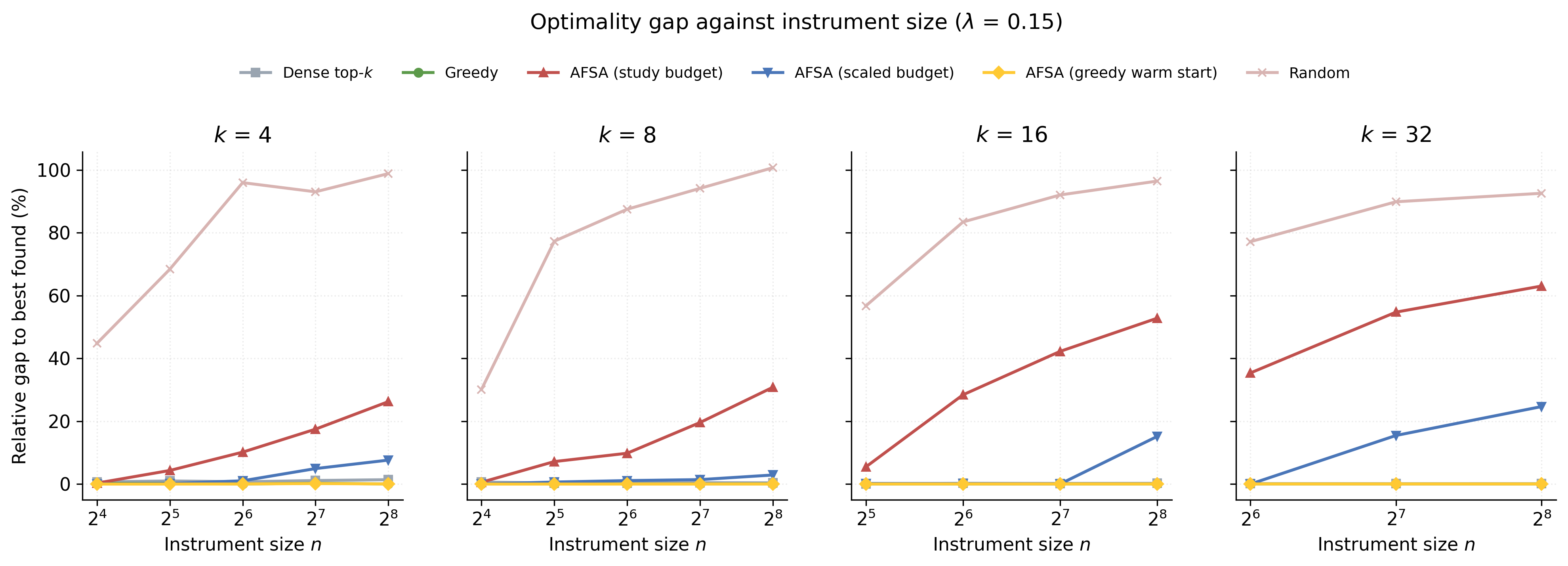}
\caption{Relative optimality gap against instrument size, one panel per
  budget $k$, at $\lambda = 0.15$. Greedy and the warm-started swarm are
  indistinguishable from the best solution found at every scale; the
  fixed-budget swarm used in the simulation study degrades steeply.}
\label{fig:scalinggap}
\end{figure}

\paragraph{Forward greedy is very hard to beat.}
Across the entire sweep, greedy stayed within 0.10\% of the best
solution found (Table~\ref{tab:scalinggap},
Figure~\ref{fig:scalinggap}). Where exhaustive search was tractable, it
returned the exact optimum in 77.7\% of instances. It is not a straw
man, and the fact that Equation~\ref{eq:objective} is not submodular in
general \citep{nemhauser1978analysis} does not appear to cost it much in
practice on instances with this cluster structure.

\paragraph{The study's swarm budget, not the swarm, is what failed.}
AFSA at the fixed budget of Section~\ref{sec:method} degraded from a
0.44\% gap at $n = 16$ to 43.23\% at $n = 256$, and found the exact
optimum in only 46.3\% of the tractable instances against greedy's
77.7\%. It is a poor optimiser at any $n$ beyond the smallest, and it
was already worse than greedy at the $n = 32$ operating point the
simulation used. This is the direct explanation of the null in
Section~\ref{sec:nulls}: the swarm did not lose because population
methods are unsuited to this problem, it lost because it was given a
budget that does not grow with the search space.

\paragraph{Growing the budget helps, up to a point.}
The scaled-budget swarm stayed within 0.6\% up to $n = 64$ and found the
exact optimum in 84.7\% of tractable instances, exceeding greedy's
77.7\%. Beyond that it degrades again (5.46\% at $n = 128$, 12.59\% at
$n = 256$), because even a budget growing with $\log \binom{n}{k}$ is a
vanishing fraction of the search space.

\paragraph{Warm-starting from greedy dominates.}
The warm-started swarm stayed within 0.05\% at every scale tested and
found the exact optimum in 96.3\% of tractable instances. Against greedy
head to head it won or tied in 100\% of instances (16.0\% strict wins,
84.0\% ties, mean $\Delta = +0.001$), which is what one expects of a
method that starts from greedy's answer and keeps the better of the two.
It costs roughly seven times greedy's evaluations
(Table~\ref{tab:scalingcost}).

\begin{table}[t]
\centering
\caption{Objective evaluations at $k = 8$, $\lambda = 0.15$, averaged
  over 25 instances. Dense top-$k$ and random require a single
  evaluation. Exhaustive search is tractable only at $n = 16$.}
\label{tab:scalingcost}
\small
\renewcommand{\arraystretch}{1.2}
\begin{tabular}{@{}r r r r r r@{}}
\toprule
\textbf{$n$} & \textbf{Greedy} & \textbf{AFSA study} &
\textbf{AFSA scaled} & \textbf{AFSA warm} & \textbf{Exhaustive} \\
\midrule
16  & \phantom{0,}101 & 417 & \phantom{0,}981 & \phantom{0,}1{,}083 & 12{,}871 \\
32  & \phantom{0,}229 & 468 & 3{,}112 & \phantom{0,}3{,}375 & n/a \\
64  & \phantom{0,}485 & 521 & 6{,}556 & \phantom{0,}7{,}112 & n/a \\
128 & \phantom{0,}997 & 479 & 10{,}933 & 11{,}926 & n/a \\
256 & 2{,}021 & 375 & 13{,}963 & 16{,}002 & n/a \\
\bottomrule
\end{tabular}
\end{table}

\begin{figure}[t]
\centering
\includegraphics[width=0.84\textwidth]{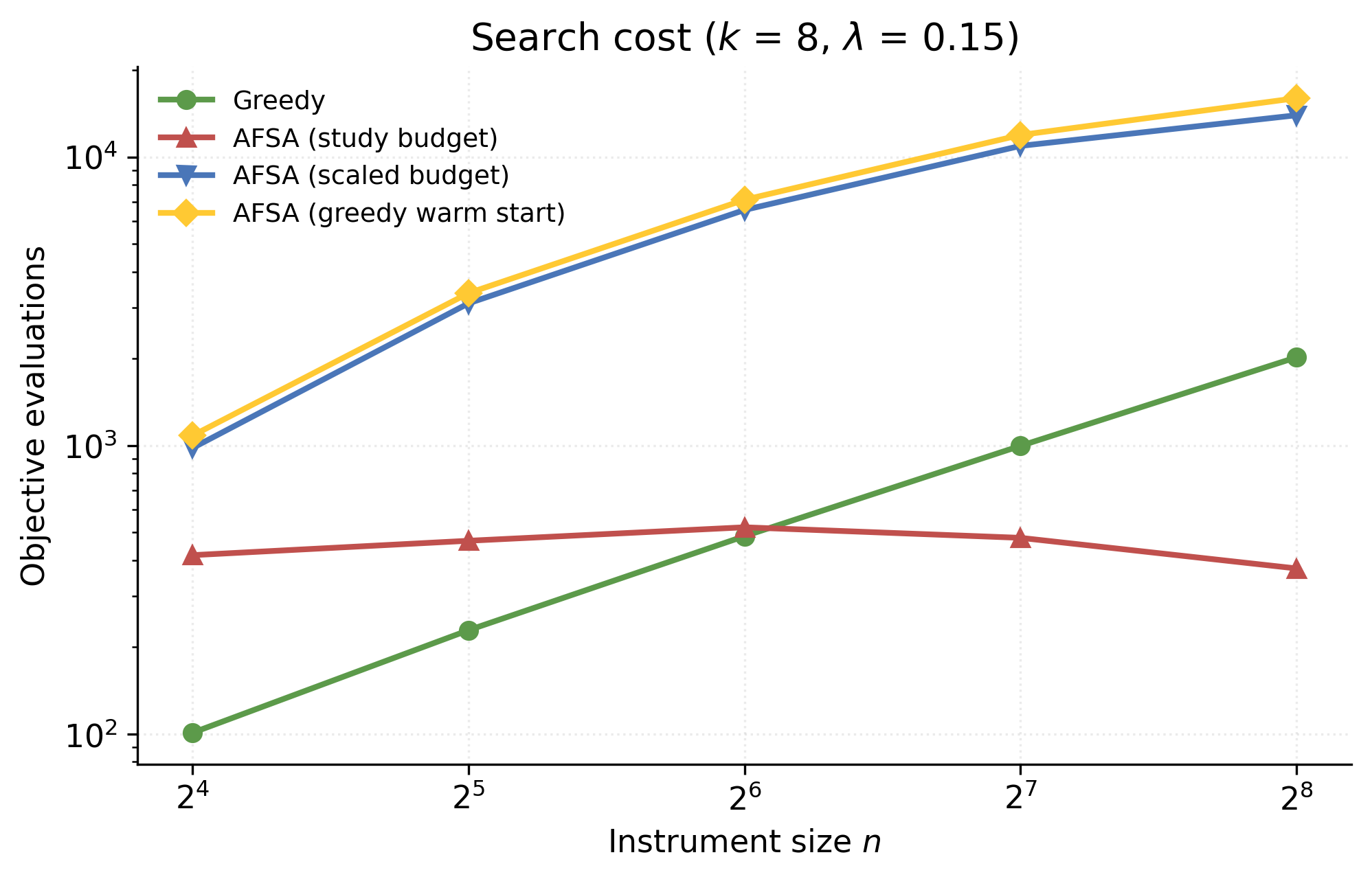}
\caption{Search cost in objective evaluations at $k = 8$,
  $\lambda = 0.15$. The fixed-budget swarm spends a roughly constant
  amount regardless of instrument size, which is exactly why its
  solution quality collapses as $n$ grows.}
\label{fig:scalingcost}
\end{figure}

\begin{figure}[t]
\centering
\includegraphics[width=0.84\textwidth]{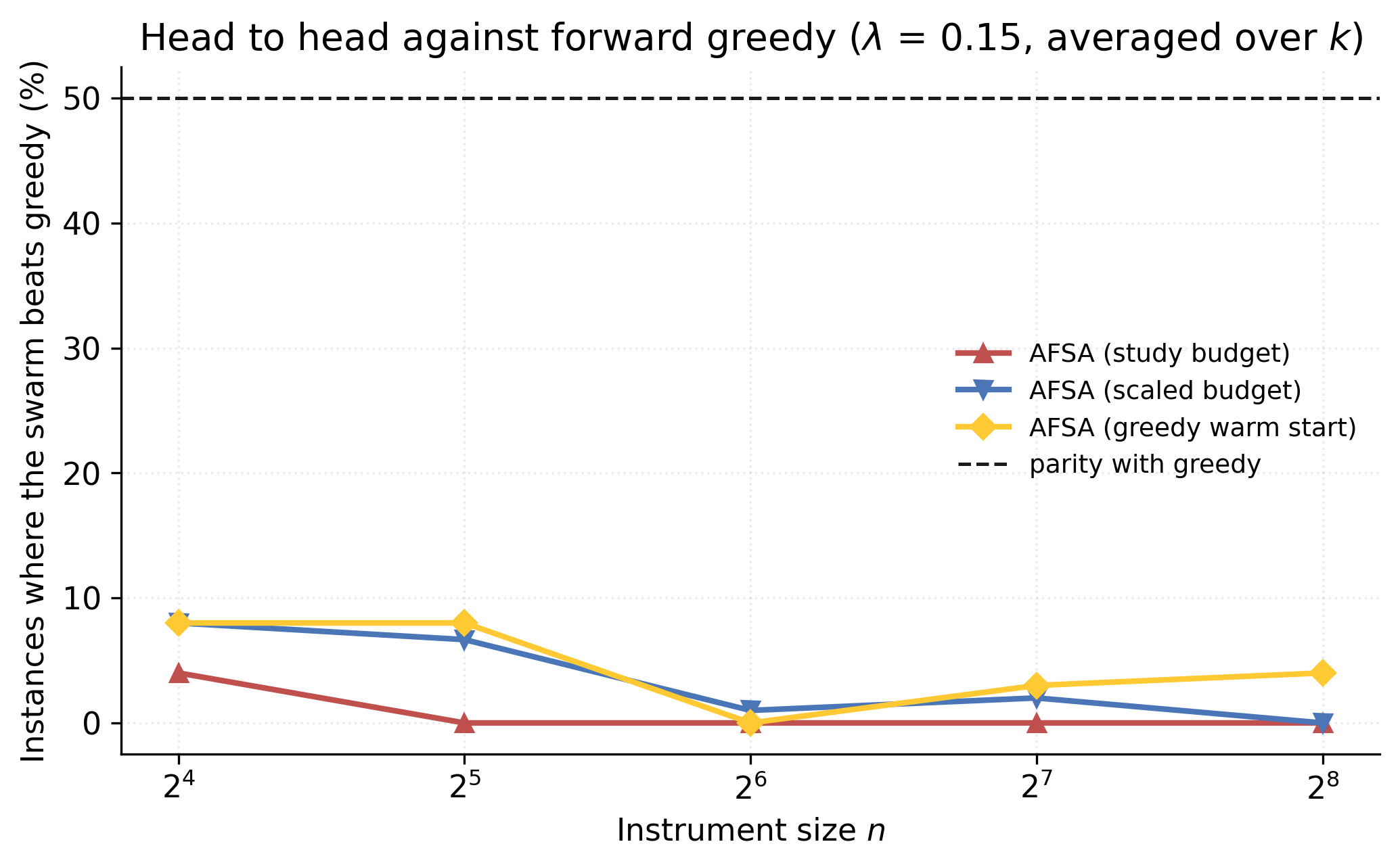}
\caption{Head-to-head strict win rate against forward greedy at
  $\lambda = 0.15$, averaged over $k$. Strict wins are rare because ties
  are common: greedy and the warm-started swarm frequently return the
  same subset.}
\label{fig:scalingwin}
\end{figure}

\begin{figure}[t]
\centering
\includegraphics[width=0.84\textwidth]{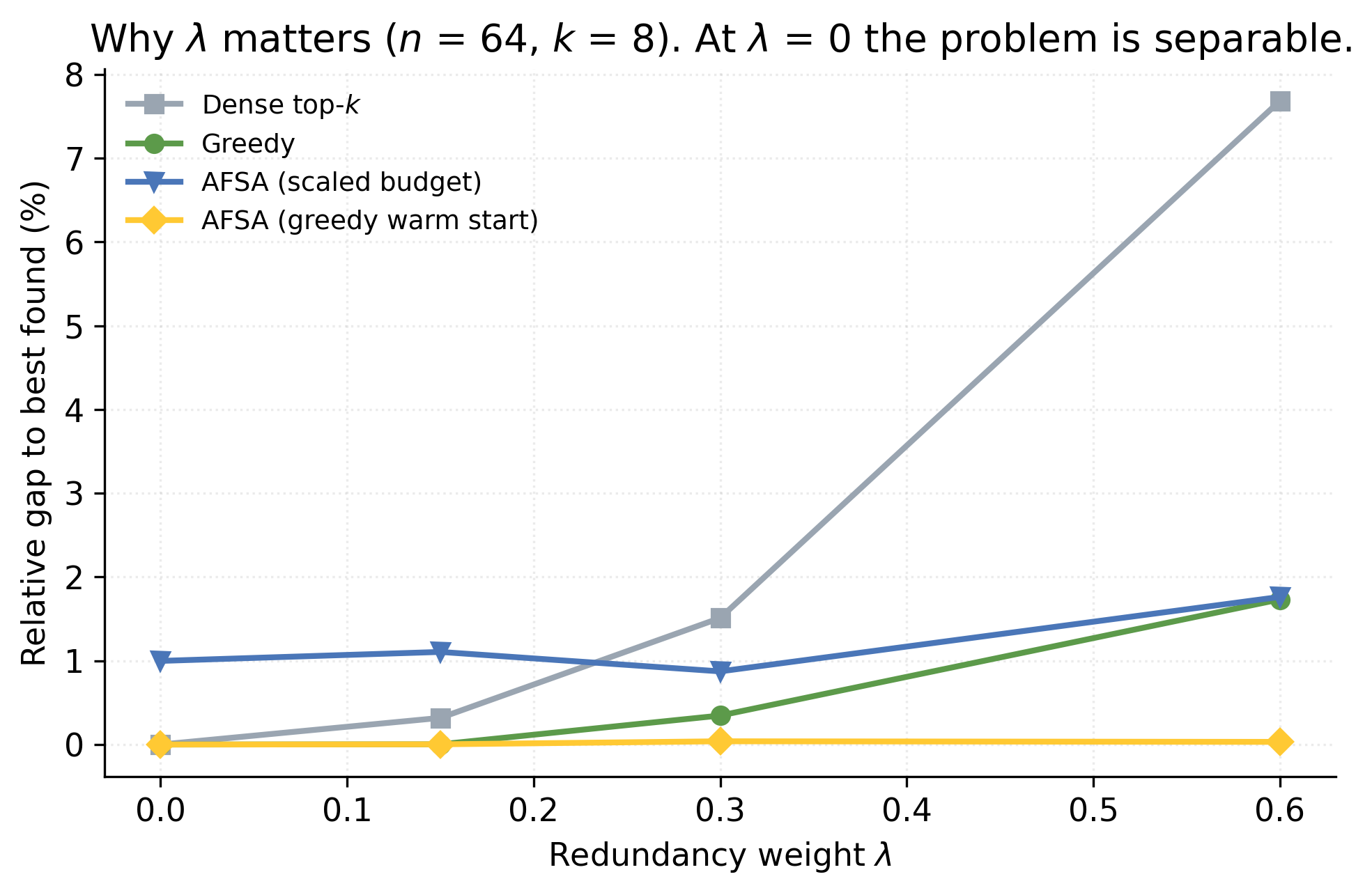}
\caption{Effect of the redundancy weight at $n = 64$, $k = 8$. At
  $\lambda = 0$ the problem is a sort; as $\lambda$ rises,
  relevance-only selection loses ground and combinatorial search starts
  to earn its cost.}
\label{fig:scalinglambda}
\end{figure}

\paragraph{The redundancy weight decides whether any of this matters.}
Figure~\ref{fig:scalinglambda} and Table~\ref{tab:scalinglambda} isolate
$\lambda$ at $n = 64$, $k = 8$. At $\lambda = 0$ the objective is
separable, dense top-$k$ is exactly optimal by construction, and every
search method is wasted effort. As $\lambda$ rises the relevance-only
solution degrades: a 0.32\% gap at $\lambda = 0.15$, 1.51\% at
$\lambda = 0.30$ and 7.68\% at $\lambda = 0.60$, at which point greedy
also begins to lose ground (1.73\%). This locates the regime in which
combinatorial search over a payload is worth performing at all: it is
governed by how much a system penalises near-duplicate fields, and at
the $\lambda = 0.15$ we used, the penalty is mild enough that dense
top-$k$ is nearly optimal. That is the second half of the explanation
for the null against dense top-$k$ in Section~\ref{sec:nulls}.

\begin{table}[t]
\centering
\caption{Relative gap to the best solution found, in percent, against
  the redundancy weight $\lambda$ at $n = 64$, $k = 8$. At
  $\lambda = 0$ the objective is separable and sorting is optimal.}
\label{tab:scalinglambda}
\small
\renewcommand{\arraystretch}{1.2}
\begin{tabular}{@{}r r r r r@{}}
\toprule
\textbf{$\lambda$} & \textbf{Dense top-$k$} & \textbf{Greedy} &
\textbf{AFSA scaled} & \textbf{AFSA warm} \\
\midrule
0.00 & 0.00 & 0.00 & 1.00 & 0.00 \\
0.15 & 0.32 & 0.00 & 1.11 & 0.00 \\
0.30 & 1.51 & 0.35 & 0.87 & 0.04 \\
0.60 & 7.68 & 1.73 & 1.76 & 0.03 \\
\bottomrule
\end{tabular}
\end{table}

\subsection{What this means for deployment}

Three recommendations follow, and we state them as recommendations
because they change what we would build.

\begin{enumerate}
\item \textbf{Use forward greedy.} At any instrument size we tested it
  is within 0.1\% of the best solution found, at a cost of
  $O(k n)$ evaluations. For an on-device retriever over sixty fields
  this is a few hundred dot products and is not worth improving on.

\item \textbf{If a population method is used, warm-start it from
  greedy.} A cold-started swarm is strictly worse than greedy at every
  scale beyond the smallest. A warm-started one never loses, reaches the
  exhaustive optimum in 96\% of tractable instances, and costs about
  seven times greedy. Whether that is worth paying depends on whether
  the objective value is what one actually cares about, which
  Section~\ref{sec:nulls} suggests it may not be.

\item \textbf{Set $\lambda$ deliberately, or drop the redundancy term.}
  Below roughly $\lambda = 0.3$ the redundancy penalty rarely changes
  which fields win, and a system that is not going to tune $\lambda$
  should use dense top-$k$ and save the search entirely. Above it, the
  penalty matters and greedy is the right tool.
\end{enumerate}

\noindent The broader methodological point is that the null in
Section~\ref{sec:nulls} was not uninformative. Reading it required
taking the optimisation question out of the language-model setting,
where a downstream metric with a moderate effect size cannot resolve a
0.2\% difference in objective value, and asking it where it could be
answered.

\section{Discussion}
\label{sec:discussion}

\subsection{A context result, and why it is a privacy result}

The finding we would put first is not the effect size but the ratio.
Eight retrieved fields, 211 tokens, matched the fidelity of injecting
the entire 32-field user model, 915 tokens. Nothing in the study
suggests that the remaining twenty-four fields were doing work that the
eight did not.

For efficiency this is a modest saving. For architecture it is the whole
argument. Section~\ref{sec:aum} committed \aum{} to local residence: the
store sits on the user's device or in a user-controlled vault, never in
a third-party log. That commitment is easy to state and hard to keep if
the whole model must be transmitted on every complex request, because a
model that is transmitted in full on every turn has, for practical
purposes, been uploaded. A budgeted payload changes the character of the
commitment. If eight fields suffice, disclosure per query is bounded by
construction, the bound is measurable, and the fields that never leave
are disproportionately the inner-shell ones, which are exactly the ones
whose exposure carries the greatest cost.

\begin{figure}[t]
\centering
\includegraphics[width=0.90\textwidth]{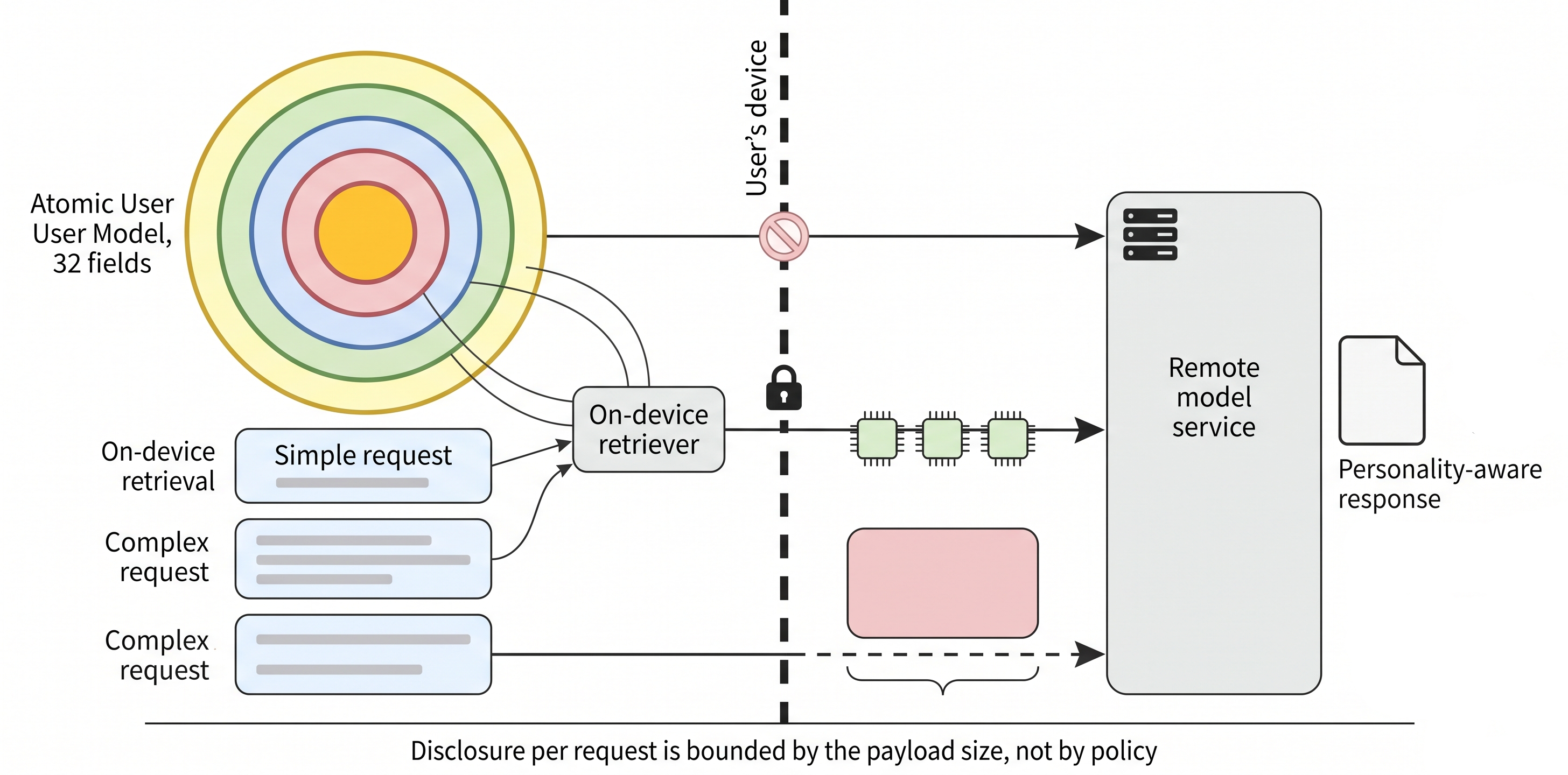}
\caption{What crosses the device boundary. The full 32-field model, 915
  tokens in the study, never leaves the device. A simple request leaves
  nothing. A complex request sends a payload of eight fields, 211
  tokens, which the user can be shown before it is sent. The bound on
  disclosure per query is a measured quantity rather than a policy
  promise.}
\label{fig:dataflow}
\end{figure}

In the vocabulary of contextual integrity
\citep{nissenbaum2011contextual}, the design question is not whether the
user model is secret but whether each flow out of it matches the norms
of the context that produced it. A retrieval stage that returns a small
task-conditioned slice is a mechanism for making flows contextual, and
the size of the slice is the parameter that governs it. This is what we
mean by privacy by architecture rather than privacy by policy: the
former is a property one can measure in tokens, and we measured it.

There is a second consequence for scrutability. Eight short
natural-language strings can be shown to a user before they are sent;
thirty-two cannot, and a JSON blob certainly cannot. The
scrutable-personalisation literature has long argued that users should
be able to inspect and correct the model a system holds of them
\citep{kay2012scrutinize,czarkowski2002scrutable,bakalov2010introspective},
and evidence that transparency and control change acceptance and
behaviour is not new \citep{cramer2008transparency,harper2015control}.
The obstacle has usually been that the model is too large or too
statistical to show. A budgeted payload of readable sentences removes
that obstacle for the per-request case: a user can be shown exactly what
the assistant was told about them for this draft.

\subsection{What the null results bound, and what they do not}

Four controls returned null, and Section~\ref{sec:scaling} explained
why. We think the joint reading is more interesting than any of them
individually.

The positive results establish that injecting a structured,
task-plausible slice of a user model changes generated style
substantially relative to both a generic baseline and the flat
preference notes deployed systems store. The nulls establish that, at
this scale, nothing about \emph{which} slice or about \emph{how hard one
searches for it} adds anything further. Together they locate the effect
in the representation rather than in the retrieval machinery. That is a
narrower claim than the one the architecture invites, and it has a
practical implication we would act on: a first deployment should use
forward greedy over a well-specified schema, and should not spend
engineering effort on the selection algorithm until the schema is
larger and the redundancy penalty is doing real work.

Two boundaries on the nulls deserve emphasis. First, they are nulls at
an operating point with a pool of about seventeen fields, and
Section~\ref{sec:scaling} shows the optimisation landscape is genuinely
flat there: greedy is within 0.1\% of the exhaustive optimum, so there
was nothing for a better search to recover. They are not evidence that
selection is irrelevant at the full sixty-field specification, and we do
not present them as such. Second, a null against random selection within
a $\sigma(t)$-filtered pool is not a null against random selection from
the whole model. The pool was already restricted to shells the task
touches, so ``random'' here means random among plausible fields, which
is a considerably stronger baseline than it sounds.

\subsection{Personalisation is worth most to the people the default serves least}

The result with the clearest design consequence is the one in
Section~\ref{sec:priordistance}: the benefit of the user model was
largest exactly for those participants whose voice the un-personalised
assistant reproduced worst ($\rho = -0.61$, $p = 0.013$ at participant
level). The blunt minimalist gained $+0.611$; the meticulous planner,
whose numbered and dated register a default assistant already produces,
gained nothing.

Read as a statistic this is a moderate correlation on sixteen points
that did not replicate on the automatic style metric. Read as a design
observation it says something sharper. A default assistant voice is not
neutral. It is a particular register, and being well served by an
un-personalised assistant means happening to write the way that register
writes. Personalisation, on this reading, is not a uniform improvement
distributed evenly across users; it is a correction whose size depends
on how far a person sits from the model's prior.

This connects to a finding that motivates the whole project.
\citet{jakesch2023opinionated} showed that co-writing with an
opinionated language model shifts users' own expressed views. If a
default assistant voice can move what a person says, then users whose
register is furthest from the default face the largest pull towards it,
and are also the users for whom a model of their own voice does the most
work. Personalisation in that framing is not a convenience feature but a
countermeasure, and the task-level pattern in
Table~\ref{tab:bytask} says the same thing at a different granularity:
the effect was near zero on the deadline-extension email, the most
conventionalised genre in the battery, where the assistant's default
already is the correct answer.

\subsection{Implications for the design of personalised assistants}

Four implications follow that we would apply to a system.

\begin{enumerate}
\item \textbf{Store the person, retrieve the preference.} The unit of
  storage should be the stable layer, and the situational preference
  should be derived at generation time. This is what makes an unseen
  task addressable and a stale preference re-interpretable rather than
  merely old.

\item \textbf{Budget the payload, and show it.} A small
  task-conditioned payload costs nothing measurable in quality, bounds
  disclosure per request, and is small enough to display. A system that
  shows the user the eight sentences it used has made scrutability
  concrete rather than aspirational.

\item \textbf{Route simple requests around the model.} The classifier
  stage routed factual lookups to bypass with 98.8\% accuracy. This is
  cheap to implement and eliminates a class of disclosure that buys
  nothing, and we would regard omitting it as a design error rather than
  a missed optimisation.

\item \textbf{Expect uneven benefit, and design for the tail.} If the
  gain concentrates on users whose voice the default serves worst, then
  average-case evaluation will systematically understate the value of
  personalisation to the users who need it most. Reporting a mean effect
  across a user population is not sufficient; the distribution is the
  result.
\end{enumerate}

\subsection{Limitations}
\label{sec:limitations}

\paragraph{Simulated participants.}
The most important limitation is the one stated first in
Section~\ref{sec:whysimulate}. Every participant here is
language-model-simulated. A judge model rating on behalf of a persona
description has an easier task than a person recognising their own
voice, and the absolute fidelity means sit near the middle of a
five-point scale for every condition, including the best. The critical
literature on synthetic participants
\citep{bisbee2024synthetic,wang2025flatten} applies to this study in
full, and no result here should be transferred to a claim about human
users.

\paragraph{The personas are not a sample.}
Sixteen personas written to be distinguishable are a friendlier
population than any real one. The idiolect markers were authored, so the
style differences the study measures are differences an author put
there. In particular, conclusions about \emph{which} users benefit
should not be transferred to demographic groups, since the personas
carry an author's assumptions about how people with a given trait
profile write.

\paragraph{Judge reliability is moderate.}
ICC(2,1) of 0.57 across seeds is moderate rather than good
\citep{koo2016icc}. Agreement within one scale point was 86\%, so the
judge is consistent about direction and imprecise about magnitude, which
is adequate for paired contrasts and inadequate for absolute claims.
Position bias was not detected but the test was close to threshold
($p = 0.061$).

\paragraph{Acquisition is unspecified.}
\aum{} specifies the fields a user representation should contain and not
how they are populated. Self report is unreliable for inner-shell
entries \citep{vazire2010soka,connelly2010other,baumeister2007selfreports};
passive observation raises the privacy concerns the architecture exists
to avoid; and clinical assessment does not scale. We treat this as an
open problem on which \aum{} as a target representation is agnostic, but
it is the largest gap between this paper and a deployable system.

\paragraph{Update is unspecified.}
Fields carry a nominal update period $\tau$, but no mechanism is given
for detecting that a field has become stale or for revising it without
re-running the whole instrument. Given that the motivating example in
Section~\ref{sec:intro} is precisely a staleness failure, this is a
conspicuous omission.

\paragraph{Text only, and one language.}
The specification is text-based. Many personality cues are non-verbal:
tone of voice, facial expression, pacing, response latency. The study
also runs entirely in English, and the personality frameworks \aum{}
draws on were developed predominantly in specific populations, so
cross-cultural validity of the inner tiers is untested
\citep{mccrae2002crosscultural}.

\paragraph{Single generator and judge.}
One generator model and one judge model were used. Whether the effect
sizes transfer across model families is unknown, and a judge from a
single family may carry systematic preferences that a panel would
average out.

\paragraph{Lifespan and capacity.}
\aum{} as specified assumes an adult user with an articulable identity.
It does not represent young children, or persons with significant
cognitive impairment, and we do not think it should be extended to them
without a separate ethical analysis.

\subsection{Future work}
\label{sec:future}

The necessary next study is the human one this simulation substitutes
for. Its shape is determined by what the simulation cannot answer:
participants populate their own \aum{}, write their own references for
the same six tasks, and rate outputs from the same conditions blind,
with the primary measure being self-rated fidelity and forced-choice
identification of their own style. Section~\ref{sec:ethics} states the
conditions under which we think that study can be run.

Three further directions follow from the results rather than from the
gaps. First, acquisition: whether an \aum{} can be populated
incrementally from ordinary interaction without the passive observation
the architecture is meant to avoid, perhaps by asking the user a small
number of well-chosen questions at moments when the answer is cheap to
give. Second, update: detecting staleness by watching for conflict
between a stored field and new evidence, which is what the cross-shell
internal-conflicts entry exists to support. Third, scale: whether the
selection nulls of Section~\ref{sec:nulls} persist at the full
sixty-field specification, where Section~\ref{sec:scaling} suggests the
optimisation landscape is no longer flat.

\section{Ethical Considerations}
\label{sec:ethics}

A structured user model that records inner-shell content such as fears,
difficult episodes and core values is among the most sensitive artefacts
a person could hold. Building one creates risks that a preference list
does not, and we do not think those risks are adequately answered by
noting that the model is opt-in.

\paragraph{Three binding constraints.}
We bind \aum{} to three. \textbf{Local residence}: the store sits on the
user's device or in a user-controlled vault, never in a third-party log,
and only the retrieved payload leaves it. Section~\ref{sec:results}
makes this quantitative rather than aspirational: a payload of eight
fields reaches the fidelity of the whole model, so budgeted retrieval
bounds what any single query can disclose at no measured cost in
quality. \textbf{User authorship}: every field is inspectable, editable
and deletable, and eight short strings are few enough to display before
they are sent. \textbf{Purpose restriction}: \aum{} is intended for
task-style personalisation under the user's direction, not for
profiling, advertising, surveillance, or persuasion against the user's
interest.

\paragraph{The misuse the architecture cannot prevent.}
A high-resolution model of a person is valuable to anyone who wants to
influence that person, and an operator who ignores the local-residence
constraint gains a profiling asset of a kind that does not currently
exist at scale. Our own results sharpen this rather than softening it:
Section~\ref{sec:priordistance} found the effect largest for
participants whose voice the default serves worst, and those are also
the people for whom an accurate model would be most identifying. Users
of LLM-based conversational agents already disclose a great deal and
reason about the risks only partially \citep{zhang2024fairgame}, and
language models can leak personal information present in training data
\citep{huang2022leaking}.

We do not think this argues against specifying the representation.
Opaque, implicit user models already sit inside large-scale systems with
weaker user control, and an explicit schema at least makes the contents
auditable and contestable. But mitigation has to be architectural, in
where the data lives, as well as procedural, in who may query it and for
what, and a paper that proposes the representation without saying so
would be incomplete.

\paragraph{Human participants.}
No human participants were involved in this work and no human-subjects
data was collected. Every participant in Sections~\ref{sec:method} and
\ref{sec:results} is language-model-simulated and every persona is
fictional and declared in code (Appendix~\ref{app:personas}). The human
study described in Section~\ref{sec:future} would require informed
consent, the right to withdraw with deletion of all collected \aum{}
fields, and institutional ethics approval, since participants would
disclose identity-level content about themselves. Inner-shell fields
should be optional by default and skippable without penalty, and
participants should be able to review their populated instrument before
any of it is used.

\paragraph{A concern specific to simulation.}
Personas written by an author carry that author's assumptions about how
people of a given trait profile write. Ours are deliberately varied but
are not a sample of any population. Conclusions about which users
benefit from personalisation should not be transferred to demographic
groups on the basis of this study, and we have tried to phrase
Section~\ref{sec:priordistance} so that it cannot be read that way.

\section{Conclusion}
\label{sec:conclusion}

Personalisation of language-model assistants currently stores what a
user has done. We argued that it should store the person from whom those
doings come, and that the two are not interchangeable because
personality is comparatively stable while preferences are situational.
We named the phenomenon that makes the difference visible, personality
seepage, in which a prompt carries a fingerprint the assistant mirrors
without being able to read. We specified the Atomic User Model, a
five-tier structured and human-readable representation whose metaphor
encodes differential stability and differential observability, and a
personality-aware retrieval pipeline that treats it as an index rather
than a prompt prefix.

In a simulation study with sixteen language-model-simulated
participants, six style-sensitive tasks and three seeds, retrieving
eight fields matched the style fidelity of injecting the whole user
model at 23\% of the context, improved on the flat preference notes
deployed systems store by 0.24 points on a five-point scale, and raised
forced-choice identification of the participant's own voice from 14.9\%
to 42.7\% against 25\% chance. Four controls returned null, and a
synthetic scaling study of the retriever explained why: at the operating
point used, forward greedy is already within 0.1\% of the exhaustive
optimum, so there was nothing for a population method to recover. The
effect lives in the representation, not in the search over it.

The result we would carry forward is neither the effect size nor the
null. It is that a small, readable, task-conditioned payload is enough,
because that is what makes a user model something a person can keep on
their own device and read before it is sent. And it is that the benefit
was largest for the participants whose voice the default assistant
served worst, which suggests that personalisation is least valuable to
the users who resemble the model's prior and most valuable to everyone
else.

\ifblind\else
\section*{CRediT authorship contribution statement}

\textbf{B. Sankar}: Conceptualization, Methodology, Software,
Formal analysis, Investigation, Writing, original draft,
Writing, review and editing, Supervision, Project administration.
\textbf{Deepthika S}: Methodology, Software, Validation,
Investigation, Data curation, Writing, review and editing.
\textbf{Pawni Yadav}: Methodology, Software, Validation,
Investigation, Visualization, Writing, review and editing.
\textbf{Amogh A S}: Software, Validation, Data curation,
Visualization, Writing, review and editing.
\fi


\section*{Declaration of competing interest}

The authors declare that they have no known competing financial
interests or personal relationships that could have appeared to
influence the work reported in this paper.

\section*{Declaration of generative AI in the writing process}

The studies reported in Sections~\ref{sec:method} to \ref{sec:scaling}
use large language models as the object of study and as simulated
participants; this use is described in full in those sections. In
preparing this manuscript the authors additionally used a large language
model to assist with drafting and editing of prose. The authors reviewed
and edited all content and take full responsibility for the content of
the publication.

\ifblind\else
\section*{Data and code availability}

The simulation code, the persona declarations, every prompt, and the run manifest and ledger, all per-trial records, the analysis scripts, and the scaling benchmark are available at \url{https://github.com/REPOSITORY-URL-HERE}.
Each number reported in Sections~\ref{sec:results} and \ref{sec:scaling} is reproducible from the released artefacts and a seed.
\fi
\section*{Funding declaration}

This research did not receive any specific grant from funding agencies in the public, commercial, or not-for-profit sectors.




\bibliographystyle{elsarticle-harv}
\bibliography{references}

\newpage
\appendix

\section{The Full Atomic User Model Specification}
\label{app:spec}


Table~\ref{tab:fullspec} gives the full specification. Fields marked
$\dagger$ form the reduced 32-field instrument used in both studies and
listed in Table~\ref{tab:instrument}; they are the only fields for which
this paper reports evidence. The remainder are part of the
specification but were not exercised, and we flag that distinction
rather than presenting the whole schema as validated.

\begin{table}[!ht]
\centering
\caption{The full \aum{} specification, 60 fields across five tiers plus
  cross-shell integration. $\dagger$ marks the 32 fields in the reduced
  instrument used in both studies. $\tau$ is the nominal update period
  for the tier.}
\label{tab:fullspec}
\footnotesize
\renewcommand{\arraystretch}{1.1}
\begin{tabular}{@{}p{0.17\textwidth} p{0.07\textwidth} p{0.70\textwidth}@{}}
\toprule
\textbf{Tier} & \textbf{$\tau$} & \textbf{Fields} \\
\midrule
\textbf{Nucleus} \newline core identity &
years &
Core Values$\dagger$; Fundamental Beliefs$\dagger$; Identity
Anchors$\dagger$; Moral Framework$\dagger$; Existential
Orientation$\dagger$; Life Purpose$\dagger$; Self-Concept;
Non-Negotiables; Sources of Meaning; Legacy Orientation \\
\addlinespace
\textbf{\shellone{}} \newline psychological core &
months &
Big Five Profile$\dagger$; Emotional Architecture$\dagger$; Attachment
Style$\dagger$; Motivational Drivers$\dagger$; Fear Map$\dagger$; Stress
Response$\dagger$; Type Inventory (e.g. MBTI, where known); Emotional
Regulation Strategy; Risk Disposition; Self-Efficacy Beliefs \\
\addlinespace
\textbf{\shelltwo{}} \newline cognitive and experiential &
months &
Formative Experiences$\dagger$; Difficult Episodes$\dagger$; Learning
Style$\dagger$; Cognitive Style$\dagger$; Life Narrative$\dagger$;
Knowledge Base$\dagger$; Domain Expertise; Skill Inventory; Memory
Salience; Attention Profile \\
\addlinespace
\textbf{\shellthree{}} \newline behavioural patterns &
weeks &
Daily Habits$\dagger$; Communication Patterns$\dagger$; Decision
Behaviour$\dagger$; Coping Mechanisms$\dagger$; Relationship
Behaviour$\dagger$; Work Habits$\dagger$; Conflict Behaviour;
Time and Planning Behaviour; Consumption Patterns; Information Seeking
Behaviour \\
\addlinespace
\textbf{\shellfour{}} \newline social and contextual &
weeks &
Social Roles$\dagger$; Cultural Conditioning$\dagger$; Digital
Identity$\dagger$; Reputation$\dagger$; Social Mask$\dagger$;
Register Range$\dagger$; Physiological Traits; Language Repertoire;
Network Position; Situational Constraints \\
\addlinespace
\textbf{CrossShell} \newline integration &
months &
Internal Conflicts$\dagger$; Authenticity Index$\dagger$; Growth
Trajectory; Shell Alignment Map; Change Readiness; Contradiction Log;
Self-Other Divergence; Stability Estimate; Confidence Annotation;
Provenance Record \\
\bottomrule
\end{tabular}
\end{table}

Four properties of the schema are worth restating in one place.

\begin{enumerate}
\item Every field value is a short first-person natural-language
  sentence, not a score or a vector.
\item The $(\mathrm{shell}, \mathrm{name})$ pairs are fixed, so two
  users' models are structurally comparable and $\sigma(t)$ is
  definable.
\item Update period $\tau$ is a tier property, so a refresh policy can
  be written without consulting field values.
\item The CrossShell tier is not a sixth shell but a set of relations
  over the other five. \emph{Internal Conflicts} and
  \emph{Contradiction Log} record where a \shellthree{} behaviour
  contradicts a Nucleus value; \emph{Authenticity Index} and
  \emph{Shell Alignment Map} measure agreement between what the inner
  shells hold and what the outer shells present;
  \emph{Confidence Annotation} and \emph{Provenance Record} track how a
  field was populated and how much weight it should carry.
\end{enumerate}

\section{Prompts}
\label{app:prompts}

Every prompt used in the simulation study is reproduced here verbatim.
Placeholders in braces are substituted at run time.

\subsection{Persona conditioning block}

Prepended as the system message for every generation the simulated
participant performs.

\begin{quote}\ttfamily\footnotesize\raggedright
You are role-playing a specific person. Stay in character at all times.\\
Personality (Big Five z-scores, -2 low to +2 high): \{traits\}\\
Life context: \{context\}\\
Formative background: \{narrative\}\\
How you actually write: \{idiolect\}\\
Write the way this person writes, including its rough edges. Do not
describe the persona; be it.
\end{quote}

\subsection{Instrument population}

\begin{quote}\ttfamily\footnotesize\raggedright
You are completing a guided self-report instrument called the Atomic
User Model.\\[4pt]
Fill in EVERY field below with one short first-person sentence (roughly
10 to 25 words) describing yourself truthfully in character. Be specific
and concrete; avoid generic self-description that could apply to
anyone.\\[4pt]
The JSON MUST contain all six shells (Nucleus, Shell1\_Psychological,
Shell2\_Cognitive, Shell3\_Behavioural, Shell4\_Social, CrossShell) and
all 32 fields as non-empty strings. Do not leave any shell empty. Do not
nest extra objects inside a field; each field value is a single sentence
string.\\[4pt]
Return ONLY a valid JSON object with exactly this structure and these
keys: \{schema\}\\[4pt]
No commentary, no markdown fences.
\end{quote}

\noindent On a failed parse the following corrective is appended and the
call retried, up to six attempts:

\begin{quote}\ttfamily\footnotesize\raggedright
Your previous JSON filled too few fields. Fill ALL 32 fields in ALL six
shells. Every value must be a non-empty first-person sentence. Empty
shells are not acceptable.
\end{quote}

\subsection{Prompt surface}

The only user-derived text the retriever ever sees. Generated with the
persona block plus the participant's completed instrument as system
context.

\begin{quote}\ttfamily\footnotesize\raggedright
You are typing a request to an AI assistant. Write only the message you
would type, in your own voice, with your own level of detail and
formality. Do not write the assistant's answer.\\[4pt]
What you want help with: \{task\}\\[4pt]
Keep it under 60 words. Output the message only.
\end{quote}

\subsection{Reference text}

The ground truth. Generated without sight of the prompt surface.

\begin{quote}\ttfamily\footnotesize\raggedright
Write your own response to the task below, exactly as you would actually
write it yourself, with no assistant helping you. Match your usual
length, structure, formality and habits.\\[4pt]
Task: \{task\}\\[4pt]
Output the text only, with no preamble.
\end{quote}

\subsection{Prior conversations for the preference-notes baseline}

Four transcripts per participant, on topics held out from the evaluation
battery.

\begin{quote}\ttfamily\footnotesize\raggedright
Write a short transcript of a past conversation you had with an AI
assistant, four to six turns, labelled 'User:' and 'Assistant:'. The
topic was: \{topic\}. Write your own turns in your own voice. Let your
preferences and habits show through naturally rather than stating them
outright.
\end{quote}

\noindent The six held-out topics are: choosing between two laptops for
everyday work; sorting out a recurring problem with a household
appliance; putting together a week of simple meals; deciding what to do
with a free Saturday; picking a book to read next; working out how to
reply to a slightly awkward group message.

\subsection{Preference-note distillation}

Run by a separate summariser with no access to the persona or the
\aum{}.

\begin{quote}\ttfamily\footnotesize\raggedright
\textbf{System:} You are the memory component of an AI assistant. You
read past conversations and store durable user preferences as a flat
list of short statements, the way a production memory feature does. You
have no other information about the user.\\[6pt]
\textbf{User:} From the transcripts below, extract the durable
preferences worth remembering about this user. Write at most 12 short
statements, one per line, each starting with '- '. Record preferences
and habits only, not events. No headings, no commentary.\\[4pt]
\{transcripts\}
\end{quote}

\subsection{Task classifier}

\begin{quote}\ttfamily\footnotesize\raggedright
\textbf{System:} You route requests for a personalised assistant. A
request is 'simple' if a correct answer is the same regardless of who
asked (a factual lookup, a unit conversion, an arithmetic result). It is
'complex' if the form of an acceptable answer depends on who is asking.
Answer with JSON only.\\[6pt]
\textbf{User:} Classify this user request.\\[4pt]
Request: \{prompt\_surface\}\\[4pt]
Return JSON: \{"complexity": "simple" or "complex", "task\_type": one of
[communication, self\_presentation, relational\_evaluative,
planning\_decision, explanatory, creative, factual]\}
\end{quote}

\subsection{Generation}

\begin{quote}\ttfamily\footnotesize\raggedright
\textbf{System:} You are a helpful AI assistant. Complete the user's
task and return only the finished text they asked for, with no preamble,
no explanation of your choices, and no options.\\[6pt]
\textbf{User:} \{context\_block\}User request:\\
\{prompt\_surface\}
\end{quote}

\noindent The context block is empty for the generic condition, and
otherwise a delimited section:
\texttt{[Selected user model fields]} followed by one bulleted field per
line and a closing tag, or \texttt{[Remembered user preferences]} for
the preference-notes condition, or \texttt{[User model]} followed by the
full instrument as JSON for the Full \aum{} condition. Context tokens
are counted on exactly this string.

\subsection{Fidelity judge}

\begin{quote}\ttfamily\footnotesize\raggedright
\textbf{System:} You are judging, on behalf of a specific person,
whether a piece of text reads like something that person would have
written. You are given that person's self-description and one text they
genuinely wrote. Judge style, voice, length, structure and habits, not
whether the text is good. Answer with a single integer and nothing
else.\\[6pt]
\textbf{User:} The person:\\
- Life context: \{context\}\\
- How they write: \{idiolect\}\\[4pt]
A text this person genuinely wrote for the task "\{task\}":\\
--- \{reference\} ---\\[4pt]
Now rate this other text for the same task on a 1 to 5 scale, where 5
means 'this reads exactly like something they would have written' and 1
means 'this does not read like them at all'.\\
--- \{candidate\} ---\\[4pt]
Return ONLY a single integer from 1 to 5.
\end{quote}

\noindent The judge never receives the \aum{}. Only the persona's life
context and idiolect line are shown, plus the participant's own
reference.

\subsection{Identification judge}

\begin{quote}\ttfamily\footnotesize\raggedright
\textbf{System:} You are shown one text a specific person genuinely
wrote, and four candidate texts for the same task. Exactly one candidate
was produced by a system that had access to a model of that person.
Choose the candidate whose style most closely matches the person's own
writing. Answer with a single letter.\\[6pt]
\textbf{User:} The person:\\
- Life context: \{context\}\\
- How they write: \{idiolect\}\\[4pt]
A text this person genuinely wrote for the task "\{task\}":\\
--- \{reference\} ---\\[4pt]
Candidates:\\
\{A, B, C, D in a per-trial shuffled order\}\\[4pt]
Return ONLY the letter of the candidate that best matches this person's
own style.
\end{quote}

\section{The Simulated Participants}
\label{app:personas}

The sixteen personas are declared in code rather than sampled, so the
population is fixed across seeds and auditable. Each carries Big Five
$z$-scores, a life context, a formative-experience seed and idiolect
markers. Table~\ref{tab:personas} in the body gives the abridged form;
Table~\ref{tab:personasfull} gives the life context and formative
background in full.

\begin{longtable}{@{}p{0.16\textwidth} p{0.38\textwidth} p{0.40\textwidth}@{}}
\caption{Life context and formative background for each simulated
  participant.}
\label{tab:personasfull} \\
\toprule
\textbf{Label} & \textbf{Life context} & \textbf{Formative background} \\
\midrule
\endfirsthead
\toprule
\textbf{Label} & \textbf{Life context} & \textbf{Formative background} \\
\midrule
\endhead
\bottomrule
\endfoot
meticulous planner & Operations lead at a mid-size logistics firm; runs everything from checklists. & A shipment error early in their career cost the company a client; they have double-checked everything since. \\
warm over-explainer & Secondary school teacher; the person colleagues go to when something needs smoothing over. & Grew up as the eldest of four and became the household mediator before they were twelve. \\
blunt minimalist & Backend engineer; treats most meetings as an unpriced tax on the day. & Learned to code alone at fourteen from documentation and has never quite unlearned the register. \\
associative creative & Freelance illustrator; several projects always half-open at once. & Dropped out of an economics degree to go to art school and treats that as the decision that defines them. \\
analytical flat & Quantitative analyst; distrusts any claim without an interval attached. & A badly designed undergraduate experiment taught them how easy it is to fool yourself with data. \\
bright energetic & Community manager for a hobbyist platform; genuinely likes the people they work with. & Found their footing running a fan forum as a teenager and never lost the habit of talking to a crowd. \\
dry sceptic & Investigative reporter turned freelance editor; assumes the press release is lying. & Watched a story they had championed fall apart under fact-checking and now leads with the caveat. \\
formal correct & Civil servant in a planning department; twenty years of writing that must survive scrutiny. & Trained in an office where a misplaced comma in a notice could invalidate it. \\
scattered fast & Startup founder wearing four hats; replies from three time zones in one week. & Left a stable job on two weeks' notice and has been running on that momentum for three years. \\
nurturing relational & Paediatric nurse; the emotional temperature of a room is the first thing they read. & A long stretch caring for a grandparent taught them that people mostly want to be heard first. \\
competitive driver & Sales director; keeps a personal scoreboard nobody asked to see. & Came up through a commission-only role and still measures the week in closed deals. \\
reflective introvert & Archivist; the working day is mostly quiet and they prefer it that way. & Spent a year abroad where they spoke the language badly and learned to say less, better. \\
storytelling extrovert & Corporate trainer; cannot make a point without an anecdote attached. & Grew up in a family where dinner was competitive storytelling and brevity lost. \\
pragmatic efficient & Site foreman; the day is measured in what got finished. & Twenty years on sites where the person who over-explained held everyone else up. \\
quirky metaphorical & Research librarian with a sideline in cryptic crosswords. & Was the child who read the encyclopaedia for fun and has never found a better hobby. \\
stoic calm & Emergency dispatcher; twelve-hour shifts where panic is the one useless response. & Learned early that the calmest voice on the line is the one that gets things done. \\
\end{longtable}

\section{Additional Statistical Detail}
\label{app:stats}

\subsection{Judge reliability across seeds}

\begin{table}[!ht]
\centering
\caption{Agreement between the fidelity judge's ratings across seeds,
  over 960 (condition, participant, task) targets. ICC(2,1) over all
  three seeds is 0.573.}
\label{tab:judgeagree}
\small
\renewcommand{\arraystretch}{1.2}
\begin{tabular}{@{}c c r r r r@{}}
\toprule
\textbf{Seed A} & \textbf{Seed B} & \textbf{$n$} &
\textbf{Exact} & \textbf{Within 1} & \textbf{Spearman $\rho$} \\
\midrule
0 & 1 & 960 & 43.1\% & 87.4\% & 0.589 \\
0 & 2 & 960 & 42.0\% & 85.6\% & 0.546 \\
1 & 2 & 960 & 43.2\% & 86.9\% & 0.599 \\
\bottomrule
\end{tabular}
\end{table}

\subsection{Use of the rating scale, and position of the chosen option}

\begin{table}[!ht]
\centering
\caption{Left: distribution of fidelity ratings over all 2{,}880
  trials. Right: share of identification trials in which the judge chose
  the option in each slot, against the uniform 25\% expectation, with
  Wilson intervals. A chi-square test over slots gives
  $\chi^2 = 7.37$, 3 d.f., $p = 0.061$.}
\label{tab:judgedist}
\small
\renewcommand{\arraystretch}{1.2}
\begin{tabular}{@{}c r @{\hspace{2cm}} c r r c@{}}
\toprule
\textbf{Rating} & \textbf{\% of trials} &
\textbf{Slot} & \textbf{Chosen} & \textbf{\%} & \textbf{95\% CI} \\
\midrule
1 & 13.8\% & A & \phantom{0}66 & 22.9\% & [18.4, 28.1] \\
2 & 25.2\% & B & \phantom{0}84 & 29.2\% & [24.2, 34.7] \\
3 & 25.8\% & C & \phantom{0}79 & 27.4\% & [22.6, 32.9] \\
4 & 30.9\% & D & \phantom{0}59 & 20.5\% & [16.2, 25.5] \\
5 & \phantom{0}4.3\% & & & & \\
\bottomrule
\end{tabular}
\end{table}

\clearpage
\subsection{Payload composition by task type}

\begin{table}[!ht]
\centering
\caption{Share of the retrieved payload drawn from each shell, by task
  type, over all \aumr{} trials. Rows sum to 100\% up to rounding. This
  is the tabular form of Figure~\ref{fig:shellusage}.}
\label{tab:shellusage}
\small
\renewcommand{\arraystretch}{1.2}
\begin{tabular}{@{}l r r r r r r@{}}
\toprule
\textbf{Task type} & \textbf{Nucleus} & \textbf{\shellone{}} &
\textbf{\shelltwo{}} & \textbf{\shellthree{}} & \textbf{\shellfour{}} &
\textbf{Cross} \\
\midrule
communication         & \phantom{0}0.3 & \phantom{0}0.5 & \phantom{0}0.3 & 44.4 & 42.3 & 12.2 \\
self-presentation     & 37.2 & 11.5 & \phantom{0}1.0 & \phantom{0}1.3 & 41.2 & \phantom{0}7.8 \\
relational-evaluative & 24.7 & 13.3 & \phantom{0}0.0 & 45.6 & \phantom{0}1.6 & 14.8 \\
planning-decision     & \phantom{0}0.0 & 21.3 & 21.6 & 49.5 & \phantom{0}0.0 & \phantom{0}7.6 \\
explanatory           & \phantom{0}0.0 & \phantom{0}0.0 & 30.0 & 40.1 & 17.5 & 12.5 \\
\bottomrule
\end{tabular}
\end{table}

\subsection{Optimality of each retriever where exhaustive search is tractable}

\begin{table}[!ht]
\centering
\caption{Share of instances in which each method returned the exhaustive
  optimum, over the twelve cells of the scaling sweep where
  $\binom{n}{k} \le 60{,}000$ and exhaustive search was therefore
  computed. 25 instances per cell.}
\label{tab:optimality}
\small
\renewcommand{\arraystretch}{1.2}
\begin{tabular}{@{}l r@{}}
\toprule
\textbf{Method} & \textbf{Exact optimum found} \\
\midrule
AFSA (greedy warm start) & 96.3\% \\
AFSA (scaled budget)     & 84.7\% \\
Forward greedy           & 77.7\% \\
AFSA (study budget)      & 46.3\% \\
Dense top-$k$            & 41.3\% \\
Random                   & \phantom{0}0.0\% \\
\bottomrule
\end{tabular}
\end{table}

\end{document}